\documentclass[aps,prl,twocolumn,superscriptaddress,a4paper]{revtex4-1}
\usepackage[dvipdfmx]{graphicx}

\usepackage[normalem]{ulem}

\usepackage{amsmath,amsthm,amssymb}
\usepackage{amsmath,bm}
\usepackage{color}
\usepackage{ifthen}

\usepackage{multirow,bigdelim}

\newcount \commentout
\newcommand{\1}{\mbox{1}\hspace{-0.25em}\mbox{l}}

\newcommand{\cyan}[1]{{{#1}}}

\newcommand{\blue}[1]{{{#1}}}
\newcommand{\magenta}[1]{{{#1}}}
\newcommand{\green}[1]{{{#1}}}

\newlength{\figwidth}
\newlength{\figlarge}
\begin{document}
\title{
Mirror skin effect and its electric circuit simulation
}

\author{Tsuneya Yoshida} 
\author{Tomonari Mizoguchi} 
\author{Yasuhiro Hatsugai} 
\affiliation{Department of Physics, University of Tsukuba, Ibaraki 305-8571, Japan}
\date{\today}
\begin{abstract}
We analyze impacts of crystalline symmetry on the non-Hermitian skin effects. Focusing on mirror symmetry, we propose a novel type of skin effects, a mirror skin effect, which results in significant dependence of energy spectrum on the boundary condition only for the mirror invariant line in the two-dimensional Brillouin zone. This effect arises from the topological properties characterized by a mirror winding number.
We further reveal that the mirror skin effect can be observed for an electric circuit composed of negative impedance converters with current inversion where switching the boundary condition significantly changes the admittance eigenvalues only along the mirror invariant lines.
Furthermore, we demonstrate that extensive localization of the eigenstates for each mirror sector result\magenta{s} in an anomalous voltage response.
\end{abstract}
\maketitle

\textit{Introduction.--}
Topological properties~\cite{TI_review_Hasan10,TI_review_Qi10,Hatsugai_PRL93} of the systems have become a central issue in condensed matter systems because of their remarkable ubiquity. The topological phenomena can be observed even for classical systems (e.g., photonic systems~\cite{Haldane_chiralPHC_PRL08,Raghu_chiralPHC_PRA08,Wang_chiralPHC_Nature09}, mechanical systems~\cite{Kane_NatPhys13,Kariyado_SR15,Susstrunk_TopoMech_Sci15,Huber_TopoMech_NatPhys16,Delplace_topoEq_Science17}, electric circuits~\cite{Albert_Topoelecircit_PRL15,Lee_Topoelecircit_CommPhys18} etc.), which are mathematically described by an eigenvalue problem. 
Among the extensive studies of the topological physics, one of the significant progresses is the discovery of the topological crystalline insulators~\cite{JTeo_PRB08,Fu_TCI_PRL2011,Hsieh_TCH_SnTe_2012,Tanaka_TCI_SnTe2012} which has elucidated that the topological properties can be enriched by the crystalline symmetry~\blue{\cite{Hashimoto_HOTI_PRB17,Benalcazar_HOTI_Science17,Schindler_HOTI_Science18,Benalcazar_HOTI_PRB17,Hayashi_HOTI_PRB18,Imhof_HOTI_NatPhys18,Araki_HOTI_PRB19,Ghorashi_HOTI_PRB19,Kudo_HOTMI_PRL19,Mizoguchi_HOTI_PRM19}}.
A prime example of the topological crystalline insulators is SnTe~\cite{Hsieh_TCH_SnTe_2012,Tanaka_TCI_SnTe2012} where the mirror Chern number topologically protects two surface Dirac cones. 

Along with the above significant progresses, non-Hermitian topological systems have been extensively studied, which has discovered a variety of novel phenomena~\cite{Hatano_PRL96,Hu_nH_PRB11,Esaki_nH_PRB11,Guo_nHExp_PRL09,Ruter_nHExp_NatPhys10,Regensburger_nHExp_Nat12,Zhen_AcciEP_Nat15,Hassan_EP_PRL17,TELeePRL16_Half_quantized,YXuPRL17_exceptional_ring,Yoshida_nHFQH19,VKozii_nH_arXiv17,Yoshida_EP_DMFT_PRB18,Kimura_SPERs_PRB19,Zyuzin_nHEP_PRB18,Papaji_nHEP_PRB19,Matsushita_ER_arXiv19,Yoshida_nHRev_arXiv20}. 
An important difference of non-Hermitian topological systems is that there exist two types of gaps~\cite{Kawabata_gapped_PRX19}, a line gap~\cite{HShen2017_non-Hermi,KKawabata_TopoUni_NatComm19} and a point gap~\cite{Gong_class_PRX18,Zhou_gapped_class_PRB19}. 
The line gap topology indicates the presence of the Hermitian counterpart.
The point gap topology protects \cyan{the non-Hermitian band touching} in the bulk~\cite{TKato_EP_book1966,Rotter_EP_JPA09,Budich_SPERs_PRB19,Okugawa_SPERs_PRB19,Zhou_SPERs_Optica19,Yoshida_SPERs_PRB19,Kawabata_gapless_PRL19,Yoshida_SPERs_mech19,Carlstrom_nHknot_PRA18}, \cyan{such as exceptional points etc.}
Other unique phenomena induced by the non-Hermiticity\cyan{, on which we focus in this paper,} can be observed for the system with boundaries~\blue{\cite{Alvarez_nHSkin_PRB18,KFlore_nHSkin_PRL18,SYao_nHSkin-1D_PRL18,SYao_nHSkin-2D_PRL18,EElizabet_PRBnHSkinHOTI_PRB19,Rui_nH_PRB19,Yokomizo_BBC_PRL19,Okuma_nHBBCpg_PRL19,Xiao_nHSkin_Exp_arXiv19}}.
\cyan{In particular, it has been elucidated that the non-Hermitian skin effect is induced by the nontrivial point gap topology}; the winding number characterizes the non-Hermitian skin effect of class A (no symmetry)~\cite{Lee_Skin19,Zhang_BECskin19,Okuma_BECskin19}.
The mathematically rigorous proof of the above relation has been obtained in Ref.~\onlinecite{Okuma_BECskin19} for class A. In addition, the $\mathbb{Z}_2$ skin effect with time-reversal symmetry has been proposed~\cite{Okuma_BECskin19}.

The above progresses for Hermitian and non-Hermitian systems lead us to the following issue\blue{:} understanding impacts of crystalline symmetry on non-Hermitian topological properties.
\blue{This issue} is crucial because a variety of \blue{novel topological phenomena} are expected as \blue{are the cases} for Hermitian systems.
\blue{
In spite of its significance, the above issue has not been sufficiently explored.
In particular, there are few works elucidating novel skin effects induced by the crystalline topology in the bulk.
}

Therefore, in this \blue{paper}, we analyze effects of mirror symmetry on non-Hermitian skin effects, shedding new light on the interplay between crystalline symmetry and non-Hermitian topology. 
Our analysis discovers a novel type of skin effects, a mirror skin effect\magenta{,} which results in significant dependence of energy spectrum on the boundary condition only along mirror invariant lines in the two-dimensional Brillouin zone. 
\blue{
Here, by energy spectrum, we denote spectrum of the non-Hermitian Hamiltonian.
}
We also elucidate that a mirror winding number characterizes this skin effect. 
We verify the mirror skin effect by numerically diagonalizing a tight-binding model with the mirror winding number taking one. 
\blue{
We note that the mirror skin effect cannot be observed in the absence of mirror symmetry, meaning that mirror symmetry is essential.
}
\begin{figure}[!h]
\begin{minipage}{1\hsize}
\begin{center}
\includegraphics[width=1\hsize,clip]{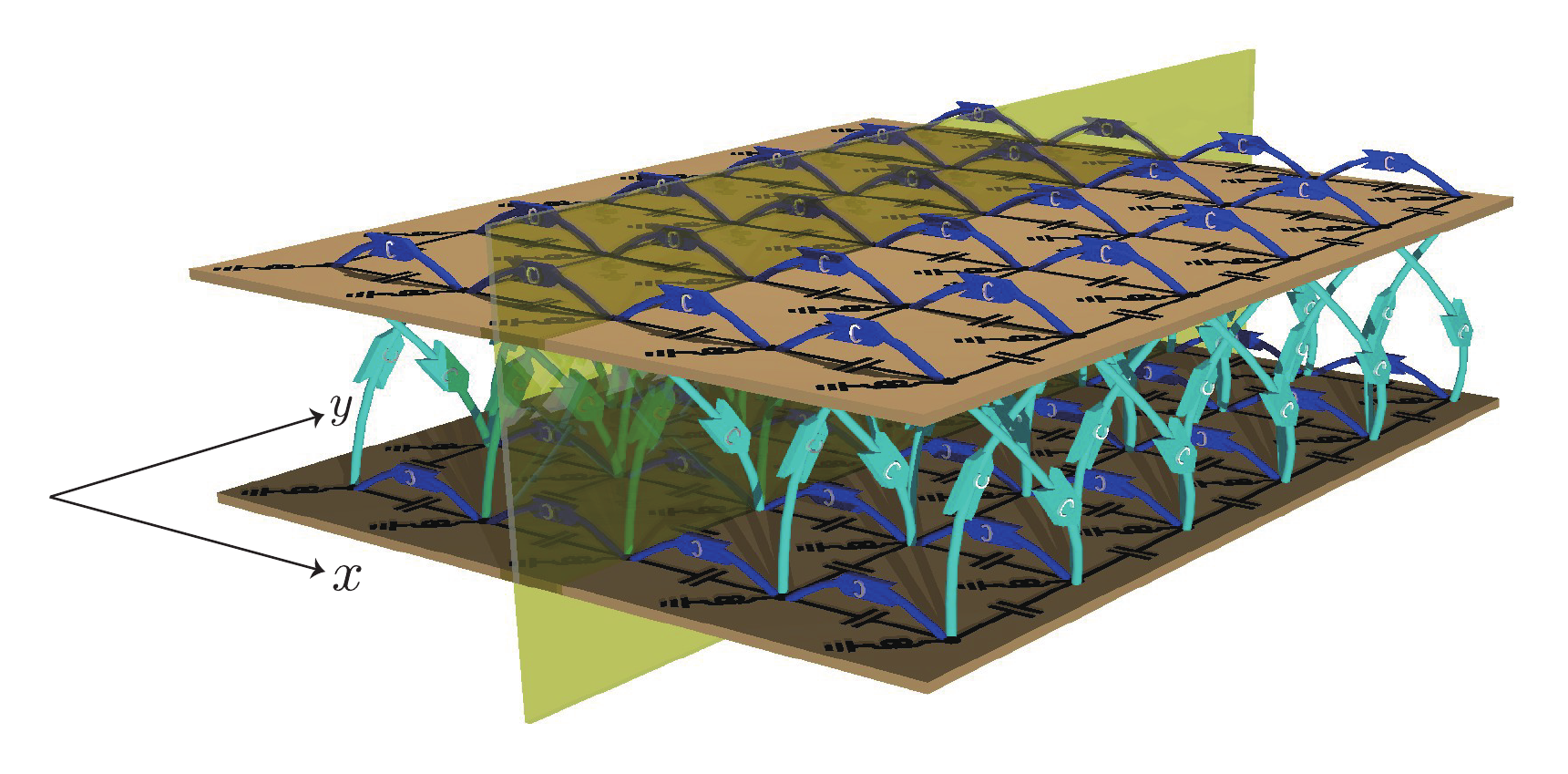}
\end{center}
\end{minipage}
\caption{(Color Online).
Sketch of the electric circuit showing the mirror skin effect. The blue (cyan) symbols represent negative impedance converters with current inversion connecting intra-layer (inter-layer) nodes, respectively. 
We refer to upper (lower) layer as A (B), respectively.
The black symbols denotes capacitors. Dots represent nodes. This system preserves the mirror symmetry whose mirror plane is illustrated with the vertical yellow plane.
}
\label{fig: Emodel}
\end{figure}
Furthermore, by making use of the ubiquity of the topological phenomenon, we theoretically suggest that the mirror skin effect can be observed for an electric circuit composed of negative impedance converters with current inversion (see Fig.~\ref{fig: Emodel}). In this system, switching the boundary conditions drastically changes the impedance for the mirror invariant lines, which serves as a distinct evidence of the mirror skin effect for the electric circuit.

\textit{
Theory of mirror skin effect.--
}
Let us first elucidate that the topology protected by mirror symmetry induces a novel skin effect.

For comparison, we start with a brief review of the ordinary skin effect for symmetry class A. Consider a two-dimensional system under the periodic boundary condition for the $x$-direction which can be regarded as a set of one-dimensional systems aligned along the $x$-direction in the momentum space. 
When the winding number $\nu_\mathrm{tot}(k_x)$ takes a finite value for the subsystem with a given momentum $k_x$, the skin effect occurs; the energy spectrum significantly changes by switching the boundary condition for the $y$-direction, [i.e., the periodic boundary condition (PBC) to the open boundary condition (OBC)]. This relation can be understood with topological deformation; each subsystem for given momentum $k_x$ is topologically deformed into the Hatano-Nelson model \blue{[or its variant showing a higher value of winding number $\nu_\mathrm{tot}(k_x)$]} \blue{exhibiting} the skin effect. 
The above fact indicates that the ordinary skin effect of class A is induced by the winding number $\nu_\mathrm{tot}(k_x)$. (For later use, we call it total winding number.)

\blue{Now, we elucidate a novel skin effect whose topological properties are protected by mirror symmetry;} in contrast to the ordinary skin effect mentioned above, the mirror skin effect elucidated below occurs even when the total winding number is zero for arbitrary momenta $k_x$. 
\blue{We call this skin effect mirror skin effect.}
In the rest of this paper, we assume $\nu_\mathrm{tot}(k_x)=0$ unless otherwise stated.

Firstly, we note that the presence of mirror symmetry results in an additional topological invariant. 
Consider the Hamiltonian which is invariant under applying the mirror operator $M_x$;
\begin{subequations}
\begin{eqnarray}
M_x H(\bm{k}) M^{-1}_x &=& H(M_x\bm{k}),
\end{eqnarray}
\begin{eqnarray}
M_x &=&U_mP_x,
\end{eqnarray}
\end{subequations}
where $P_x$ flips the momentum $\bm{k}:=(k_x,k_y)\to M_x\bm{k}:=(-k_x,k_y)$. $U_m$ is an unitary matrix satisfying $U^2_m=1$.
Along the mirror invariant line specified by $k^*_x$, the Hamiltonian can be block-diagonalized for the plus and the minus sectors of the operator $M_x$.
Thus, besides the total winding number $\nu_\mathrm{tot}$, the following mirror winding number can be defined
\begin{eqnarray}
\nu_M &=&(\nu_+-\nu_-)/2.
\end{eqnarray}
Here, $\nu_\pm(k^*_x)$ denotes the winding number computed with the block-diagonalized Hamiltonian $H_{\pm}(k^*_x,k_y)$ for each sector
\begin{eqnarray}
\label{eq: def vpm}
\nu_\pm(k^*_x)
&=& 
\int \frac{dk_y}{2\pi i} \partial_{k_y} \log \mathrm{det} [H_{\pm}(k^*_x,k_y)-E_{\mathrm{pg}}],
\end{eqnarray}
where $E_{\mathrm{pg}}$ is the reference energy for the point gap~\cite{Liu_MirrClassifi_PRB19,Rclassfi_footnote}. 
We note that the total winding number is computed with $\nu_\mathrm{tot}(k^*_x) =\nu_+(k^*_x)+\nu_-(k^*_x)$ for the mirror invariant lines.
\blue{
For a nontrivial value of the winding number $\nu_\pm(k^*_x)$, the non-Hermiticity is essential.
}

The mirror winding number taking a nontrivial value results in a \blue{novel} skin effect, \blue{which is a main result of this paper}; in spite of $\nu_{\mathrm{tot}}=0$, the energy eigenvalues significantly depend on the boundary condition for the mirror invariant line in the Brillouin zone.
\blue{
This skin effect should generically occur when the mirror winding number is finite because each subsector of the Hamiltonian corresponds to the one belonging to class A which shows the ordinary skin effect characterized by the winding number $\nu_{\pm}$.
}
\blue{
Here, we stress that the mirror symmetry protects the topology inducing the skin effect~\cite{Rui_Rskin_PRB19,Rskin_footnote}.
}

In the following, we verify that \blue{the nontrivial topology characterized by} the mirror winding number results in the above significant dependence by numerically analyzing a tight-binding model.
The Hamiltonian reads,
\begin{eqnarray}
\label{eq: toy Hami}
H(\bm{k})&=& [2t(\cos k_x +\cos k_y)-\mu] \rho_0+i\Delta \sin k_x \rho_3\nonumber \\
&& \quad +i\Delta \sin k_y \rho_2,
\end{eqnarray}
where $\rho_i$ ($i=1,2,3$) are the Pauli matrices and $\rho_0$ is the $2\times 2$ identity matrix.
\blue{
As discussed in Sec.~\ref{sec: rel yaoryu} of the supplemental material~\cite{supplemental}, this model is related to a Hermitian system~\cite{Yao_ZtoZ8_PRB13}.
}
\blue{
Furthermore, as discussed below, a system having the same topological properties of the above model can be realized for an electric circuit.
}
The above \blue{non-Hermitian} Hamiltonian preserves the mirror symmetry with $M_x=\rho_2P_x$.
Therefore, for $k^*_x=0$ or $\pi$, the Hamiltonian can be block-diagonalized with $\rho_2$. For $k^*_x=0$ ($k^*_x=\pi$), the mirror winding number takes $\nu_\mathrm{M}=1$ with $E_{\mathrm{pg}}=2t-\mu$ ($E_{\mathrm{pg}}=-2t-\mu$), while the total winding number is zero for the arbitrary value of $k_x$.

\begin{figure}[!h]
\begin{minipage}{0.49\hsize}
\begin{center}
\includegraphics[width=1\hsize,clip]{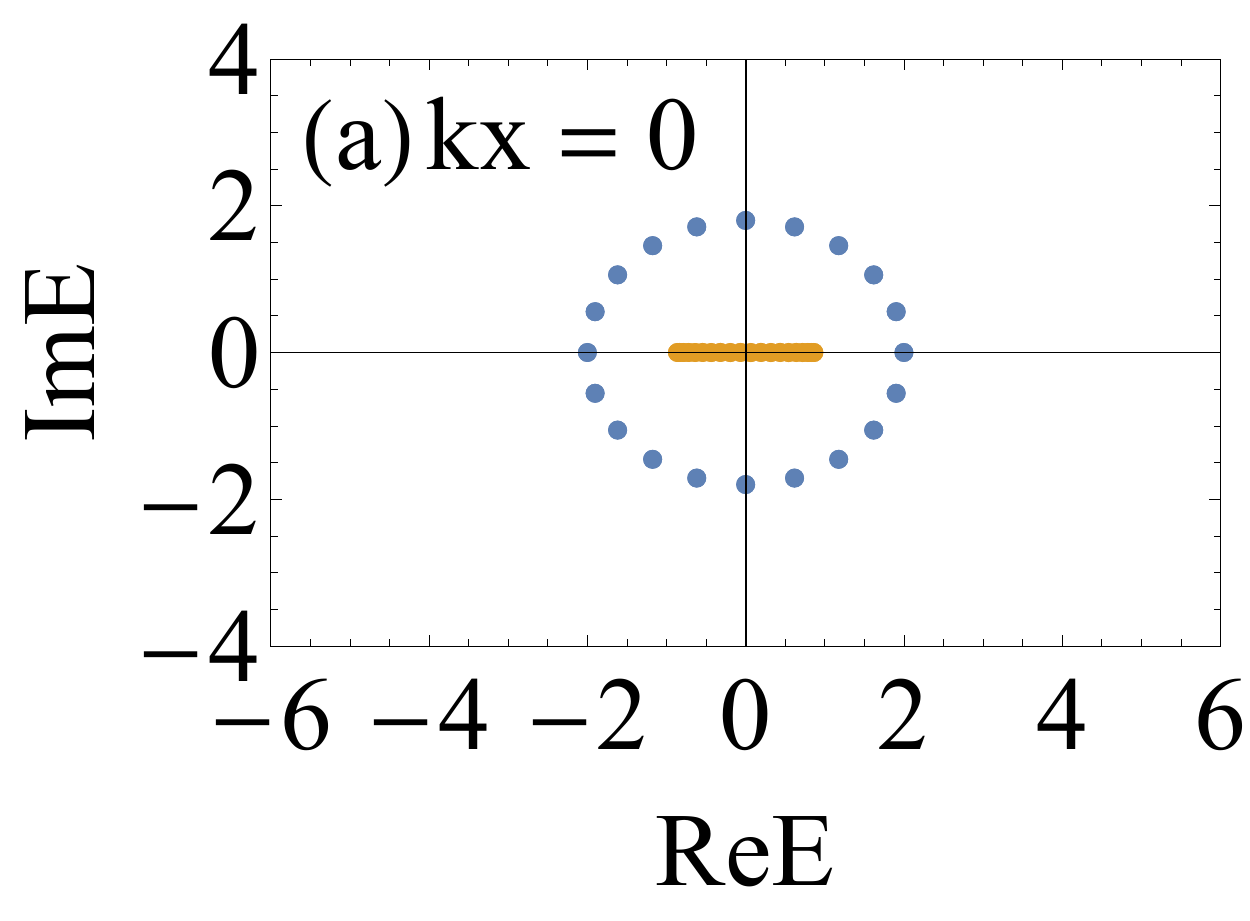}
\end{center}
\end{minipage}
\begin{minipage}{0.49\hsize}
\begin{center}
\includegraphics[width=1\hsize,clip]{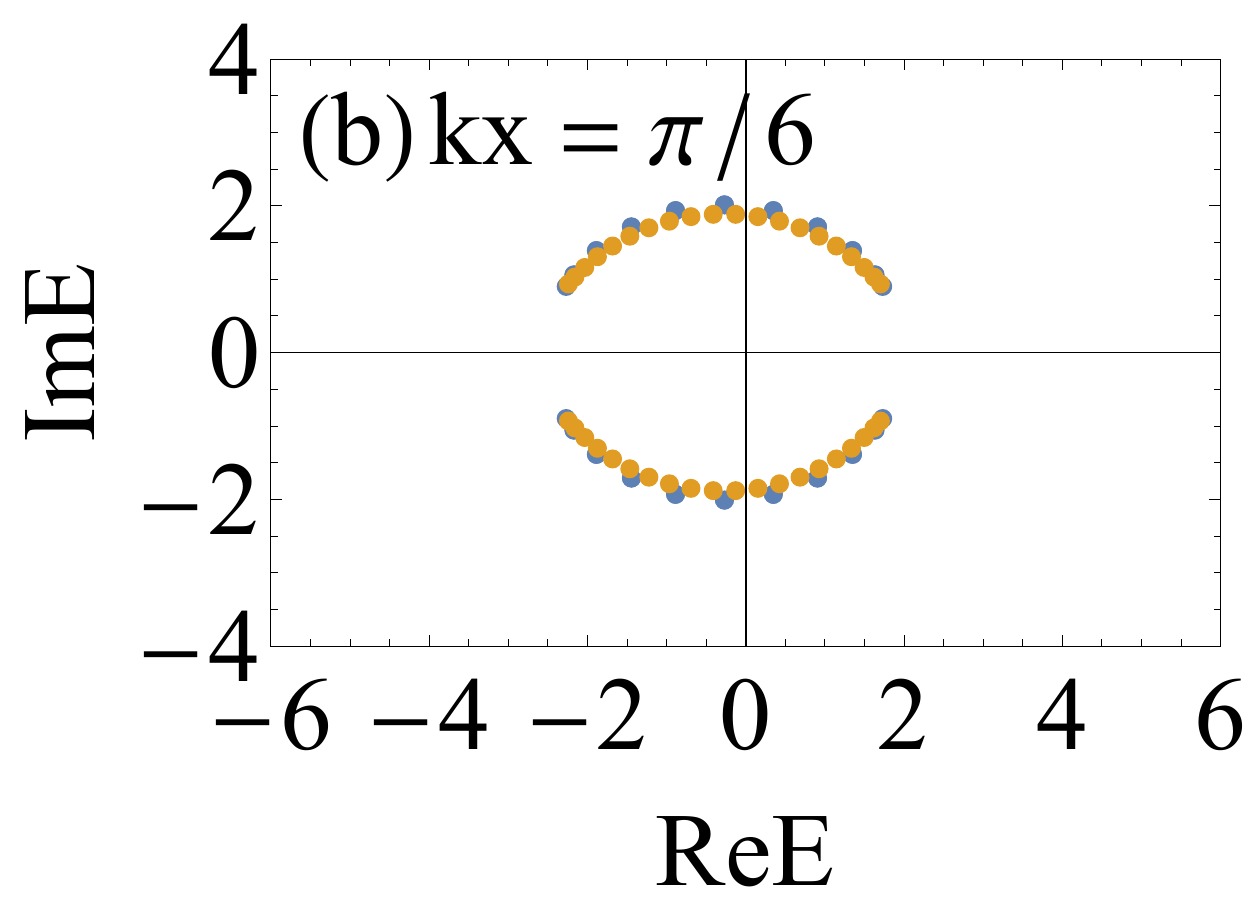}
\end{center}
\end{minipage}
\begin{minipage}{0.49\hsize}
\begin{center}
\includegraphics[width=1\hsize,clip]{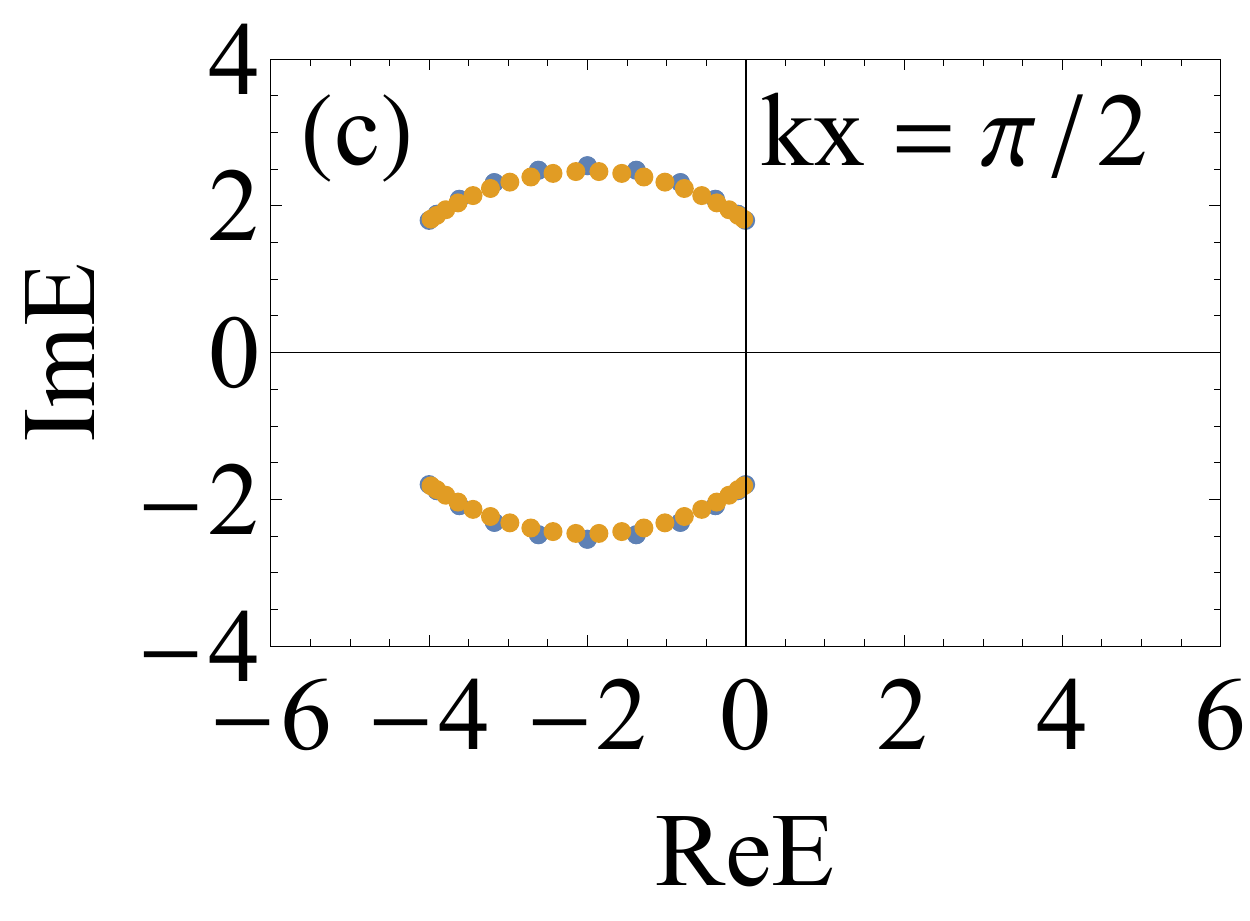}
\end{center}
\end{minipage}
\begin{minipage}{0.49\hsize}
\begin{center}
\includegraphics[width=1\hsize,clip]{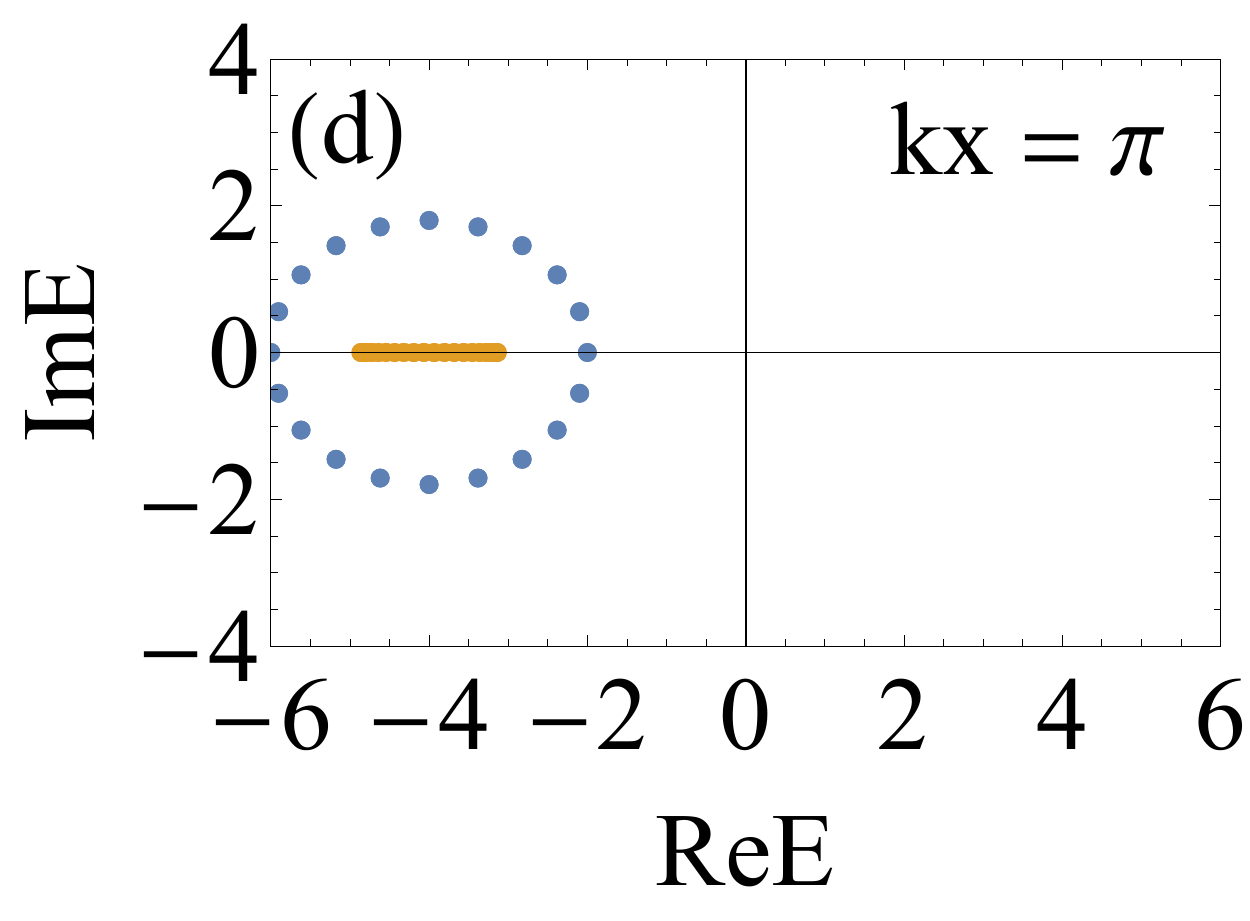}
\end{center}
\end{minipage}
\caption{(Color Online).
Energy spectrum of the Hamiltonian~(\ref{eq: toy Hami}) for $(t,\mu,\Delta)=(1,2,1.8)$.
Blue (orange) dots are data obtained under the (open) periodic boundary condition for the $y$-direction.
For the $x$-direction, the periodic boundary condition is imposed. The number of sites for the $y$-direction is set to $L_y=20$.
}
\label{fig: Engtoy}
\end{figure}

In Fig.~\ref{fig: Engtoy} the energy spectrum of the Hamiltonian~(\ref{eq: toy Hami}) \blue{is} plotted for $(t,\mu,\Delta)=(1,2,1.8)$ at $k_x=0,\pi/6,\pi/2,\pi$.
The data denoted with blue (orange) dots represent the energy eigenvalues for the PBC (OBC) along the $y$-direction, respectively.
Figure~\ref{fig: Engtoy}(a) indicates that the energy spectrum under the PBC form a circle enclosing the origin of the complex plane which is consistent with the relation, $\nu_{\mathrm{M}}=1$ for $k_x=0$.
Imposing the OBC along the $y$-direction significantly changes the spectrum; energy eigenvalues are aligned along the real axis (i.e., $\mathrm{Im} E_n\sim 0$ with $n=1,2, \cdots, \mathrm{dim}H$). This striking dependence of the energy spectrum is a signal of the skin effects.
Here, we note that the mirror symmetry plays an essential role; away from the mirror invariant line, the spectra obtained for the two distinct boundary conditions coincide with each other [see Figs.~\ref{fig: Engtoy}(b)~and~(c)]. At $k_x=\pi$, the mirror symmetry is preserved which again induces the skin effect [see Fig.~\ref{fig: Engtoy}(d)].

\begin{figure}[!h]
\begin{minipage}{0.49\hsize}
\begin{center}
\includegraphics[width=1\hsize,clip]{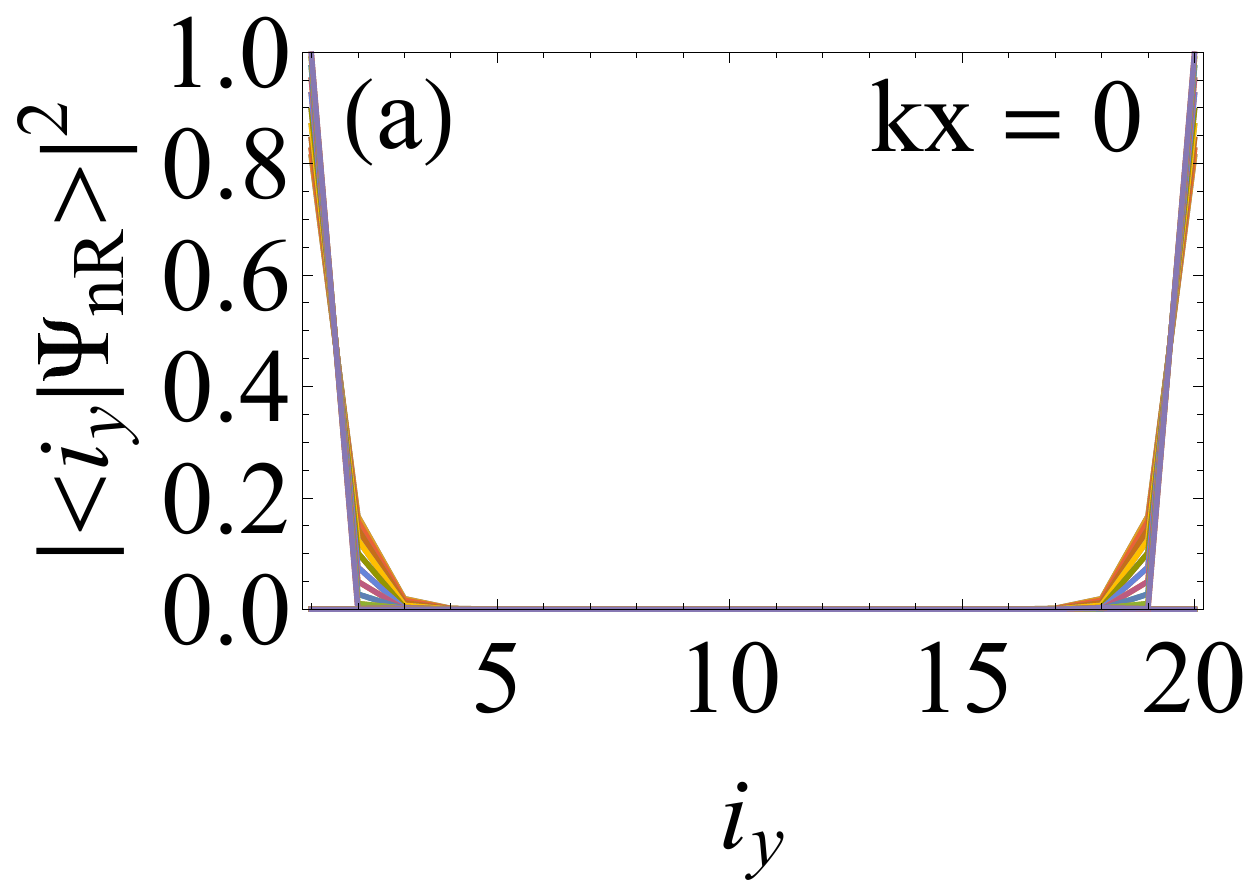}
\end{center}
\end{minipage}
\begin{minipage}{0.49\hsize}
\begin{center}
\includegraphics[width=1\hsize,clip]{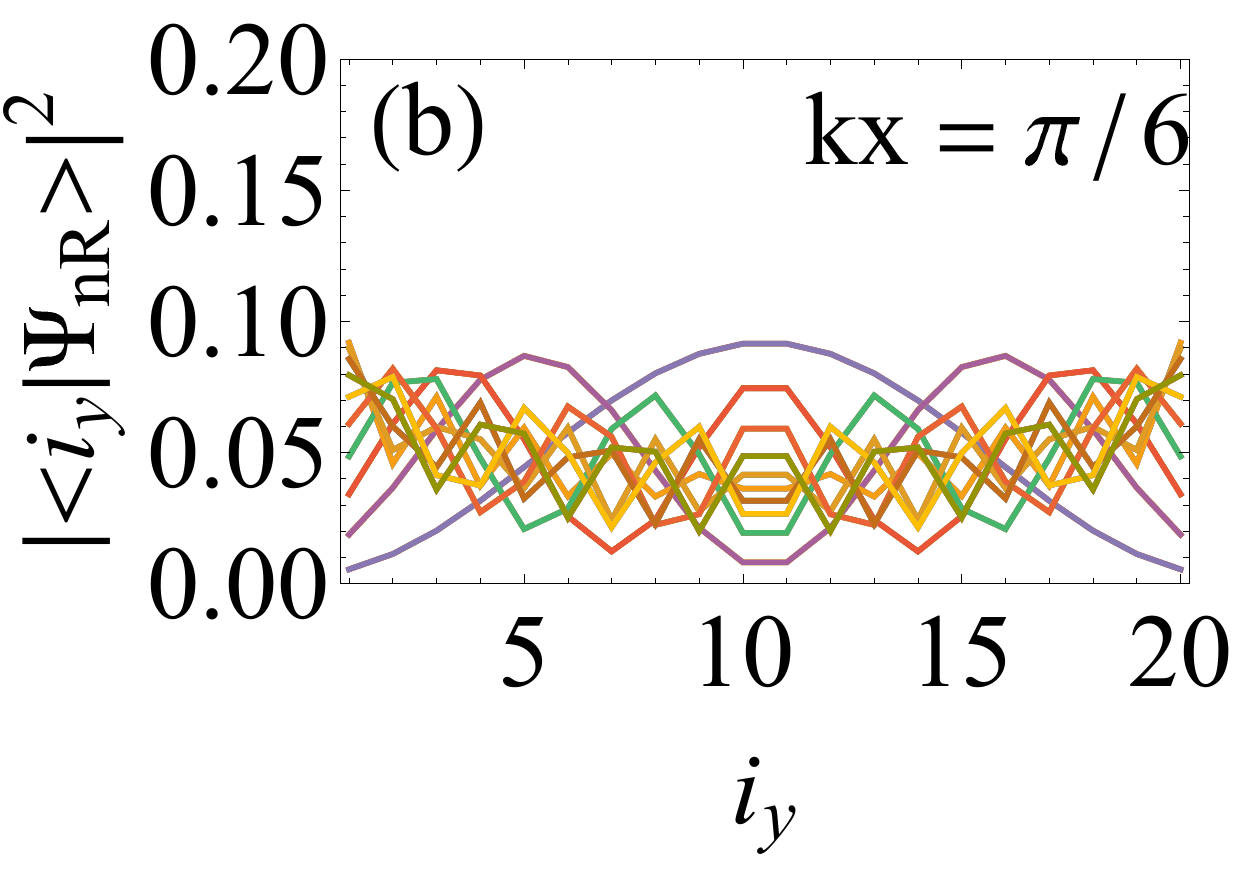}
\end{center}
\end{minipage}
\caption{(Color Online).
(a) [(b)]: Amplitude of the right eigenvectors of the Hamiltonian~(\ref{eq: toy Hami}) for $(t,\mu,\Delta)=(1,2,1.8)$ at $k_x=0$ $(\pi/6)$, respectively.
The data \blue{are} obtained under the periodic (open) boundary condition for the $x$- ($y$-) direction\blue{, respectively}. The number of the sites along the $y$-direction is set to $L_y=20$.
}
\label{fig: Amptoy_OBC}
\end{figure}

In association with the significant change of the energy eigenvalues, the eigenvectors shows extensive localization. Figure~\ref{fig: Amptoy_OBC} plots amplitude of the right eigenvectors $|\langle i_y |\Psi_{nR}\rangle |^2$ for $k_x=0, \pi/6$. Here, $|\Psi_{nR}\rangle $ denotes the right eigenvector of the Hamiltonian~(\ref{eq: toy Hami}) (i.e., $H|\Psi_{nR}\rangle =|\Psi_{nR}\rangle E_n$ with $n=1,\cdots,\mathrm{dim}H$), and $i_y$ labels the sites along the $y$-direction. We note that the eigenstates are extended in the bulk under the PBC along \blue{the} $y$-direction (see Sec.~\ref{sec: toy PBC app} in the supplemental material~\cite{supplemental}).
Imposing the OBC for \blue{the} $y$-direction results in extensive localization of eigenstates.
Figure~\ref{fig: Amptoy_OBC}(a) shows that at the mirror invariant line $k_x=0$, all of the eigenstates for the Hamiltonian $H_+(k_x=0)$ [$H_-(k_x=0)$] are localized around $i_y=0$ ($i_y=L_y$), respectively, while the states are extended under the PBC.
The anomalous localization under the OBC is another characteristic feature of the skin effect. 
We note that away from the mirror invariant line, the states localized around each edge are mixed and extend to the bulk [see Fig.~\ref{fig: Amptoy_OBC}(b)], which also indicates that the mirror symmetry is essential.

Based on the above significant difference of energy spectra and the eigenstates, we can conclude that a nontrivial value of the mirror winding number $\nu_\mathrm{M}$ results in the mirror skin effect.
\blue{
We note that adding a mirror-symmetry breaking term extinguishes the skin effect, which explicitly demonstrates that mirror symmetry protects the skin effect.
For further detailed numerical analysis, see Secs.~\ref{sec: toy PBC app}~and~\ref{sec: nu2 noM app} of the supplemental material~\cite{supplemental}.
}
We furthermore propose that this mirror skin effect can be observed for an electric circuit composed of negative impedance converters (see Fig.~\ref{fig: Emodel}), the details of which are discussed in the rest of this paper.

\textit{
Mirror skin effect in an electric circuit.--
}
%
%
Before detailed proposal for the implementation of the circuit showing the mirror skin effect, let us briefly review how an electric circuit mimics a generic tight-binding Hamiltonian. 
Consider an electric circuit where the voltage $V_a(\omega)$ is applied at nodes $a=1,2,\cdots$ with angular frequency $\omega$.
In this case, based on the Kirchhoff's law, the current $I_b(\omega)$ at node $b$ is given by
\begin{eqnarray}
I_b(\omega) &=& \sum_{a} J_{ba}(\omega) V_a(\omega).
\end{eqnarray}
Thus, the admittance matrix $J_{ba}(\omega)$ serves as a Hamiltonian for the corresponding tight-binding model, which means that topological phenomena can also be observed for electric circuits~\cite{Lee_Topoelecircit_CommPhys18}.
For instance, the Su-Schrieffer-Heeger model can be realized for an electric circuit composed only with capacitors and inductors. 
The energy conservation of the electric circuit implies the Hermiticity of the matrix $J_{ba}(\omega)$ up to the global phase factor $i$.

\begin{figure}[!h]
\begin{minipage}{0.49\hsize}
\begin{center}
\includegraphics[width=1\hsize,clip]{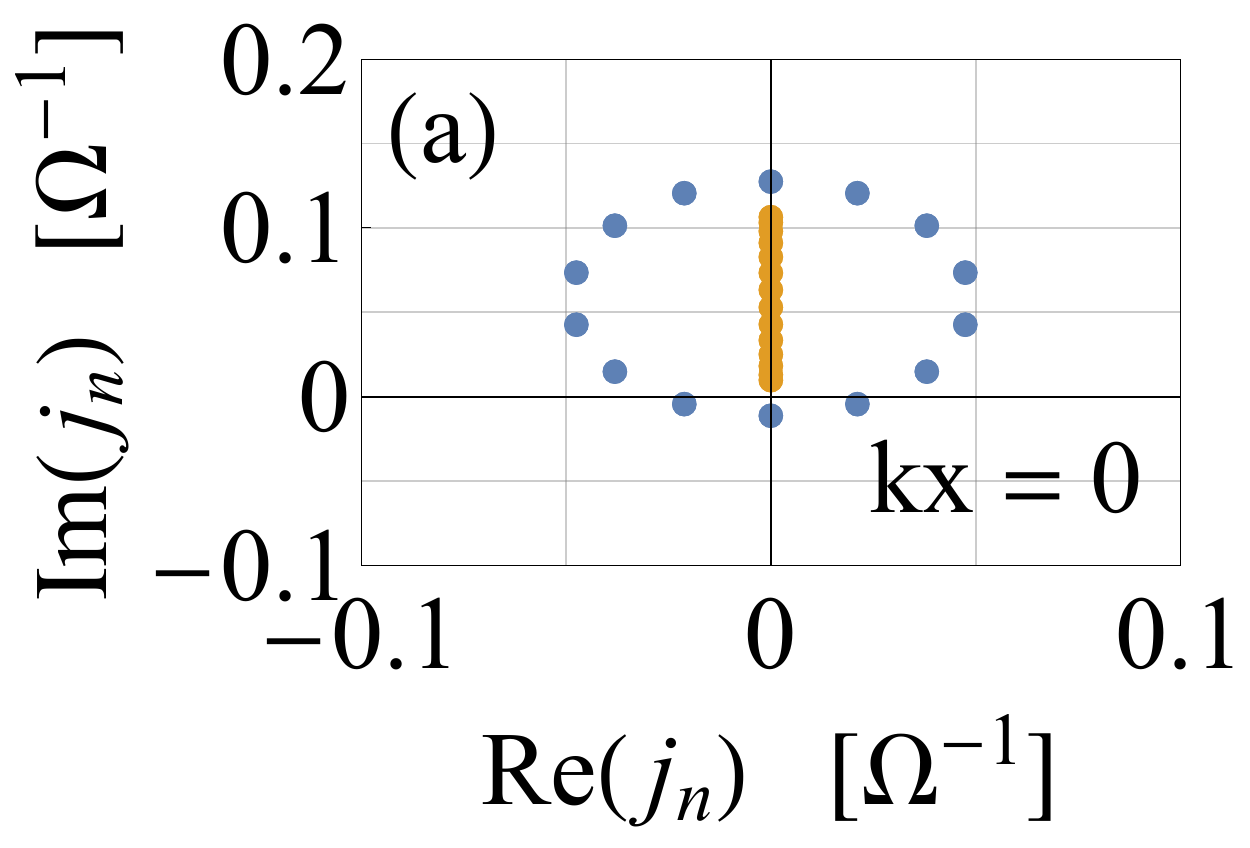}
\end{center}
\end{minipage}
\begin{minipage}{0.49\hsize}
\begin{center}
\includegraphics[width=1\hsize,clip]{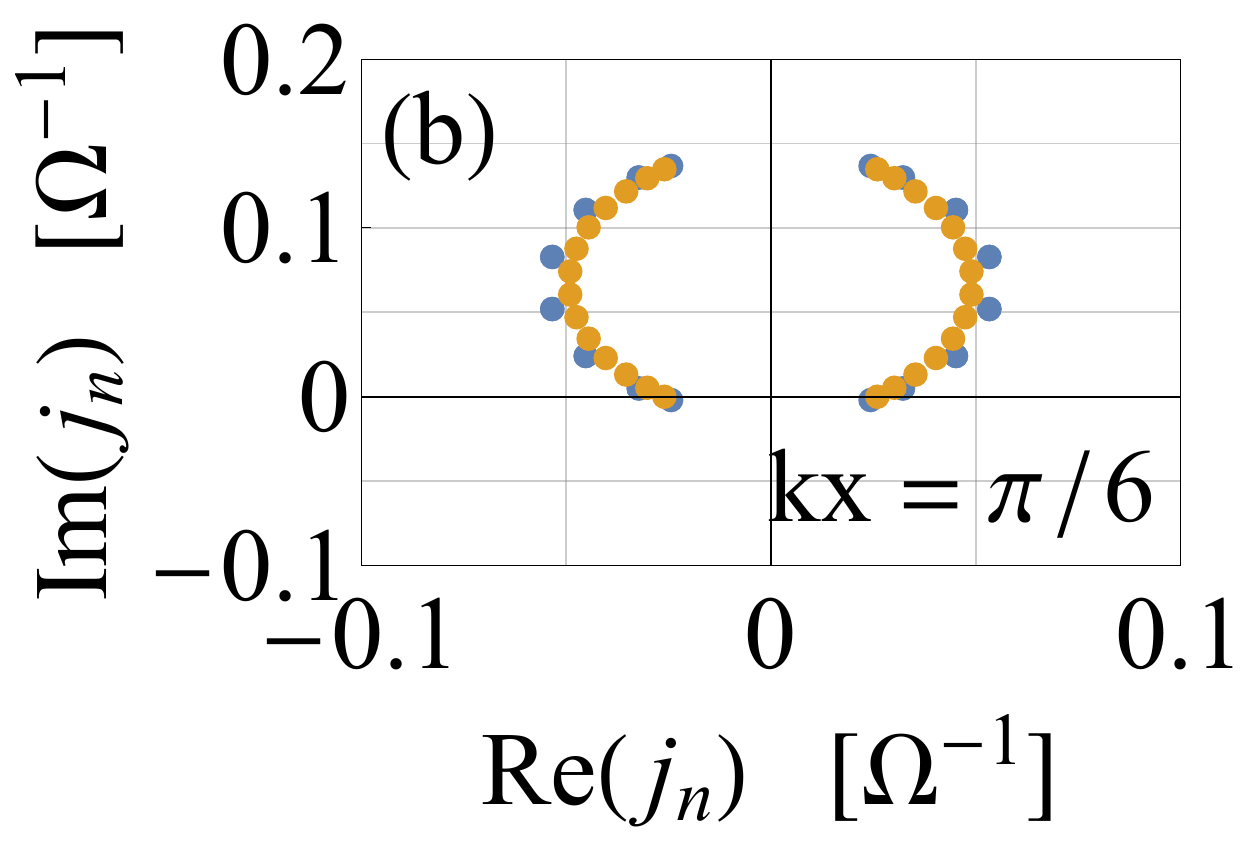}
\end{center}
\end{minipage}
\begin{minipage}{0.49\hsize}
\begin{center}
\includegraphics[width=1\hsize,clip]{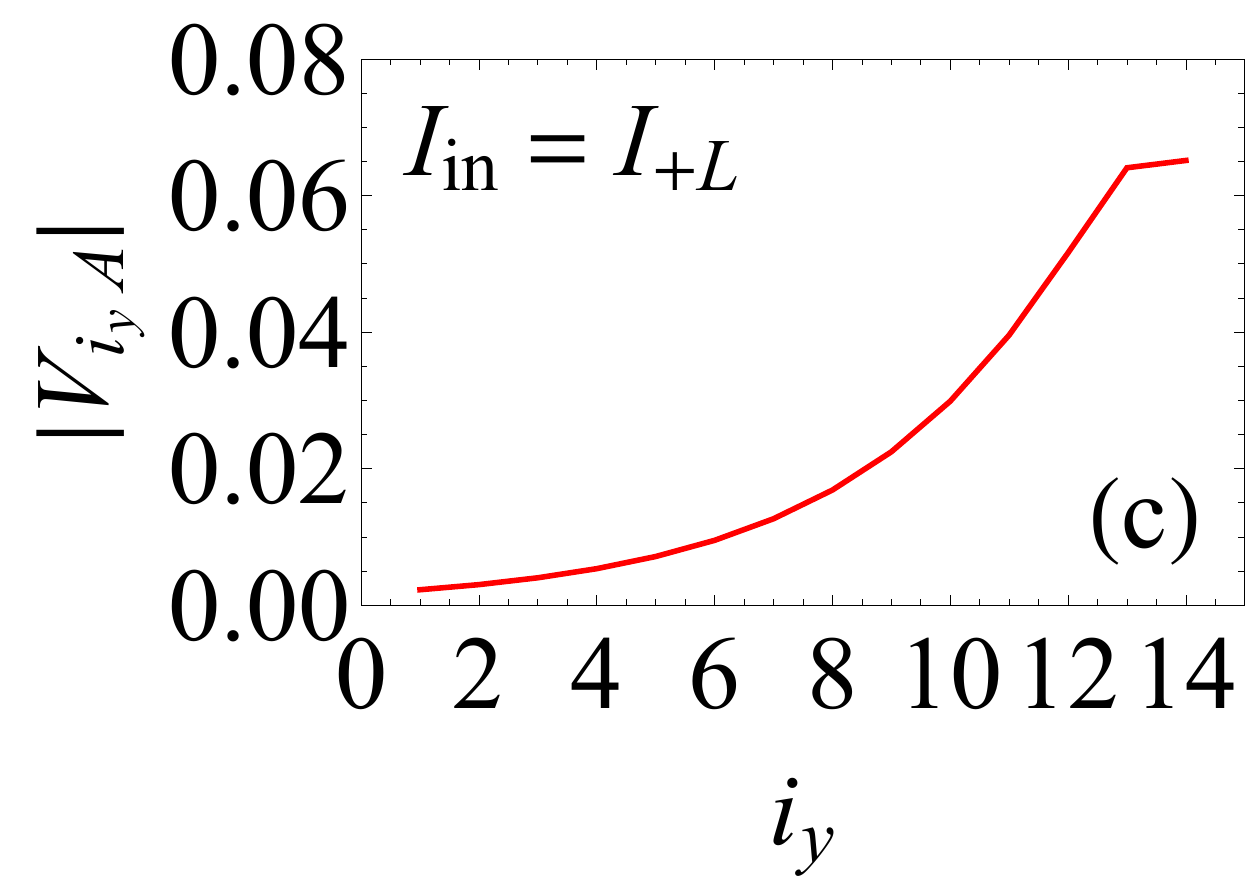}
\end{center}
\end{minipage}
\begin{minipage}{0.49\hsize}
\begin{center}
\includegraphics[width=1\hsize,clip]{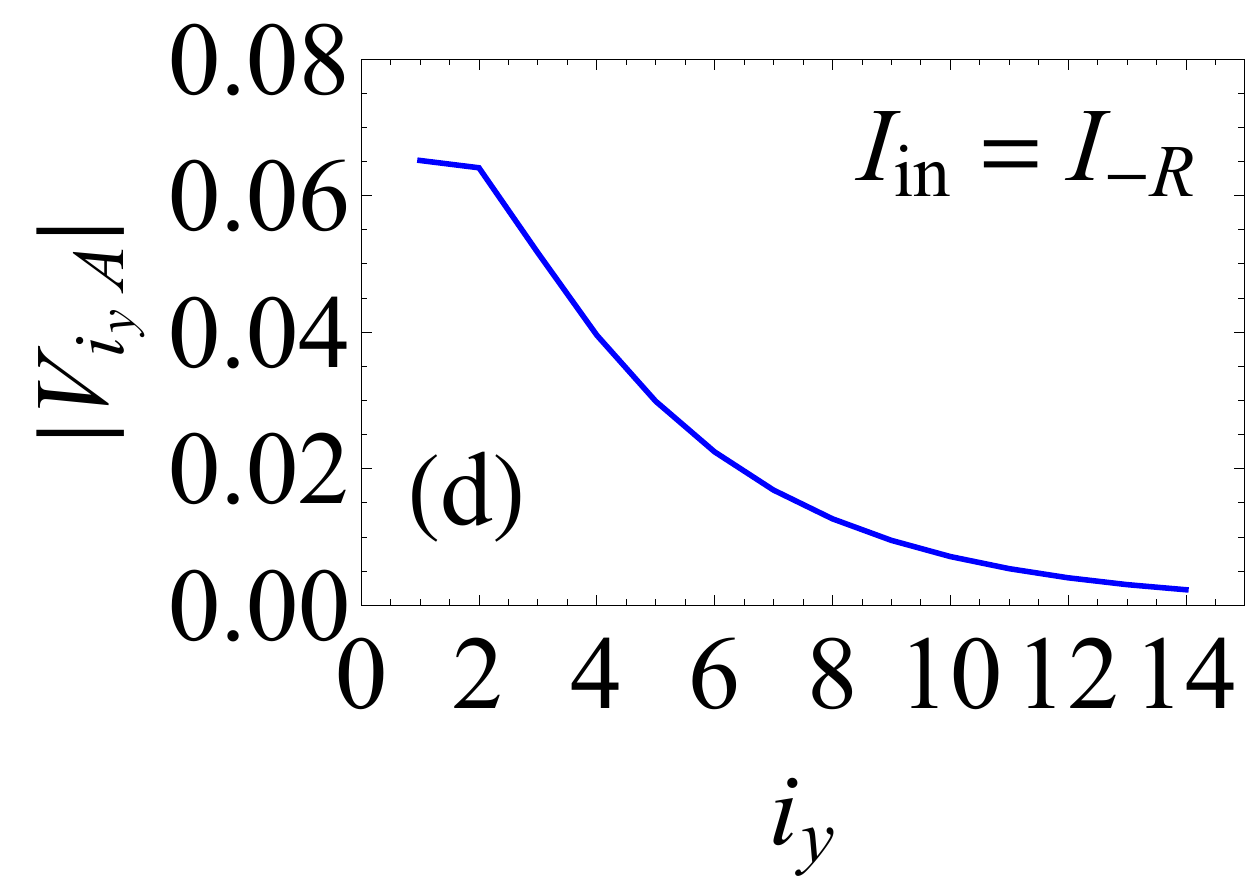}
\end{center}
\end{minipage}
\begin{minipage}{1\hsize}
\begin{center}
\includegraphics[width=1\hsize,clip]{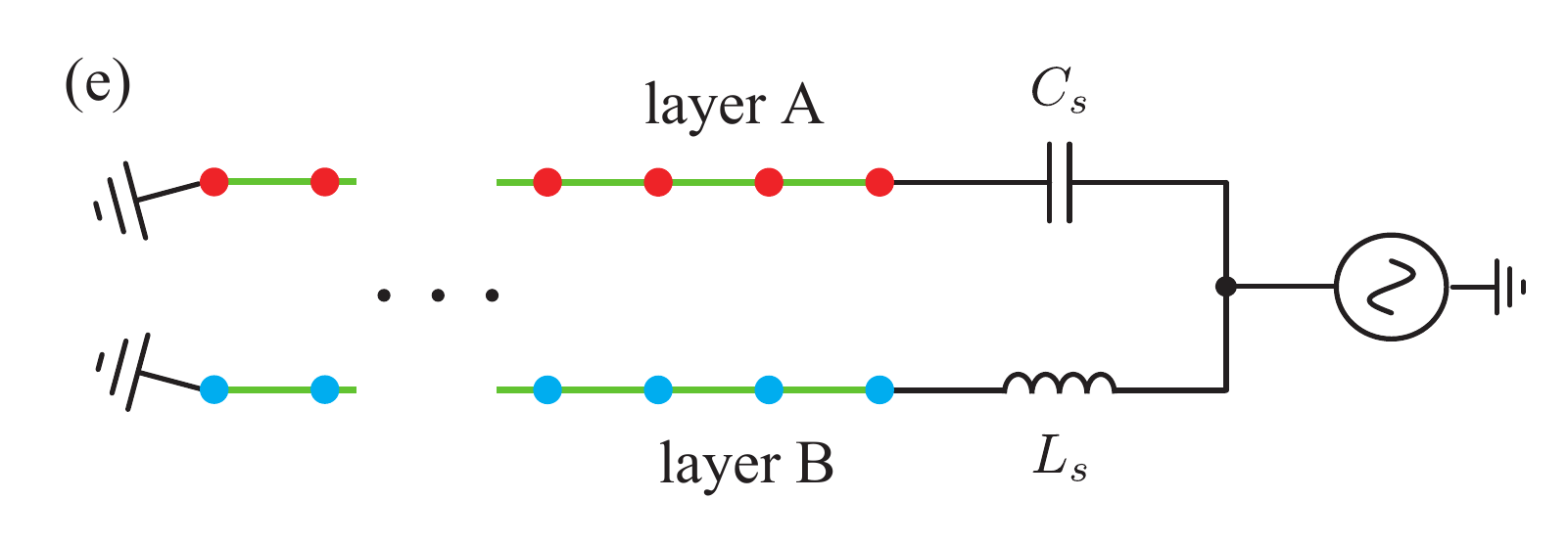}
\end{center}
\end{minipage}
\caption{(Color Online).
(a) [(b)]: Admittance spectrum for each case of the boundary condition for $k_x=0$ ($k_x=\pi/6$), respectively.
Blue (orange) dots denote the admittance eigenvalue $j_n$ under the periodic (open) boundary condition for the $y$-direction with $L_y=14$.
The periodic boundary condition is imposed for the $x$-direction. 
The data are obtained for $L_0=120\mathrm{\mu}\mathrm{H}$, $C_0=47\mathrm{n}\mathrm{F}$, $C_1=33\mathrm{n}\mathrm{F}$, and $f=117.4\mathrm{kHz}$, where $f$ denotes frequency $\omega=2\pi f$.
(c) [(d)]: Voltage profile of layer A for the case where the current with plus (minus) sector of reflection is fed. 
(e) Illustration of setup to observe the anomalous voltage response with $I_\mathrm{in}=I_{\mathrm{+L}}$. For the above parameter sets, the fed current is $I_s=0.0001\mathrm{A}$ with the fed voltage $V_s=0.002\mathrm{V}$ for $C_s=0.068\mathrm{\mu F}$, and $L_s=27\mathrm{\mu H}$.
}
\label{fig: jn}
\end{figure}

Now, let us discuss how to experimentally verify the mirror skin effect for electric circuits.
In order to implement a circuit showing the mirror skin effect, we need to reproduce non-Hermitian terms of the Hamiltonian [i.e., the second and the third term of Eq.~(\ref{eq: toy Hami})], which can be accomplished by employing the negative impedance converters~\cite{Hofmann_EleCirChern_PRL19,Helbig_ExpSkin_19}. 
Specifically, we propose that an electric circuit shown in Fig.~\ref{fig: Emodel} serves as a platform of the mirror skin effect. 
The corresponding admittance matrix is given by
\begin{subequations}
\label{eq: I=JVskin tot}
\begin{eqnarray}
\label{eq: I=JVskin}
\left(
\begin{array}{c}
I_A(\omega,\bm{k})  \\
I_B(\omega,\bm{k}) 
\end{array}
\right)
&=&J(\omega,\bm{k})  
\left(
\begin{array}{c}
V_A(\omega,\bm{k})  \\
V_B(\omega,\bm{k})
\end{array}
\right),
\end{eqnarray}
%
\begin{eqnarray}
\label{eq: Jskin}
J(\omega,\bm{k}) &=& i\omega [
-2C_0(\cos k_x+\cos k_y)\rho_0+(4C_0+\frac{1}{\omega^2 L_0})\rho_0 \nonumber \\
&&\quad + 2iC_1\sin k_x \rho_3 +2iC_1 \sin k_y \rho_1 ],
\end{eqnarray}
\end{subequations}
in the momentum space.
$I_\alpha(\omega,\bm{k})$ and $V_\alpha(\omega,\bm{k})$ ($\alpha=A,B$) denotes the Fourier transformed current and voltage, respectively~\cite{Diode_footnote,Ezawa_elecirPRB19}. 
The detailed derivation of Eq.~(\ref{eq: I=JVskin tot}) is given in Sec.~\ref{sec: derivation of J app} of the supplemental material~\cite{supplemental}.
This model preserves the mirror symmetry with $M_x=\rho_1P_x$. 
In addition, its topology is characterized by the mirror winding number taking $\nu_{\mathrm{M}}(k^*_x)= -1$ ($k^*_x=0,\pi$) for the parameter set summarized in caption of Fig.~\ref{fig: jn}.
We note that the relation $\nu_\mathrm{tot}=0$ holds for arbitrary momentum $k_x$.
We mention here that the circuit elements of the above parameters are commercially available.
Numerical data elucidating the above topological properties are shown in Sec.~\ref{sec: vm of J app} of the supplemental material~\cite{supplemental}.

In the following, we see that the above model shows the mirror skin effect.
For the electric circuit, the skin effect can be experimentally observed by measuring the admittance eigenvalues $j_n$ with $n=1,\cdots,\mathrm{dim}J$ 
[i.e., eigenvalues of the admittance matrix~(\ref{eq: Jskin})]. 
One can access the admittance eigenvalues by the impedance measurement [$J^{-1}_{ab}(\omega)$]~\cite{Helbig_ExpSkin_19}.
When the skin effect occurs, the admittance spectrum significantly depends on the boundary condition as we have seen in Fig.~\ref{fig: Engtoy}(a).
Figure~\ref{fig: jn}(a)[(b)] shows the admittance spectrum for $k_x=0$ ($k_x=\pi/6$), respectively.
For the momentum invariant line $k_x=0$, the admittance spectrum significantly changes depending on the boundary condition [see Fig.~\ref{fig: jn}(a)]; the eigenvalues form a circle for the PBC, while they form a straight line for the OBC. 
Away from the mirror invariant line, the mirror operation is not closed for each momentum $k_x$, which results in the absence of the skin effect [see Fig.~\ref{fig: jn}(b)].

Extensive localized states, observed in Fig.~\ref{fig: Amptoy_OBC}(a), are also observed through the voltage profile with the current feed for each mirror sector.
In Fig.~\ref{fig: jn}(c) [(d)], we plot the voltage profile with the current feed $I_{\mathrm{in}}=I_{+\mathrm{{L}}}$ ($I_{\mathrm{in}}=I_{\mathrm{-{R}}}$), respectively.
Here, $I_{+\mathrm{L}i_y \alpha}=iI_{\mathrm{s}} \delta_{i_y L_y}(\delta_{\alpha A}+i\delta_{\alpha B})$, $I_{-\mathrm{L}i_y\alpha}=iI_{\mathrm{s}}\delta_{i_y 0}(\delta_{\alpha A}-i\delta_{\alpha B})$, and $I_{\mathrm{s}}=0.0001\mathrm{A}$.
Tuning the phase of the feed current is accomplished with the setup illustrated in Fig.~\ref{fig: jn}(e).
In Fig.~\ref{fig: jn}(c), we observe that the voltage response becomes large around the right edge $i_y=L_y$ although the current is fed at the left edge $i_y=0$.
The essentially same result can also be observed for the other mirror sector [see Fig.~\ref{fig: jn}(d)].

The above anomalous response for the feed current arises from the extensively localized states.
Firstly, we note that the voltage profile for inputted current can be obtained as $V_a=\sum_b J^{-1}_{ab}I_{\mathrm{in}b}$ with the impedance matrix $J^{-1}_{ab}=\sum_{n} R_{an}j^{-1}_n L^\dagger_{nb}$~\cite{Helbig_ExpSkin_19,Hofmann_ExpRecipSkin_19}. Here the matrix $R$ ($L^\dagger$) denotes the set of the right (left) eigenvectors $\sum_{b}J_{ab}R_{bn}= R_{an} j_n$ ($\sum_{a} L^\dagger_{na} J_{ab}= j_n L^\dagger_{nb} $), respectively. 
$a$ and $b$ detote the set of the labels $i_y$ and $\alpha=A,B$.
We note that $\sum_n L^\dagger_{nb}I_{\mathrm{+L}b}=0$ when $n$ labels the states for the minus sector.
Thus, we can see that the only eigenvectors for the plus sector contribute to the voltage response.
In addition, when one of the states ($n=n_0$) for the plus sector is dominant~\cite{L0_footnote}, the voltage is estimated as
$|V_{i_yA}| \sim \left| R_{i_yAn_0}j^{-1}_{n_0} \sum_{b}L^\dagger_{n_0 b}  I_{\mathrm{in}b} \right|$. Therefore, we observe that the extensive localized states induce the anomalous voltage response for $I_{\mathrm{in}}= I_{+L}$. The anomalous response shown in Fig.~\ref{fig: jn}(d) can be understood in a similar way.

The above results indicate that the mirror skin effect can be observed for the circuit shown in Fig.~\ref{fig: Emodel}. 
\blue{
Specifically, by measuring the elements of the admittance matrix in the real space and by applying the Fourier transformation, one can experimentally observe the significant change of admittance eigenvalues for $k_x=0,\pi$ [Figs.~\ref{fig: jn}(a)~and~(b)]~\cite{Helbig_ExpSkin_19,Hofmann_ExpRecipSkin_19}. 
In a similar way, one can experimentally observe the anomalous voltage response [Figs.~\ref{fig: jn}(c)~and~(d)] by extracting the contribution of $k_x=0,\pi$ form data measured in the real space. We note that it can be another approach to measure the above quantities for the system where only a few sites are aligned along the $x$-direction.
}

\textit{Summary.--}
In this \blue{paper}, we have analyzed interplay between mirror symmetry and skin effects, shedding new light on crystalline symmetry and non-Hermitian topology. Our analysis has clarified a novel type of skin effects, a mirror skin effect which results in the significant dependence both of the energy spectrum and the states on the boundary condition only along mirror invariant lines in the two-dimensional Brillouin zone. 
The topological characterization of this skin effect can be done with the mirror winding number. 
The mirror skin effect has been verified by numerically diagonalizing a tight-binding Hamiltonian with the mirror winding number taking one.
\blue{Here, we stress that the discovery of the skin effect protected by mirror symmetry is a key result rather than introducing the mirror winding number.}

Furthermore, we have proposed how to implement the electric circuit for the experimental observation of the mirror skin effect. In this system, switching the boundary condition significantly changes the admittance eigenvalues, which serves as a distinct evidence of the mirror skin effect.

\textit{Acknowledgement.--}
This work is partly supported by JSPS KAKENHI Grants No.~JP16K13845, No.~JP17H06138, No.~JP18H05842, and No.~JP20H04627.
A part of numerical calculations were performed on the supercomputer at the ISSP in the University of Tokyo.


\begin{thebibliography}{83}%
\makeatletter
\providecommand \@ifxundefined [1]{%
 \@ifx{#1\undefined}
}%
\providecommand \@ifnum [1]{%
 \ifnum #1\expandafter \@firstoftwo
 \else \expandafter \@secondoftwo
 \fi
}%
\providecommand \@ifx [1]{%
 \ifx #1\expandafter \@firstoftwo
 \else \expandafter \@secondoftwo
 \fi
}%
\providecommand \natexlab [1]{#1}%
\providecommand \enquote  [1]{``#1''}%
\providecommand \bibnamefont  [1]{#1}%
\providecommand \bibfnamefont [1]{#1}%
\providecommand \citenamefont [1]{#1}%
\providecommand \href@noop [0]{\@secondoftwo}%
\providecommand \href [0]{\begingroup \@sanitize@url \@href}%
\providecommand \@href[1]{\@@startlink{#1}\@@href}%
\providecommand \@@href[1]{\endgroup#1\@@endlink}%
\providecommand \@sanitize@url [0]{\catcode `\\12\catcode `\$12\catcode
  `\&12\catcode `\#12\catcode `\^12\catcode `\_12\catcode `\%12\relax}%
\providecommand \@@startlink[1]{}%
\providecommand \@@endlink[0]{}%
\providecommand \url  [0]{\begingroup\@sanitize@url \@url }%
\providecommand \@url [1]{\endgroup\@href {#1}{\urlprefix }}%
\providecommand \urlprefix  [0]{URL }%
\providecommand \Eprint [0]{\href }%
\providecommand \doibase [0]{http://dx.doi.org/}%
\providecommand \selectlanguage [0]{\@gobble}%
\providecommand \bibinfo  [0]{\@secondoftwo}%
\providecommand \bibfield  [0]{\@secondoftwo}%
\providecommand \translation [1]{[#1]}%
\providecommand \BibitemOpen [0]{}%
\providecommand \bibitemStop [0]{}%
\providecommand \bibitemNoStop [0]{.\EOS\space}%
\providecommand \EOS [0]{\spacefactor3000\relax}%
\providecommand \BibitemShut  [1]{\csname bibitem#1\endcsname}%
\let\auto@bib@innerbib\@empty
\bibitem [{\citenamefont {Hasan}\ and\ \citenamefont
  {Kane}(2010)}]{TI_review_Hasan10}%
  \BibitemOpen
  \bibfield  {author} {\bibinfo {author} {\bibfnamefont {M.~Z.}\ \bibnamefont
  {Hasan}}\ and\ \bibinfo {author} {\bibfnamefont {C.~L.}\ \bibnamefont
  {Kane}},\ }\href {\doibase 10.1103/RevModPhys.82.3045} {\bibfield  {journal}
  {\bibinfo  {journal} {Rev. Mod. Phys.}\ }\textbf {\bibinfo {volume} {82}},\
  \bibinfo {pages} {3045} (\bibinfo {year} {2010})}\BibitemShut {NoStop}%
\bibitem [{\citenamefont {Qi}\ and\ \citenamefont
  {Zhang}(2011)}]{TI_review_Qi10}%
  \BibitemOpen
  \bibfield  {author} {\bibinfo {author} {\bibfnamefont {X.-L.}\ \bibnamefont
  {Qi}}\ and\ \bibinfo {author} {\bibfnamefont {S.-C.}\ \bibnamefont {Zhang}},\
  }\href {\doibase 10.1103/RevModPhys.83.1057} {\bibfield  {journal} {\bibinfo
  {journal} {Rev. Mod. Phys.}\ }\textbf {\bibinfo {volume} {83}},\ \bibinfo
  {pages} {1057} (\bibinfo {year} {2011})}\BibitemShut {NoStop}%
\bibitem [{\citenamefont {Hatsugai}(1993)}]{Hatsugai_PRL93}%
  \BibitemOpen
  \bibfield  {author} {\bibinfo {author} {\bibfnamefont {Y.}~\bibnamefont
  {Hatsugai}},\ }\href {\doibase 10.1103/PhysRevLett.71.3697} {\bibfield
  {journal} {\bibinfo  {journal} {Phys. Rev. Lett.}\ }\textbf {\bibinfo
  {volume} {71}},\ \bibinfo {pages} {3697} (\bibinfo {year}
  {1993})}\BibitemShut {NoStop}%
\bibitem [{\citenamefont {Haldane}\ and\ \citenamefont
  {Raghu}(2008)}]{Haldane_chiralPHC_PRL08}%
  \BibitemOpen
  \bibfield  {author} {\bibinfo {author} {\bibfnamefont {F.~D.~M.}\
  \bibnamefont {Haldane}}\ and\ \bibinfo {author} {\bibfnamefont
  {S.}~\bibnamefont {Raghu}},\ }\href {\doibase 10.1103/PhysRevLett.100.013904}
  {\bibfield  {journal} {\bibinfo  {journal} {Phys. Rev. Lett.}\ }\textbf
  {\bibinfo {volume} {100}},\ \bibinfo {pages} {013904} (\bibinfo {year}
  {2008})}\BibitemShut {NoStop}%
\bibitem [{\citenamefont {Raghu}\ and\ \citenamefont
  {Haldane}(2008)}]{Raghu_chiralPHC_PRA08}%
  \BibitemOpen
  \bibfield  {author} {\bibinfo {author} {\bibfnamefont {S.}~\bibnamefont
  {Raghu}}\ and\ \bibinfo {author} {\bibfnamefont {F.~D.~M.}\ \bibnamefont
  {Haldane}},\ }\href {\doibase 10.1103/PhysRevA.78.033834} {\bibfield
  {journal} {\bibinfo  {journal} {Phys. Rev. A}\ }\textbf {\bibinfo {volume}
  {78}},\ \bibinfo {pages} {033834} (\bibinfo {year} {2008})}\BibitemShut
  {NoStop}%
\bibitem [{\citenamefont {Wang}\ \emph {et~al.}(2009)\citenamefont {Wang},
  \citenamefont {Chong}, \citenamefont {Joannopoulos},\ and\ \citenamefont
  {Soljacic}}]{Wang_chiralPHC_Nature09}%
  \BibitemOpen
  \bibfield  {author} {\bibinfo {author} {\bibfnamefont {Z.}~\bibnamefont
  {Wang}}, \bibinfo {author} {\bibfnamefont {Y.}~\bibnamefont {Chong}},
  \bibinfo {author} {\bibfnamefont {J.~D.}\ \bibnamefont {Joannopoulos}}, \
  and\ \bibinfo {author} {\bibfnamefont {M.}~\bibnamefont {Soljacic}},\ }\href
  {https://doi.org/10.1038/nature08293} {\bibfield  {journal} {\bibinfo
  {journal} {Nature}\ }\textbf {\bibinfo {volume} {461}},\ \bibinfo {pages}
  {772 EP } (\bibinfo {year} {2009})}\BibitemShut {NoStop}%
\bibitem [{\citenamefont {Kane}\ and\ \citenamefont
  {Lubensky}(2013)}]{Kane_NatPhys13}%
  \BibitemOpen
  \bibfield  {author} {\bibinfo {author} {\bibfnamefont {C.~L.}\ \bibnamefont
  {Kane}}\ and\ \bibinfo {author} {\bibfnamefont {T.~C.}\ \bibnamefont
  {Lubensky}},\ }\href {https://doi.org/10.1038/nphys2835} {\bibfield
  {journal} {\bibinfo  {journal} {Nature Physics}\ }\textbf {\bibinfo {volume}
  {10}},\ \bibinfo {pages} {39 EP } (\bibinfo {year} {2013})},\ \bibinfo {note}
  {article}\BibitemShut {NoStop}%
\bibitem [{\citenamefont {Kariyado}\ and\ \citenamefont
  {Hatsugai}(2015)}]{Kariyado_SR15}%
  \BibitemOpen
  \bibfield  {author} {\bibinfo {author} {\bibfnamefont {T.}~\bibnamefont
  {Kariyado}}\ and\ \bibinfo {author} {\bibfnamefont {Y.}~\bibnamefont
  {Hatsugai}},\ }\href {https://doi.org/10.1038/srep18107} {\bibfield
  {journal} {\bibinfo  {journal} {Scientific Reports}\ }\textbf {\bibinfo
  {volume} {5}},\ \bibinfo {pages} {18107 EP } (\bibinfo {year} {2015})},\
  \bibinfo {note} {article}\BibitemShut {NoStop}%
\bibitem [{\citenamefont {S{\"u}sstrunk}\ and\ \citenamefont
  {Huber}(2015)}]{Susstrunk_TopoMech_Sci15}%
  \BibitemOpen
  \bibfield  {author} {\bibinfo {author} {\bibfnamefont {R.}~\bibnamefont
  {S{\"u}sstrunk}}\ and\ \bibinfo {author} {\bibfnamefont {S.~D.}\ \bibnamefont
  {Huber}},\ }\href {\doibase 10.1126/science.aab0239} {\bibfield  {journal}
  {\bibinfo  {journal} {Science}\ }\textbf {\bibinfo {volume} {349}},\ \bibinfo
  {pages} {47} (\bibinfo {year} {2015})}\BibitemShut {NoStop}%
\bibitem [{\citenamefont {Huber}(2016)}]{Huber_TopoMech_NatPhys16}%
  \BibitemOpen
  \bibfield  {author} {\bibinfo {author} {\bibfnamefont {S.~D.}\ \bibnamefont
  {Huber}},\ }\href {\doibase 10.1038/nphys3801} {\bibfield  {journal}
  {\bibinfo  {journal} {Nature Physics}\ }\textbf {\bibinfo {volume} {12}},\
  \bibinfo {pages} {621} (\bibinfo {year} {2016})}\BibitemShut {NoStop}%
\bibitem [{\citenamefont {Delplace}\ \emph {et~al.}(2017)\citenamefont
  {Delplace}, \citenamefont {Marston},\ and\ \citenamefont
  {Venaille}}]{Delplace_topoEq_Science17}%
  \BibitemOpen
  \bibfield  {author} {\bibinfo {author} {\bibfnamefont {P.}~\bibnamefont
  {Delplace}}, \bibinfo {author} {\bibfnamefont {J.~B.}\ \bibnamefont
  {Marston}}, \ and\ \bibinfo {author} {\bibfnamefont {A.}~\bibnamefont
  {Venaille}},\ }\href {\doibase 10.1126/science.aan8819} {\bibfield  {journal}
  {\bibinfo  {journal} {Science}\ }\textbf {\bibinfo {volume} {358}},\ \bibinfo
  {pages} {1075} (\bibinfo {year} {2017})}\BibitemShut {NoStop}%
\bibitem [{\citenamefont {Albert}\ \emph {et~al.}(2015)\citenamefont {Albert},
  \citenamefont {Glazman},\ and\ \citenamefont
  {Jiang}}]{Albert_Topoelecircit_PRL15}%
  \BibitemOpen
  \bibfield  {author} {\bibinfo {author} {\bibfnamefont {V.~V.}\ \bibnamefont
  {Albert}}, \bibinfo {author} {\bibfnamefont {L.~I.}\ \bibnamefont {Glazman}},
  \ and\ \bibinfo {author} {\bibfnamefont {L.}~\bibnamefont {Jiang}},\ }\href
  {\doibase 10.1103/PhysRevLett.114.173902} {\bibfield  {journal} {\bibinfo
  {journal} {Phys. Rev. Lett.}\ }\textbf {\bibinfo {volume} {114}},\ \bibinfo
  {pages} {173902} (\bibinfo {year} {2015})}\BibitemShut {NoStop}%
\bibitem [{\citenamefont {Lee}\ \emph {et~al.}(2018)\citenamefont {Lee},
  \citenamefont {Imhof}, \citenamefont {Berger}, \citenamefont {Bayer},
  \citenamefont {Brehm}, \citenamefont {Molenkamp}, \citenamefont {Kiessling},\
  and\ \citenamefont {Thomale}}]{Lee_Topoelecircit_CommPhys18}%
  \BibitemOpen
  \bibfield  {author} {\bibinfo {author} {\bibfnamefont {C.~H.}\ \bibnamefont
  {Lee}}, \bibinfo {author} {\bibfnamefont {S.}~\bibnamefont {Imhof}}, \bibinfo
  {author} {\bibfnamefont {C.}~\bibnamefont {Berger}}, \bibinfo {author}
  {\bibfnamefont {F.}~\bibnamefont {Bayer}}, \bibinfo {author} {\bibfnamefont
  {J.}~\bibnamefont {Brehm}}, \bibinfo {author} {\bibfnamefont {L.~W.}\
  \bibnamefont {Molenkamp}}, \bibinfo {author} {\bibfnamefont {T.}~\bibnamefont
  {Kiessling}}, \ and\ \bibinfo {author} {\bibfnamefont {R.}~\bibnamefont
  {Thomale}},\ }\href {\doibase 10.1038/s42005-018-0035-2} {\bibfield
  {journal} {\bibinfo  {journal} {Communications Physics}\ }\textbf {\bibinfo
  {volume} {1}},\ \bibinfo {pages} {39} (\bibinfo {year} {2018})}\BibitemShut
  {NoStop}%
\bibitem [{\citenamefont {Teo}\ \emph {et~al.}(2008)\citenamefont {Teo},
  \citenamefont {Fu},\ and\ \citenamefont {Kane}}]{JTeo_PRB08}%
  \BibitemOpen
  \bibfield  {author} {\bibinfo {author} {\bibfnamefont {J.~C.~Y.}\
  \bibnamefont {Teo}}, \bibinfo {author} {\bibfnamefont {L.}~\bibnamefont
  {Fu}}, \ and\ \bibinfo {author} {\bibfnamefont {C.~L.}\ \bibnamefont
  {Kane}},\ }\href {\doibase 10.1103/PhysRevB.78.045426} {\bibfield  {journal}
  {\bibinfo  {journal} {Phys. Rev. B}\ }\textbf {\bibinfo {volume} {78}},\
  \bibinfo {pages} {045426} (\bibinfo {year} {2008})}\BibitemShut {NoStop}%
\bibitem [{\citenamefont {Fu}(2011)}]{Fu_TCI_PRL2011}%
  \BibitemOpen
  \bibfield  {author} {\bibinfo {author} {\bibfnamefont {L.}~\bibnamefont
  {Fu}},\ }\href {\doibase 10.1103/PhysRevLett.106.106802} {\bibfield
  {journal} {\bibinfo  {journal} {Phys. Rev. Lett.}\ }\textbf {\bibinfo
  {volume} {106}},\ \bibinfo {pages} {106802} (\bibinfo {year}
  {2011})}\BibitemShut {NoStop}%
\bibitem [{\citenamefont {Hsieh}\ \emph {et~al.}(2012)\citenamefont {Hsieh},
  \citenamefont {Lin}, \citenamefont {Liu}, \citenamefont {Duan}, \citenamefont
  {Bansil},\ and\ \citenamefont {Fu}}]{Hsieh_TCH_SnTe_2012}%
  \BibitemOpen
  \bibfield  {author} {\bibinfo {author} {\bibfnamefont {T.~H.}\ \bibnamefont
  {Hsieh}}, \bibinfo {author} {\bibfnamefont {H.}~\bibnamefont {Lin}}, \bibinfo
  {author} {\bibfnamefont {J.}~\bibnamefont {Liu}}, \bibinfo {author}
  {\bibfnamefont {W.}~\bibnamefont {Duan}}, \bibinfo {author} {\bibfnamefont
  {A.}~\bibnamefont {Bansil}}, \ and\ \bibinfo {author} {\bibfnamefont
  {L.}~\bibnamefont {Fu}},\ }\href@noop {} {\bibfield  {journal} {\bibinfo
  {journal} {Nat. Commun.}\ }\textbf {\bibinfo {volume} {3}},\ \bibinfo {pages}
  {982} (\bibinfo {year} {2012})}\BibitemShut {NoStop}%
\bibitem [{\citenamefont {Tanaka}\ \emph {et~al.}(2012)\citenamefont {Tanaka},
  \citenamefont {Ren}, \citenamefont {Sato}, \citenamefont {Nakayama},
  \citenamefont {Souma}, \citenamefont {Takahashi}, \citenamefont {Segawa},\
  and\ \citenamefont {Ando}}]{Tanaka_TCI_SnTe2012}%
  \BibitemOpen
  \bibfield  {author} {\bibinfo {author} {\bibfnamefont {Y.}~\bibnamefont
  {Tanaka}}, \bibinfo {author} {\bibfnamefont {Z.}~\bibnamefont {Ren}},
  \bibinfo {author} {\bibfnamefont {T.}~\bibnamefont {Sato}}, \bibinfo {author}
  {\bibfnamefont {K.}~\bibnamefont {Nakayama}}, \bibinfo {author}
  {\bibfnamefont {S.}~\bibnamefont {Souma}}, \bibinfo {author} {\bibfnamefont
  {T.}~\bibnamefont {Takahashi}}, \bibinfo {author} {\bibfnamefont
  {K.}~\bibnamefont {Segawa}}, \ and\ \bibinfo {author} {\bibfnamefont
  {Y.}~\bibnamefont {Ando}},\ }\href@noop {} {\bibfield  {journal} {\bibinfo
  {journal} {Nat. Phys.}\ }\textbf {\bibinfo {volume} {8}},\ \bibinfo {pages}
  {800} (\bibinfo {year} {2012})}\BibitemShut {NoStop}%
\bibitem [{\citenamefont {Hashimoto}\ \emph {et~al.}(2017)\citenamefont
  {Hashimoto}, \citenamefont {Wu},\ and\ \citenamefont
  {Kimura}}]{Hashimoto_HOTI_PRB17}%
  \BibitemOpen
  \bibfield  {author} {\bibinfo {author} {\bibfnamefont {K.}~\bibnamefont
  {Hashimoto}}, \bibinfo {author} {\bibfnamefont {X.}~\bibnamefont {Wu}}, \
  and\ \bibinfo {author} {\bibfnamefont {T.}~\bibnamefont {Kimura}},\ }\href
  {\doibase 10.1103/PhysRevB.95.165443} {\bibfield  {journal} {\bibinfo
  {journal} {Phys. Rev. B}\ }\textbf {\bibinfo {volume} {95}},\ \bibinfo
  {pages} {165443} (\bibinfo {year} {2017})}\BibitemShut {NoStop}%
\bibitem [{\citenamefont {Benalcazar}\ \emph
  {et~al.}(2017{\natexlab{a}})\citenamefont {Benalcazar}, \citenamefont
  {Bernevig},\ and\ \citenamefont {Hughes}}]{Benalcazar_HOTI_Science17}%
  \BibitemOpen
  \bibfield  {author} {\bibinfo {author} {\bibfnamefont {W.~A.}\ \bibnamefont
  {Benalcazar}}, \bibinfo {author} {\bibfnamefont {B.~A.}\ \bibnamefont
  {Bernevig}}, \ and\ \bibinfo {author} {\bibfnamefont {T.~L.}\ \bibnamefont
  {Hughes}},\ }\href {\doibase 10.1126/science.aah6442} {\bibfield  {journal}
  {\bibinfo  {journal} {Science}\ }\textbf {\bibinfo {volume} {357}},\ \bibinfo
  {pages} {61} (\bibinfo {year} {2017}{\natexlab{a}})}\BibitemShut {NoStop}%
\bibitem [{\citenamefont {Schindler}\ \emph {et~al.}(2018)\citenamefont
  {Schindler}, \citenamefont {Cook}, \citenamefont {Vergniory}, \citenamefont
  {Wang}, \citenamefont {Parkin}, \citenamefont {Bernevig},\ and\ \citenamefont
  {Neupert}}]{Schindler_HOTI_Science18}%
  \BibitemOpen
  \bibfield  {author} {\bibinfo {author} {\bibfnamefont {F.}~\bibnamefont
  {Schindler}}, \bibinfo {author} {\bibfnamefont {A.~M.}\ \bibnamefont {Cook}},
  \bibinfo {author} {\bibfnamefont {M.~G.}\ \bibnamefont {Vergniory}}, \bibinfo
  {author} {\bibfnamefont {Z.}~\bibnamefont {Wang}}, \bibinfo {author}
  {\bibfnamefont {S.~S.~P.}\ \bibnamefont {Parkin}}, \bibinfo {author}
  {\bibfnamefont {B.~A.}\ \bibnamefont {Bernevig}}, \ and\ \bibinfo {author}
  {\bibfnamefont {T.}~\bibnamefont {Neupert}},\ }\href {\doibase
  10.1126/sciadv.aat0346} {\bibfield  {journal} {\bibinfo  {journal} {Science
  Advances}\ }\textbf {\bibinfo {volume} {4}} (\bibinfo {year} {2018}),\
  10.1126/sciadv.aat0346}\BibitemShut {NoStop}%
\bibitem [{\citenamefont {Benalcazar}\ \emph
  {et~al.}(2017{\natexlab{b}})\citenamefont {Benalcazar}, \citenamefont
  {Bernevig},\ and\ \citenamefont {Hughes}}]{Benalcazar_HOTI_PRB17}%
  \BibitemOpen
  \bibfield  {author} {\bibinfo {author} {\bibfnamefont {W.~A.}\ \bibnamefont
  {Benalcazar}}, \bibinfo {author} {\bibfnamefont {B.~A.}\ \bibnamefont
  {Bernevig}}, \ and\ \bibinfo {author} {\bibfnamefont {T.~L.}\ \bibnamefont
  {Hughes}},\ }\href {\doibase 10.1103/PhysRevB.96.245115} {\bibfield
  {journal} {\bibinfo  {journal} {Phys. Rev. B}\ }\textbf {\bibinfo {volume}
  {96}},\ \bibinfo {pages} {245115} (\bibinfo {year}
  {2017}{\natexlab{b}})}\BibitemShut {NoStop}%
\bibitem [{\citenamefont {Hayashi}(2018)}]{Hayashi_HOTI_PRB18}%
  \BibitemOpen
  \bibfield  {author} {\bibinfo {author} {\bibfnamefont {S.}~\bibnamefont
  {Hayashi}},\ }\href {\doibase 10.1007/s00220-018-3229-2} {\bibfield
  {journal} {\bibinfo  {journal} {Communications in Mathematical Physics}\
  }\textbf {\bibinfo {volume} {364}},\ \bibinfo {pages} {343} (\bibinfo {year}
  {2018})}\BibitemShut {NoStop}%
\bibitem [{\citenamefont {Imhof}\ \emph {et~al.}(2018)\citenamefont {Imhof},
  \citenamefont {Berger}, \citenamefont {Bayer}, \citenamefont {Brehm},
  \citenamefont {Molenkamp}, \citenamefont {Kiessling}, \citenamefont
  {Schindler}, \citenamefont {Lee}, \citenamefont {Greiter}, \citenamefont
  {Neupert},\ and\ \citenamefont {Thomale}}]{Imhof_HOTI_NatPhys18}%
  \BibitemOpen
  \bibfield  {author} {\bibinfo {author} {\bibfnamefont {S.}~\bibnamefont
  {Imhof}}, \bibinfo {author} {\bibfnamefont {C.}~\bibnamefont {Berger}},
  \bibinfo {author} {\bibfnamefont {F.}~\bibnamefont {Bayer}}, \bibinfo
  {author} {\bibfnamefont {J.}~\bibnamefont {Brehm}}, \bibinfo {author}
  {\bibfnamefont {L.~W.}\ \bibnamefont {Molenkamp}}, \bibinfo {author}
  {\bibfnamefont {T.}~\bibnamefont {Kiessling}}, \bibinfo {author}
  {\bibfnamefont {F.}~\bibnamefont {Schindler}}, \bibinfo {author}
  {\bibfnamefont {C.~H.}\ \bibnamefont {Lee}}, \bibinfo {author} {\bibfnamefont
  {M.}~\bibnamefont {Greiter}}, \bibinfo {author} {\bibfnamefont
  {T.}~\bibnamefont {Neupert}}, \ and\ \bibinfo {author} {\bibfnamefont
  {R.}~\bibnamefont {Thomale}},\ }\href {\doibase 10.1038/s41567-018-0246-1}
  {\bibfield  {journal} {\bibinfo  {journal} {Nature Physics}\ }\textbf
  {\bibinfo {volume} {14}},\ \bibinfo {pages} {925} (\bibinfo {year}
  {2018})}\BibitemShut {NoStop}%
\bibitem [{\citenamefont {Araki}\ \emph {et~al.}(2019)\citenamefont {Araki},
  \citenamefont {Mizoguchi},\ and\ \citenamefont
  {Hatsugai}}]{Araki_HOTI_PRB19}%
  \BibitemOpen
  \bibfield  {author} {\bibinfo {author} {\bibfnamefont {H.}~\bibnamefont
  {Araki}}, \bibinfo {author} {\bibfnamefont {T.}~\bibnamefont {Mizoguchi}}, \
  and\ \bibinfo {author} {\bibfnamefont {Y.}~\bibnamefont {Hatsugai}},\ }\href
  {\doibase 10.1103/PhysRevB.99.085406} {\bibfield  {journal} {\bibinfo
  {journal} {Phys. Rev. B}\ }\textbf {\bibinfo {volume} {99}},\ \bibinfo
  {pages} {085406} (\bibinfo {year} {2019})}\BibitemShut {NoStop}%
\bibitem [{\citenamefont {Ghorashi}\ \emph {et~al.}(2019)\citenamefont
  {Ghorashi}, \citenamefont {Hu}, \citenamefont {Hughes},\ and\ \citenamefont
  {Rossi}}]{Ghorashi_HOTI_PRB19}%
  \BibitemOpen
  \bibfield  {author} {\bibinfo {author} {\bibfnamefont {S.~A.~A.}\
  \bibnamefont {Ghorashi}}, \bibinfo {author} {\bibfnamefont {X.}~\bibnamefont
  {Hu}}, \bibinfo {author} {\bibfnamefont {T.~L.}\ \bibnamefont {Hughes}}, \
  and\ \bibinfo {author} {\bibfnamefont {E.}~\bibnamefont {Rossi}},\ }\href
  {\doibase 10.1103/PhysRevB.100.020509} {\bibfield  {journal} {\bibinfo
  {journal} {Phys. Rev. B}\ }\textbf {\bibinfo {volume} {100}},\ \bibinfo
  {pages} {020509} (\bibinfo {year} {2019})}\BibitemShut {NoStop}%
\bibitem [{\citenamefont {Kudo}\ \emph {et~al.}(2019)\citenamefont {Kudo},
  \citenamefont {Yoshida},\ and\ \citenamefont {Hatsugai}}]{Kudo_HOTMI_PRL19}%
  \BibitemOpen
  \bibfield  {author} {\bibinfo {author} {\bibfnamefont {K.}~\bibnamefont
  {Kudo}}, \bibinfo {author} {\bibfnamefont {T.}~\bibnamefont {Yoshida}}, \
  and\ \bibinfo {author} {\bibfnamefont {Y.}~\bibnamefont {Hatsugai}},\ }\href
  {\doibase 10.1103/PhysRevLett.123.196402} {\bibfield  {journal} {\bibinfo
  {journal} {Phys. Rev. Lett.}\ }\textbf {\bibinfo {volume} {123}},\ \bibinfo
  {pages} {196402} (\bibinfo {year} {2019})}\BibitemShut {NoStop}%
\bibitem [{\citenamefont {Mizoguchi}\ \emph {et~al.}(2019)\citenamefont
  {Mizoguchi}, \citenamefont {Maruyama}, \citenamefont {Okada},\ and\
  \citenamefont {Hatsugai}}]{Mizoguchi_HOTI_PRM19}%
  \BibitemOpen
  \bibfield  {author} {\bibinfo {author} {\bibfnamefont {T.}~\bibnamefont
  {Mizoguchi}}, \bibinfo {author} {\bibfnamefont {M.}~\bibnamefont {Maruyama}},
  \bibinfo {author} {\bibfnamefont {S.}~\bibnamefont {Okada}}, \ and\ \bibinfo
  {author} {\bibfnamefont {Y.}~\bibnamefont {Hatsugai}},\ }\href {\doibase
  10.1103/PhysRevMaterials.3.114201} {\bibfield  {journal} {\bibinfo  {journal}
  {Phys. Rev. Materials}\ }\textbf {\bibinfo {volume} {3}},\ \bibinfo {pages}
  {114201} (\bibinfo {year} {2019})}\BibitemShut {NoStop}%
\bibitem [{\citenamefont {Hatano}\ and\ \citenamefont
  {Nelson}(1996)}]{Hatano_PRL96}%
  \BibitemOpen
  \bibfield  {author} {\bibinfo {author} {\bibfnamefont {N.}~\bibnamefont
  {Hatano}}\ and\ \bibinfo {author} {\bibfnamefont {D.~R.}\ \bibnamefont
  {Nelson}},\ }\href {\doibase 10.1103/PhysRevLett.77.570} {\bibfield
  {journal} {\bibinfo  {journal} {Phys. Rev. Lett.}\ }\textbf {\bibinfo
  {volume} {77}},\ \bibinfo {pages} {570} (\bibinfo {year} {1996})}\BibitemShut
  {NoStop}%
\bibitem [{\citenamefont {Hu}\ and\ \citenamefont
  {Hughes}(2011)}]{Hu_nH_PRB11}%
  \BibitemOpen
  \bibfield  {author} {\bibinfo {author} {\bibfnamefont {Y.~C.}\ \bibnamefont
  {Hu}}\ and\ \bibinfo {author} {\bibfnamefont {T.~L.}\ \bibnamefont
  {Hughes}},\ }\href {\doibase 10.1103/PhysRevB.84.153101} {\bibfield
  {journal} {\bibinfo  {journal} {Phys. Rev. B}\ }\textbf {\bibinfo {volume}
  {84}},\ \bibinfo {pages} {153101} (\bibinfo {year} {2011})}\BibitemShut
  {NoStop}%
\bibitem [{\citenamefont {Esaki}\ \emph {et~al.}(2011)\citenamefont {Esaki},
  \citenamefont {Sato}, \citenamefont {Hasebe},\ and\ \citenamefont
  {Kohmoto}}]{Esaki_nH_PRB11}%
  \BibitemOpen
  \bibfield  {author} {\bibinfo {author} {\bibfnamefont {K.}~\bibnamefont
  {Esaki}}, \bibinfo {author} {\bibfnamefont {M.}~\bibnamefont {Sato}},
  \bibinfo {author} {\bibfnamefont {K.}~\bibnamefont {Hasebe}}, \ and\ \bibinfo
  {author} {\bibfnamefont {M.}~\bibnamefont {Kohmoto}},\ }\href {\doibase
  10.1103/PhysRevB.84.205128} {\bibfield  {journal} {\bibinfo  {journal} {Phys.
  Rev. B}\ }\textbf {\bibinfo {volume} {84}},\ \bibinfo {pages} {205128}
  (\bibinfo {year} {2011})}\BibitemShut {NoStop}%
\bibitem [{\citenamefont {Guo}\ \emph {et~al.}(2009)\citenamefont {Guo},
  \citenamefont {Salamo}, \citenamefont {Duchesne}, \citenamefont {Morandotti},
  \citenamefont {Volatier-Ravat}, \citenamefont {Aimez}, \citenamefont
  {Siviloglou},\ and\ \citenamefont {Christodoulides}}]{Guo_nHExp_PRL09}%
  \BibitemOpen
  \bibfield  {author} {\bibinfo {author} {\bibfnamefont {A.}~\bibnamefont
  {Guo}}, \bibinfo {author} {\bibfnamefont {G.~J.}\ \bibnamefont {Salamo}},
  \bibinfo {author} {\bibfnamefont {D.}~\bibnamefont {Duchesne}}, \bibinfo
  {author} {\bibfnamefont {R.}~\bibnamefont {Morandotti}}, \bibinfo {author}
  {\bibfnamefont {M.}~\bibnamefont {Volatier-Ravat}}, \bibinfo {author}
  {\bibfnamefont {V.}~\bibnamefont {Aimez}}, \bibinfo {author} {\bibfnamefont
  {G.~A.}\ \bibnamefont {Siviloglou}}, \ and\ \bibinfo {author} {\bibfnamefont
  {D.~N.}\ \bibnamefont {Christodoulides}},\ }\href {\doibase
  10.1103/PhysRevLett.103.093902} {\bibfield  {journal} {\bibinfo  {journal}
  {Phys. Rev. Lett.}\ }\textbf {\bibinfo {volume} {103}},\ \bibinfo {pages}
  {093902} (\bibinfo {year} {2009})}\BibitemShut {NoStop}%
\bibitem [{\citenamefont {R{\"u}ter}\ \emph {et~al.}(2010)\citenamefont
  {R{\"u}ter}, \citenamefont {Makris}, \citenamefont {El-Ganainy},
  \citenamefont {Christodoulides}, \citenamefont {Segev},\ and\ \citenamefont
  {Kip}}]{Ruter_nHExp_NatPhys10}%
  \BibitemOpen
  \bibfield  {author} {\bibinfo {author} {\bibfnamefont {C.~E.}\ \bibnamefont
  {R{\"u}ter}}, \bibinfo {author} {\bibfnamefont {K.~G.}\ \bibnamefont
  {Makris}}, \bibinfo {author} {\bibfnamefont {R.}~\bibnamefont {El-Ganainy}},
  \bibinfo {author} {\bibfnamefont {D.~N.}\ \bibnamefont {Christodoulides}},
  \bibinfo {author} {\bibfnamefont {M.}~\bibnamefont {Segev}}, \ and\ \bibinfo
  {author} {\bibfnamefont {D.}~\bibnamefont {Kip}},\ }\href@noop {} {\bibfield
  {journal} {\bibinfo  {journal} {Nature physics}\ }\textbf {\bibinfo {volume}
  {6}},\ \bibinfo {pages} {192} (\bibinfo {year} {2010})}\BibitemShut {NoStop}%
\bibitem [{\citenamefont {Regensburger}\ \emph {et~al.}(2012)\citenamefont
  {Regensburger}, \citenamefont {Bersch}, \citenamefont {Miri}, \citenamefont
  {Onishchukov}, \citenamefont {Christodoulides},\ and\ \citenamefont
  {Peschel}}]{Regensburger_nHExp_Nat12}%
  \BibitemOpen
  \bibfield  {author} {\bibinfo {author} {\bibfnamefont {A.}~\bibnamefont
  {Regensburger}}, \bibinfo {author} {\bibfnamefont {C.}~\bibnamefont
  {Bersch}}, \bibinfo {author} {\bibfnamefont {M.-A.}\ \bibnamefont {Miri}},
  \bibinfo {author} {\bibfnamefont {G.}~\bibnamefont {Onishchukov}}, \bibinfo
  {author} {\bibfnamefont {D.~N.}\ \bibnamefont {Christodoulides}}, \ and\
  \bibinfo {author} {\bibfnamefont {U.}~\bibnamefont {Peschel}},\ }\href@noop
  {} {\bibfield  {journal} {\bibinfo  {journal} {Nature}\ }\textbf {\bibinfo
  {volume} {488}},\ \bibinfo {pages} {167} (\bibinfo {year}
  {2012})}\BibitemShut {NoStop}%
\bibitem [{\citenamefont {Zhen}\ \emph {et~al.}(2015)\citenamefont {Zhen},
  \citenamefont {Hsu}, \citenamefont {Igarashi}, \citenamefont {Lu},
  \citenamefont {Kaminer}, \citenamefont {Pick}, \citenamefont {Chua},
  \citenamefont {Joannopoulos},\ and\ \citenamefont
  {Soljacic}}]{Zhen_AcciEP_Nat15}%
  \BibitemOpen
  \bibfield  {author} {\bibinfo {author} {\bibfnamefont {B.}~\bibnamefont
  {Zhen}}, \bibinfo {author} {\bibfnamefont {C.~W.}\ \bibnamefont {Hsu}},
  \bibinfo {author} {\bibfnamefont {Y.}~\bibnamefont {Igarashi}}, \bibinfo
  {author} {\bibfnamefont {L.}~\bibnamefont {Lu}}, \bibinfo {author}
  {\bibfnamefont {I.}~\bibnamefont {Kaminer}}, \bibinfo {author} {\bibfnamefont
  {A.}~\bibnamefont {Pick}}, \bibinfo {author} {\bibfnamefont {S.-L.}\
  \bibnamefont {Chua}}, \bibinfo {author} {\bibfnamefont {J.~D.}\ \bibnamefont
  {Joannopoulos}}, \ and\ \bibinfo {author} {\bibfnamefont {M.}~\bibnamefont
  {Soljacic}},\ }\href {http://dx.doi.org/10.1038/nature14889} {\bibfield
  {journal} {\bibinfo  {journal} {Nature}\ }\textbf {\bibinfo {volume} {525}},\
  \bibinfo {pages} {354 EP } (\bibinfo {year} {2015})}\BibitemShut {NoStop}%
\bibitem [{\citenamefont {Hassan}\ \emph {et~al.}(2017)\citenamefont {Hassan},
  \citenamefont {Zhen}, \citenamefont {Solja\ifmmode \check{c}\else
  \v{c}\fi{}i\ifmmode~\acute{c}\else \'{c}\fi{}}, \citenamefont {Khajavikhan},\
  and\ \citenamefont {Christodoulides}}]{Hassan_EP_PRL17}%
  \BibitemOpen
  \bibfield  {author} {\bibinfo {author} {\bibfnamefont {A.~U.}\ \bibnamefont
  {Hassan}}, \bibinfo {author} {\bibfnamefont {B.}~\bibnamefont {Zhen}},
  \bibinfo {author} {\bibfnamefont {M.}~\bibnamefont {Solja\ifmmode
  \check{c}\else \v{c}\fi{}i\ifmmode~\acute{c}\else \'{c}\fi{}}}, \bibinfo
  {author} {\bibfnamefont {M.}~\bibnamefont {Khajavikhan}}, \ and\ \bibinfo
  {author} {\bibfnamefont {D.~N.}\ \bibnamefont {Christodoulides}},\ }\href
  {\doibase 10.1103/PhysRevLett.118.093002} {\bibfield  {journal} {\bibinfo
  {journal} {Phys. Rev. Lett.}\ }\textbf {\bibinfo {volume} {118}},\ \bibinfo
  {pages} {093002} (\bibinfo {year} {2017})}\BibitemShut {NoStop}%
\bibitem [{\citenamefont {Lee}(2016)}]{TELeePRL16_Half_quantized}%
  \BibitemOpen
  \bibfield  {author} {\bibinfo {author} {\bibfnamefont {T.~E.}\ \bibnamefont
  {Lee}},\ }\href {\doibase 10.1103/PhysRevLett.116.133903} {\bibfield
  {journal} {\bibinfo  {journal} {Phys. Rev. Lett.}\ }\textbf {\bibinfo
  {volume} {116}},\ \bibinfo {pages} {133903} (\bibinfo {year}
  {2016})}\BibitemShut {NoStop}%
\bibitem [{\citenamefont {Xu}\ \emph {et~al.}(2017)\citenamefont {Xu},
  \citenamefont {Wang},\ and\ \citenamefont
  {Duan}}]{YXuPRL17_exceptional_ring}%
  \BibitemOpen
  \bibfield  {author} {\bibinfo {author} {\bibfnamefont {Y.}~\bibnamefont
  {Xu}}, \bibinfo {author} {\bibfnamefont {S.-T.}\ \bibnamefont {Wang}}, \ and\
  \bibinfo {author} {\bibfnamefont {L.-M.}\ \bibnamefont {Duan}},\ }\href
  {\doibase 10.1103/PhysRevLett.118.045701} {\bibfield  {journal} {\bibinfo
  {journal} {Phys. Rev. Lett.}\ }\textbf {\bibinfo {volume} {118}},\ \bibinfo
  {pages} {045701} (\bibinfo {year} {2017})}\BibitemShut {NoStop}%
\bibitem [{\citenamefont {Yoshida}\ \emph
  {et~al.}(2019{\natexlab{a}})\citenamefont {Yoshida}, \citenamefont {Kudo},\
  and\ \citenamefont {Hatsugai}}]{Yoshida_nHFQH19}%
  \BibitemOpen
  \bibfield  {author} {\bibinfo {author} {\bibfnamefont {T.}~\bibnamefont
  {Yoshida}}, \bibinfo {author} {\bibfnamefont {K.}~\bibnamefont {Kudo}}, \
  and\ \bibinfo {author} {\bibfnamefont {Y.}~\bibnamefont {Hatsugai}},\ }\href
  {\doibase 10.1038/s41598-019-53253-8} {\bibfield  {journal} {\bibinfo
  {journal} {Scientific Reports}\ }\textbf {\bibinfo {volume} {9}},\ \bibinfo
  {pages} {16895} (\bibinfo {year} {2019}{\natexlab{a}})}\BibitemShut {NoStop}%
\bibitem [{\citenamefont {Kozii}\ and\ \citenamefont
  {Fu}(2017)}]{VKozii_nH_arXiv17}%
  \BibitemOpen
  \bibfield  {author} {\bibinfo {author} {\bibfnamefont {V.}~\bibnamefont
  {Kozii}}\ and\ \bibinfo {author} {\bibfnamefont {L.}~\bibnamefont {Fu}},\
  }\href@noop {} {\bibfield  {journal} {\bibinfo  {journal} {arXiv preprint
  arXiv:1708.05841}\ } (\bibinfo {year} {2017})}\BibitemShut {NoStop}%
\bibitem [{\citenamefont {Yoshida}\ \emph {et~al.}(2018)\citenamefont
  {Yoshida}, \citenamefont {Peters},\ and\ \citenamefont
  {Kawakami}}]{Yoshida_EP_DMFT_PRB18}%
  \BibitemOpen
  \bibfield  {author} {\bibinfo {author} {\bibfnamefont {T.}~\bibnamefont
  {Yoshida}}, \bibinfo {author} {\bibfnamefont {R.}~\bibnamefont {Peters}}, \
  and\ \bibinfo {author} {\bibfnamefont {N.}~\bibnamefont {Kawakami}},\ }\href
  {\doibase 10.1103/PhysRevB.98.035141} {\bibfield  {journal} {\bibinfo
  {journal} {Phys. Rev. B}\ }\textbf {\bibinfo {volume} {98}},\ \bibinfo
  {pages} {035141} (\bibinfo {year} {2018})}\BibitemShut {NoStop}%
\bibitem [{\citenamefont {Kimura}\ \emph {et~al.}(2019)\citenamefont {Kimura},
  \citenamefont {Yoshida},\ and\ \citenamefont
  {Kawakami}}]{Kimura_SPERs_PRB19}%
  \BibitemOpen
  \bibfield  {author} {\bibinfo {author} {\bibfnamefont {K.}~\bibnamefont
  {Kimura}}, \bibinfo {author} {\bibfnamefont {T.}~\bibnamefont {Yoshida}}, \
  and\ \bibinfo {author} {\bibfnamefont {N.}~\bibnamefont {Kawakami}},\ }\href
  {\doibase 10.1103/PhysRevB.100.115124} {\bibfield  {journal} {\bibinfo
  {journal} {Phys. Rev. B}\ }\textbf {\bibinfo {volume} {100}},\ \bibinfo
  {pages} {115124} (\bibinfo {year} {2019})}\BibitemShut {NoStop}%
\bibitem [{\citenamefont {Zyuzin}\ and\ \citenamefont
  {Zyuzin}(2018)}]{Zyuzin_nHEP_PRB18}%
  \BibitemOpen
  \bibfield  {author} {\bibinfo {author} {\bibfnamefont {A.~A.}\ \bibnamefont
  {Zyuzin}}\ and\ \bibinfo {author} {\bibfnamefont {A.~Y.}\ \bibnamefont
  {Zyuzin}},\ }\href {\doibase 10.1103/PhysRevB.97.041203} {\bibfield
  {journal} {\bibinfo  {journal} {Phys. Rev. B}\ }\textbf {\bibinfo {volume}
  {97}},\ \bibinfo {pages} {041203} (\bibinfo {year} {2018})}\BibitemShut
  {NoStop}%
\bibitem [{\citenamefont {Papaj}\ \emph {et~al.}(2019)\citenamefont {Papaj},
  \citenamefont {Isobe},\ and\ \citenamefont {Fu}}]{Papaji_nHEP_PRB19}%
  \BibitemOpen
  \bibfield  {author} {\bibinfo {author} {\bibfnamefont {M.}~\bibnamefont
  {Papaj}}, \bibinfo {author} {\bibfnamefont {H.}~\bibnamefont {Isobe}}, \ and\
  \bibinfo {author} {\bibfnamefont {L.}~\bibnamefont {Fu}},\ }\href {\doibase
  10.1103/PhysRevB.99.201107} {\bibfield  {journal} {\bibinfo  {journal} {Phys.
  Rev. B}\ }\textbf {\bibinfo {volume} {99}},\ \bibinfo {pages} {201107}
  (\bibinfo {year} {2019})}\BibitemShut {NoStop}%
\bibitem [{\citenamefont {Matsushita}\ \emph {et~al.}(2019)\citenamefont
  {Matsushita}, \citenamefont {Nagai},\ and\ \citenamefont
  {Fujimoto}}]{Matsushita_ER_arXiv19}%
  \BibitemOpen
  \bibfield  {author} {\bibinfo {author} {\bibfnamefont {T.}~\bibnamefont
  {Matsushita}}, \bibinfo {author} {\bibfnamefont {Y.}~\bibnamefont {Nagai}}, \
  and\ \bibinfo {author} {\bibfnamefont {S.}~\bibnamefont {Fujimoto}},\ }\href
  {\doibase 10.1103/PhysRevB.100.245205} {\bibfield  {journal} {\bibinfo
  {journal} {Phys. Rev. B}\ }\textbf {\bibinfo {volume} {100}},\ \bibinfo
  {pages} {245205} (\bibinfo {year} {2019})}\BibitemShut {NoStop}%
\bibitem [{\citenamefont {Yoshida}\ \emph {et~al.}(2020)\citenamefont
  {Yoshida}, \citenamefont {Peters}, \citenamefont {Kawakami},\ and\
  \citenamefont {Hatsugai}}]{Yoshida_nHRev_arXiv20}%
  \BibitemOpen
  \bibfield  {author} {\bibinfo {author} {\bibfnamefont {T.}~\bibnamefont
  {Yoshida}}, \bibinfo {author} {\bibfnamefont {R.}~\bibnamefont {Peters}},
  \bibinfo {author} {\bibfnamefont {N.}~\bibnamefont {Kawakami}}, \ and\
  \bibinfo {author} {\bibfnamefont {Y.}~\bibnamefont {Hatsugai}},\ }\href@noop
  {} {\bibfield  {journal} {\bibinfo  {journal} {arXiv preprint
  arXiv:2002.11265}\ } (\bibinfo {year} {2020})}\BibitemShut {NoStop}%
\bibitem [{\citenamefont {Kawabata}\ \emph
  {et~al.}(2019{\natexlab{a}})\citenamefont {Kawabata}, \citenamefont
  {Shiozaki}, \citenamefont {Ueda},\ and\ \citenamefont
  {Sato}}]{Kawabata_gapped_PRX19}%
  \BibitemOpen
  \bibfield  {author} {\bibinfo {author} {\bibfnamefont {K.}~\bibnamefont
  {Kawabata}}, \bibinfo {author} {\bibfnamefont {K.}~\bibnamefont {Shiozaki}},
  \bibinfo {author} {\bibfnamefont {M.}~\bibnamefont {Ueda}}, \ and\ \bibinfo
  {author} {\bibfnamefont {M.}~\bibnamefont {Sato}},\ }\href {\doibase
  10.1103/PhysRevX.9.041015} {\bibfield  {journal} {\bibinfo  {journal} {Phys.
  Rev. X}\ }\textbf {\bibinfo {volume} {9}},\ \bibinfo {pages} {041015}
  (\bibinfo {year} {2019}{\natexlab{a}})}\BibitemShut {NoStop}%
\bibitem [{\citenamefont {Shen}\ \emph {et~al.}(2017)\citenamefont {Shen},
  \citenamefont {Zhen},\ and\ \citenamefont {Fu}}]{HShen2017_non-Hermi}%
  \BibitemOpen
  \bibfield  {author} {\bibinfo {author} {\bibfnamefont {H.}~\bibnamefont
  {Shen}}, \bibinfo {author} {\bibfnamefont {B.}~\bibnamefont {Zhen}}, \ and\
  \bibinfo {author} {\bibfnamefont {L.}~\bibnamefont {Fu}},\ }\href@noop {}
  {\bibfield  {journal} {\bibinfo  {journal} {arXiv preprint arXiv:1706.07435}\
  } (\bibinfo {year} {2017})}\BibitemShut {NoStop}%
\bibitem [{\citenamefont {Kawabata}\ \emph
  {et~al.}(2019{\natexlab{b}})\citenamefont {Kawabata}, \citenamefont
  {Higashikawa}, \citenamefont {Gong}, \citenamefont {Ashida},\ and\
  \citenamefont {Ueda}}]{KKawabata_TopoUni_NatComm19}%
  \BibitemOpen
  \bibfield  {author} {\bibinfo {author} {\bibfnamefont {K.}~\bibnamefont
  {Kawabata}}, \bibinfo {author} {\bibfnamefont {S.}~\bibnamefont
  {Higashikawa}}, \bibinfo {author} {\bibfnamefont {Z.}~\bibnamefont {Gong}},
  \bibinfo {author} {\bibfnamefont {Y.}~\bibnamefont {Ashida}}, \ and\ \bibinfo
  {author} {\bibfnamefont {M.}~\bibnamefont {Ueda}},\ }\href {\doibase
  10.1038/s41467-018-08254-y} {\bibfield  {journal} {\bibinfo  {journal}
  {Nature Communications}\ }\textbf {\bibinfo {volume} {10}},\ \bibinfo {pages}
  {297} (\bibinfo {year} {2019}{\natexlab{b}})}\BibitemShut {NoStop}%
\bibitem [{\citenamefont {Gong}\ \emph {et~al.}(2018)\citenamefont {Gong},
  \citenamefont {Ashida}, \citenamefont {Kawabata}, \citenamefont {Takasan},
  \citenamefont {Higashikawa},\ and\ \citenamefont {Ueda}}]{Gong_class_PRX18}%
  \BibitemOpen
  \bibfield  {author} {\bibinfo {author} {\bibfnamefont {Z.}~\bibnamefont
  {Gong}}, \bibinfo {author} {\bibfnamefont {Y.}~\bibnamefont {Ashida}},
  \bibinfo {author} {\bibfnamefont {K.}~\bibnamefont {Kawabata}}, \bibinfo
  {author} {\bibfnamefont {K.}~\bibnamefont {Takasan}}, \bibinfo {author}
  {\bibfnamefont {S.}~\bibnamefont {Higashikawa}}, \ and\ \bibinfo {author}
  {\bibfnamefont {M.}~\bibnamefont {Ueda}},\ }\href {\doibase
  10.1103/PhysRevX.8.031079} {\bibfield  {journal} {\bibinfo  {journal} {Phys.
  Rev. X}\ }\textbf {\bibinfo {volume} {8}},\ \bibinfo {pages} {031079}
  (\bibinfo {year} {2018})}\BibitemShut {NoStop}%
\bibitem [{\citenamefont {Zhou}\ and\ \citenamefont
  {Lee}(2019)}]{Zhou_gapped_class_PRB19}%
  \BibitemOpen
  \bibfield  {author} {\bibinfo {author} {\bibfnamefont {H.}~\bibnamefont
  {Zhou}}\ and\ \bibinfo {author} {\bibfnamefont {J.~Y.}\ \bibnamefont {Lee}},\
  }\href {\doibase 10.1103/PhysRevB.99.235112} {\bibfield  {journal} {\bibinfo
  {journal} {Phys. Rev. B}\ }\textbf {\bibinfo {volume} {99}},\ \bibinfo
  {pages} {235112} (\bibinfo {year} {2019})}\BibitemShut {NoStop}%
\bibitem [{\citenamefont {Kat{\=o}}(1966)}]{TKato_EP_book1966}%
  \BibitemOpen
  \bibfield  {author} {\bibinfo {author} {\bibfnamefont {T.}~\bibnamefont
  {Kat{\=o}}},\ }\href@noop {} {\emph {\bibinfo {title} {Perturbation theory
  for linear operators}}},\ Vol.\ \bibinfo {volume} {132}\ (\bibinfo
  {publisher} {Springer},\ \bibinfo {year} {1966})\BibitemShut {NoStop}%
\bibitem [{\citenamefont {Rotter}(2009)}]{Rotter_EP_JPA09}%
  \BibitemOpen
  \bibfield  {author} {\bibinfo {author} {\bibfnamefont {I.}~\bibnamefont
  {Rotter}},\ }\href {\doibase 10.1088/1751-8113/42/15/153001} {\bibfield
  {journal} {\bibinfo  {journal} {Journal of Physics A: Mathematical and
  Theoretical}\ }\textbf {\bibinfo {volume} {42}},\ \bibinfo {pages} {153001}
  (\bibinfo {year} {2009})}\BibitemShut {NoStop}%
\bibitem [{\citenamefont {Budich}\ \emph {et~al.}(2019)\citenamefont {Budich},
  \citenamefont {Carlstr\"om}, \citenamefont {Kunst},\ and\ \citenamefont
  {Bergholtz}}]{Budich_SPERs_PRB19}%
  \BibitemOpen
  \bibfield  {author} {\bibinfo {author} {\bibfnamefont {J.~C.}\ \bibnamefont
  {Budich}}, \bibinfo {author} {\bibfnamefont {J.}~\bibnamefont {Carlstr\"om}},
  \bibinfo {author} {\bibfnamefont {F.~K.}\ \bibnamefont {Kunst}}, \ and\
  \bibinfo {author} {\bibfnamefont {E.~J.}\ \bibnamefont {Bergholtz}},\ }\href
  {\doibase 10.1103/PhysRevB.99.041406} {\bibfield  {journal} {\bibinfo
  {journal} {Phys. Rev. B}\ }\textbf {\bibinfo {volume} {99}},\ \bibinfo
  {pages} {041406} (\bibinfo {year} {2019})}\BibitemShut {NoStop}%
\bibitem [{\citenamefont {Okugawa}\ and\ \citenamefont
  {Yokoyama}(2019)}]{Okugawa_SPERs_PRB19}%
  \BibitemOpen
  \bibfield  {author} {\bibinfo {author} {\bibfnamefont {R.}~\bibnamefont
  {Okugawa}}\ and\ \bibinfo {author} {\bibfnamefont {T.}~\bibnamefont
  {Yokoyama}},\ }\href {\doibase 10.1103/PhysRevB.99.041202} {\bibfield
  {journal} {\bibinfo  {journal} {Phys. Rev. B}\ }\textbf {\bibinfo {volume}
  {99}},\ \bibinfo {pages} {041202} (\bibinfo {year} {2019})}\BibitemShut
  {NoStop}%
\bibitem [{\citenamefont {Zhou}\ \emph {et~al.}(2019)\citenamefont {Zhou},
  \citenamefont {Lee}, \citenamefont {Liu},\ and\ \citenamefont
  {Zhen}}]{Zhou_SPERs_Optica19}%
  \BibitemOpen
  \bibfield  {author} {\bibinfo {author} {\bibfnamefont {H.}~\bibnamefont
  {Zhou}}, \bibinfo {author} {\bibfnamefont {J.~Y.}\ \bibnamefont {Lee}},
  \bibinfo {author} {\bibfnamefont {S.}~\bibnamefont {Liu}}, \ and\ \bibinfo
  {author} {\bibfnamefont {B.}~\bibnamefont {Zhen}},\ }\href {\doibase
  10.1364/OPTICA.6.000190} {\bibfield  {journal} {\bibinfo  {journal} {Optica}\
  }\textbf {\bibinfo {volume} {6}},\ \bibinfo {pages} {190} (\bibinfo {year}
  {2019})}\BibitemShut {NoStop}%
\bibitem [{\citenamefont {Yoshida}\ \emph
  {et~al.}(2019{\natexlab{b}})\citenamefont {Yoshida}, \citenamefont {Peters},
  \citenamefont {Kawakami},\ and\ \citenamefont
  {Hatsugai}}]{Yoshida_SPERs_PRB19}%
  \BibitemOpen
  \bibfield  {author} {\bibinfo {author} {\bibfnamefont {T.}~\bibnamefont
  {Yoshida}}, \bibinfo {author} {\bibfnamefont {R.}~\bibnamefont {Peters}},
  \bibinfo {author} {\bibfnamefont {N.}~\bibnamefont {Kawakami}}, \ and\
  \bibinfo {author} {\bibfnamefont {Y.}~\bibnamefont {Hatsugai}},\ }\href
  {\doibase 10.1103/PhysRevB.99.121101} {\bibfield  {journal} {\bibinfo
  {journal} {Phys. Rev. B}\ }\textbf {\bibinfo {volume} {99}},\ \bibinfo
  {pages} {121101} (\bibinfo {year} {2019}{\natexlab{b}})}\BibitemShut
  {NoStop}%
\bibitem [{\citenamefont {Kawabata}\ \emph
  {et~al.}(2019{\natexlab{c}})\citenamefont {Kawabata}, \citenamefont
  {Bessho},\ and\ \citenamefont {Sato}}]{Kawabata_gapless_PRL19}%
  \BibitemOpen
  \bibfield  {author} {\bibinfo {author} {\bibfnamefont {K.}~\bibnamefont
  {Kawabata}}, \bibinfo {author} {\bibfnamefont {T.}~\bibnamefont {Bessho}}, \
  and\ \bibinfo {author} {\bibfnamefont {M.}~\bibnamefont {Sato}},\ }\href
  {\doibase 10.1103/PhysRevLett.123.066405} {\bibfield  {journal} {\bibinfo
  {journal} {Phys. Rev. Lett.}\ }\textbf {\bibinfo {volume} {123}},\ \bibinfo
  {pages} {066405} (\bibinfo {year} {2019}{\natexlab{c}})}\BibitemShut
  {NoStop}%
\bibitem [{\citenamefont {Yoshida}\ and\ \citenamefont
  {Hatsugai}(2019)}]{Yoshida_SPERs_mech19}%
  \BibitemOpen
  \bibfield  {author} {\bibinfo {author} {\bibfnamefont {T.}~\bibnamefont
  {Yoshida}}\ and\ \bibinfo {author} {\bibfnamefont {Y.}~\bibnamefont
  {Hatsugai}},\ }\href {\doibase 10.1103/PhysRevB.100.054109} {\bibfield
  {journal} {\bibinfo  {journal} {Phys. Rev. B}\ }\textbf {\bibinfo {volume}
  {100}},\ \bibinfo {pages} {054109} (\bibinfo {year} {2019})}\BibitemShut
  {NoStop}%
\bibitem [{\citenamefont {Carlstr\"om}\ and\ \citenamefont
  {Bergholtz}(2018)}]{Carlstrom_nHknot_PRA18}%
  \BibitemOpen
  \bibfield  {author} {\bibinfo {author} {\bibfnamefont {J.}~\bibnamefont
  {Carlstr\"om}}\ and\ \bibinfo {author} {\bibfnamefont {E.~J.}\ \bibnamefont
  {Bergholtz}},\ }\href {\doibase 10.1103/PhysRevA.98.042114} {\bibfield
  {journal} {\bibinfo  {journal} {Phys. Rev. A}\ }\textbf {\bibinfo {volume}
  {98}},\ \bibinfo {pages} {042114} (\bibinfo {year} {2018})}\BibitemShut
  {NoStop}%
\bibitem [{\citenamefont {Martinez~Alvarez}\ \emph {et~al.}(2018)\citenamefont
  {Martinez~Alvarez}, \citenamefont {Barrios~Vargas},\ and\ \citenamefont
  {Foa~Torres}}]{Alvarez_nHSkin_PRB18}%
  \BibitemOpen
  \bibfield  {author} {\bibinfo {author} {\bibfnamefont {V.~M.}\ \bibnamefont
  {Martinez~Alvarez}}, \bibinfo {author} {\bibfnamefont {J.~E.}\ \bibnamefont
  {Barrios~Vargas}}, \ and\ \bibinfo {author} {\bibfnamefont {L.~E.~F.}\
  \bibnamefont {Foa~Torres}},\ }\href {\doibase 10.1103/PhysRevB.97.121401}
  {\bibfield  {journal} {\bibinfo  {journal} {Phys. Rev. B}\ }\textbf {\bibinfo
  {volume} {97}},\ \bibinfo {pages} {121401} (\bibinfo {year}
  {2018})}\BibitemShut {NoStop}%
\bibitem [{\citenamefont {Kunst}\ \emph {et~al.}(2018)\citenamefont {Kunst},
  \citenamefont {Edvardsson}, \citenamefont {Budich},\ and\ \citenamefont
  {Bergholtz}}]{KFlore_nHSkin_PRL18}%
  \BibitemOpen
  \bibfield  {author} {\bibinfo {author} {\bibfnamefont {F.~K.}\ \bibnamefont
  {Kunst}}, \bibinfo {author} {\bibfnamefont {E.}~\bibnamefont {Edvardsson}},
  \bibinfo {author} {\bibfnamefont {J.~C.}\ \bibnamefont {Budich}}, \ and\
  \bibinfo {author} {\bibfnamefont {E.~J.}\ \bibnamefont {Bergholtz}},\ }\href
  {\doibase 10.1103/PhysRevLett.121.026808} {\bibfield  {journal} {\bibinfo
  {journal} {Phys. Rev. Lett.}\ }\textbf {\bibinfo {volume} {121}},\ \bibinfo
  {pages} {026808} (\bibinfo {year} {2018})}\BibitemShut {NoStop}%
\bibitem [{\citenamefont {Yao}\ and\ \citenamefont
  {Wang}(2018)}]{SYao_nHSkin-1D_PRL18}%
  \BibitemOpen
  \bibfield  {author} {\bibinfo {author} {\bibfnamefont {S.}~\bibnamefont
  {Yao}}\ and\ \bibinfo {author} {\bibfnamefont {Z.}~\bibnamefont {Wang}},\
  }\href {\doibase 10.1103/PhysRevLett.121.086803} {\bibfield  {journal}
  {\bibinfo  {journal} {Phys. Rev. Lett.}\ }\textbf {\bibinfo {volume} {121}},\
  \bibinfo {pages} {086803} (\bibinfo {year} {2018})}\BibitemShut {NoStop}%
\bibitem [{\citenamefont {Yao}\ \emph {et~al.}(2018)\citenamefont {Yao},
  \citenamefont {Song},\ and\ \citenamefont {Wang}}]{SYao_nHSkin-2D_PRL18}%
  \BibitemOpen
  \bibfield  {author} {\bibinfo {author} {\bibfnamefont {S.}~\bibnamefont
  {Yao}}, \bibinfo {author} {\bibfnamefont {F.}~\bibnamefont {Song}}, \ and\
  \bibinfo {author} {\bibfnamefont {Z.}~\bibnamefont {Wang}},\ }\href {\doibase
  10.1103/PhysRevLett.121.136802} {\bibfield  {journal} {\bibinfo  {journal}
  {Phys. Rev. Lett.}\ }\textbf {\bibinfo {volume} {121}},\ \bibinfo {pages}
  {136802} (\bibinfo {year} {2018})}\BibitemShut {NoStop}%
\bibitem [{\citenamefont {Edvardsson}\ \emph {et~al.}(2019)\citenamefont
  {Edvardsson}, \citenamefont {Kunst},\ and\ \citenamefont
  {Bergholtz}}]{EElizabet_PRBnHSkinHOTI_PRB19}%
  \BibitemOpen
  \bibfield  {author} {\bibinfo {author} {\bibfnamefont {E.}~\bibnamefont
  {Edvardsson}}, \bibinfo {author} {\bibfnamefont {F.~K.}\ \bibnamefont
  {Kunst}}, \ and\ \bibinfo {author} {\bibfnamefont {E.~J.}\ \bibnamefont
  {Bergholtz}},\ }\href {\doibase 10.1103/PhysRevB.99.081302} {\bibfield
  {journal} {\bibinfo  {journal} {Phys. Rev. B}\ }\textbf {\bibinfo {volume}
  {99}},\ \bibinfo {pages} {081302} (\bibinfo {year} {2019})}\BibitemShut
  {NoStop}%
\bibitem [{\citenamefont {Rui}\ \emph {et~al.}(2019{\natexlab{a}})\citenamefont
  {Rui}, \citenamefont {Zhao},\ and\ \citenamefont {Schnyder}}]{Rui_nH_PRB19}%
  \BibitemOpen
  \bibfield  {author} {\bibinfo {author} {\bibfnamefont {W.~B.}\ \bibnamefont
  {Rui}}, \bibinfo {author} {\bibfnamefont {Y.~X.}\ \bibnamefont {Zhao}}, \
  and\ \bibinfo {author} {\bibfnamefont {A.~P.}\ \bibnamefont {Schnyder}},\
  }\href {\doibase 10.1103/PhysRevB.99.241110} {\bibfield  {journal} {\bibinfo
  {journal} {Phys. Rev. B}\ }\textbf {\bibinfo {volume} {99}},\ \bibinfo
  {pages} {241110} (\bibinfo {year} {2019}{\natexlab{a}})}\BibitemShut
  {NoStop}%
\bibitem [{\citenamefont {Yokomizo}\ and\ \citenamefont
  {Murakami}(2019)}]{Yokomizo_BBC_PRL19}%
  \BibitemOpen
  \bibfield  {author} {\bibinfo {author} {\bibfnamefont {K.}~\bibnamefont
  {Yokomizo}}\ and\ \bibinfo {author} {\bibfnamefont {S.}~\bibnamefont
  {Murakami}},\ }\href {\doibase 10.1103/PhysRevLett.123.066404} {\bibfield
  {journal} {\bibinfo  {journal} {Phys. Rev. Lett.}\ }\textbf {\bibinfo
  {volume} {123}},\ \bibinfo {pages} {066404} (\bibinfo {year}
  {2019})}\BibitemShut {NoStop}%
\bibitem [{\citenamefont {Okuma}\ and\ \citenamefont
  {Sato}(2019)}]{Okuma_nHBBCpg_PRL19}%
  \BibitemOpen
  \bibfield  {author} {\bibinfo {author} {\bibfnamefont {N.}~\bibnamefont
  {Okuma}}\ and\ \bibinfo {author} {\bibfnamefont {M.}~\bibnamefont {Sato}},\
  }\href {\doibase 10.1103/PhysRevLett.123.097701} {\bibfield  {journal}
  {\bibinfo  {journal} {Phys. Rev. Lett.}\ }\textbf {\bibinfo {volume} {123}},\
  \bibinfo {pages} {097701} (\bibinfo {year} {2019})}\BibitemShut {NoStop}%
\bibitem [{\citenamefont {Xiao}\ \emph {et~al.}(2020)\citenamefont {Xiao},
  \citenamefont {Deng}, \citenamefont {Wang}, \citenamefont {Zhu},
  \citenamefont {Wang}, \citenamefont {Yi},\ and\ \citenamefont
  {Xue}}]{Xiao_nHSkin_Exp_arXiv19}%
  \BibitemOpen
  \bibfield  {author} {\bibinfo {author} {\bibfnamefont {L.}~\bibnamefont
  {Xiao}}, \bibinfo {author} {\bibfnamefont {T.}~\bibnamefont {Deng}}, \bibinfo
  {author} {\bibfnamefont {K.}~\bibnamefont {Wang}}, \bibinfo {author}
  {\bibfnamefont {G.}~\bibnamefont {Zhu}}, \bibinfo {author} {\bibfnamefont
  {Z.}~\bibnamefont {Wang}}, \bibinfo {author} {\bibfnamefont {W.}~\bibnamefont
  {Yi}}, \ and\ \bibinfo {author} {\bibfnamefont {P.}~\bibnamefont {Xue}},\
  }\href {\doibase 10.1038/s41567-020-0836-6} {\bibfield  {journal} {\bibinfo
  {journal} {Nature Physics}\ } (\bibinfo {year} {2020}),\
  10.1038/s41567-020-0836-6}\BibitemShut {NoStop}%
\bibitem [{\citenamefont {Lee}\ and\ \citenamefont
  {Thomale}(2019)}]{Lee_Skin19}%
  \BibitemOpen
  \bibfield  {author} {\bibinfo {author} {\bibfnamefont {C.~H.}\ \bibnamefont
  {Lee}}\ and\ \bibinfo {author} {\bibfnamefont {R.}~\bibnamefont {Thomale}},\
  }\href {\doibase 10.1103/PhysRevB.99.201103} {\bibfield  {journal} {\bibinfo
  {journal} {Phys. Rev. B}\ }\textbf {\bibinfo {volume} {99}},\ \bibinfo
  {pages} {201103} (\bibinfo {year} {2019})}\BibitemShut {NoStop}%
\bibitem [{\citenamefont {Zhang}\ \emph {et~al.}(2019)\citenamefont {Zhang},
  \citenamefont {Yang},\ and\ \citenamefont {Fang}}]{Zhang_BECskin19}%
  \BibitemOpen
  \bibfield  {author} {\bibinfo {author} {\bibfnamefont {K.}~\bibnamefont
  {Zhang}}, \bibinfo {author} {\bibfnamefont {Z.}~\bibnamefont {Yang}}, \ and\
  \bibinfo {author} {\bibfnamefont {C.}~\bibnamefont {Fang}},\ }\href@noop {}
  {\bibfield  {journal} {\bibinfo  {journal} {arXiv preprint arXiv:1910.01131}\
  } (\bibinfo {year} {2019})}\BibitemShut {NoStop}%
\bibitem [{\citenamefont {Okuma}\ \emph {et~al.}(2020)\citenamefont {Okuma},
  \citenamefont {Kawabata}, \citenamefont {Shiozaki},\ and\ \citenamefont
  {Sato}}]{Okuma_BECskin19}%
  \BibitemOpen
  \bibfield  {author} {\bibinfo {author} {\bibfnamefont {N.}~\bibnamefont
  {Okuma}}, \bibinfo {author} {\bibfnamefont {K.}~\bibnamefont {Kawabata}},
  \bibinfo {author} {\bibfnamefont {K.}~\bibnamefont {Shiozaki}}, \ and\
  \bibinfo {author} {\bibfnamefont {M.}~\bibnamefont {Sato}},\ }\href {\doibase
  10.1103/PhysRevLett.124.086801} {\bibfield  {journal} {\bibinfo  {journal}
  {Phys. Rev. Lett.}\ }\textbf {\bibinfo {volume} {124}},\ \bibinfo {pages}
  {086801} (\bibinfo {year} {2020})}\BibitemShut {NoStop}%
\bibitem [{\citenamefont {Liu}\ \emph {et~al.}(2019)\citenamefont {Liu},
  \citenamefont {Jiang},\ and\ \citenamefont {Chen}}]{Liu_MirrClassifi_PRB19}%
  \BibitemOpen
  \bibfield  {author} {\bibinfo {author} {\bibfnamefont {C.-H.}\ \bibnamefont
  {Liu}}, \bibinfo {author} {\bibfnamefont {H.}~\bibnamefont {Jiang}}, \ and\
  \bibinfo {author} {\bibfnamefont {S.}~\bibnamefont {Chen}},\ }\href {\doibase
  10.1103/PhysRevB.99.125103} {\bibfield  {journal} {\bibinfo  {journal} {Phys.
  Rev. B}\ }\textbf {\bibinfo {volume} {99}},\ \bibinfo {pages} {125103}
  (\bibinfo {year} {2019})}\BibitemShut {NoStop}%
\bibitem [{Rcl()}]{Rclassfi_footnote}%
  \BibitemOpen
  \href@noop {} {}\bibinfo {note} {{ \blue{ Topological classification with
  mirror symmetry can be carried out for the
  bulk~\cite{Liu_MirrClassifi_PRB19}. For two-dimensional systems, the
  classification result of symmetry class A is $\mathbb{Z}$. This topological
  properties are characterized by the mirror winding number. The key result of
  this paper is elucidating that the bulk topology results in the novel skin
  effect protected by mirror symmetry. } }}\BibitemShut {NoStop}%
\bibitem [{\citenamefont {Rui}\ \emph {et~al.}(2019{\natexlab{b}})\citenamefont
  {Rui}, \citenamefont {Hirschmann},\ and\ \citenamefont
  {Schnyder}}]{Rui_Rskin_PRB19}%
  \BibitemOpen
  \bibfield  {author} {\bibinfo {author} {\bibfnamefont {W.~B.}\ \bibnamefont
  {Rui}}, \bibinfo {author} {\bibfnamefont {M.~M.}\ \bibnamefont {Hirschmann}},
  \ and\ \bibinfo {author} {\bibfnamefont {A.~P.}\ \bibnamefont {Schnyder}},\
  }\href {\doibase 10.1103/PhysRevB.100.245116} {\bibfield  {journal} {\bibinfo
   {journal} {Phys. Rev. B}\ }\textbf {\bibinfo {volume} {100}},\ \bibinfo
  {pages} {245116} (\bibinfo {year} {2019}{\natexlab{b}})}\BibitemShut
  {NoStop}%
\bibitem [{Rsk()}]{Rskin_footnote}%
  \BibitemOpen
  \href@noop {} {}\bibinfo {note} {{ \blue{ The mirror skin effect discussed
  here differs from the one mentioned in Ref.~\cite{Rui_Rskin_PRB19} where
  applying the mirror operator exchanges the left and right edges. Furthermore,
  our work elucidates the relation between the skin effect and the point-gap
  topology with mirror symmetry, which is a key result of our work. }
  }}\BibitemShut {NoStop}%
\bibitem [{sup()}]{supplemental}%
  \BibitemOpen
  \href@noop {} {}\bibinfo {note} {{ Supplemental Material }}\BibitemShut
  {NoStop}%
\bibitem [{\citenamefont {Yao}\ and\ \citenamefont
  {Ryu}(2013)}]{Yao_ZtoZ8_PRB13}%
  \BibitemOpen
  \bibfield  {author} {\bibinfo {author} {\bibfnamefont {H.}~\bibnamefont
  {Yao}}\ and\ \bibinfo {author} {\bibfnamefont {S.}~\bibnamefont {Ryu}},\
  }\href {\doibase 10.1103/PhysRevB.88.064507} {\bibfield  {journal} {\bibinfo
  {journal} {Phys. Rev. B}\ }\textbf {\bibinfo {volume} {88}},\ \bibinfo
  {pages} {064507} (\bibinfo {year} {2013})}\BibitemShut {NoStop}%
\bibitem [{\citenamefont {Hofmann}\ \emph
  {et~al.}(2019{\natexlab{a}})\citenamefont {Hofmann}, \citenamefont {Helbig},
  \citenamefont {Lee}, \citenamefont {Greiter},\ and\ \citenamefont
  {Thomale}}]{Hofmann_EleCirChern_PRL19}%
  \BibitemOpen
  \bibfield  {author} {\bibinfo {author} {\bibfnamefont {T.}~\bibnamefont
  {Hofmann}}, \bibinfo {author} {\bibfnamefont {T.}~\bibnamefont {Helbig}},
  \bibinfo {author} {\bibfnamefont {C.~H.}\ \bibnamefont {Lee}}, \bibinfo
  {author} {\bibfnamefont {M.}~\bibnamefont {Greiter}}, \ and\ \bibinfo
  {author} {\bibfnamefont {R.}~\bibnamefont {Thomale}},\ }\href {\doibase
  10.1103/PhysRevLett.122.247702} {\bibfield  {journal} {\bibinfo  {journal}
  {Phys. Rev. Lett.}\ }\textbf {\bibinfo {volume} {122}},\ \bibinfo {pages}
  {247702} (\bibinfo {year} {2019}{\natexlab{a}})}\BibitemShut {NoStop}%
\bibitem [{\citenamefont {Helbig}\ \emph {et~al.}(2019)\citenamefont {Helbig},
  \citenamefont {Hofmann}, \citenamefont {Imhof}, \citenamefont {Abdelghany},
  \citenamefont {Kiessling}, \citenamefont {Molenkamp}, \citenamefont {Lee},
  \citenamefont {Szameit}, \citenamefont {Greiter},\ and\ \citenamefont
  {Thomale}}]{Helbig_ExpSkin_19}%
  \BibitemOpen
  \bibfield  {author} {\bibinfo {author} {\bibfnamefont {T.}~\bibnamefont
  {Helbig}}, \bibinfo {author} {\bibfnamefont {T.}~\bibnamefont {Hofmann}},
  \bibinfo {author} {\bibfnamefont {S.}~\bibnamefont {Imhof}}, \bibinfo
  {author} {\bibfnamefont {M.}~\bibnamefont {Abdelghany}}, \bibinfo {author}
  {\bibfnamefont {T.}~\bibnamefont {Kiessling}}, \bibinfo {author}
  {\bibfnamefont {L.~W.}\ \bibnamefont {Molenkamp}}, \bibinfo {author}
  {\bibfnamefont {C.~H.}\ \bibnamefont {Lee}}, \bibinfo {author} {\bibfnamefont
  {A.}~\bibnamefont {Szameit}}, \bibinfo {author} {\bibfnamefont
  {M.}~\bibnamefont {Greiter}}, \ and\ \bibinfo {author} {\bibfnamefont
  {R.}~\bibnamefont {Thomale}},\ }\href@noop {} {\bibfield  {journal} {\bibinfo
   {journal} {arXiv preprint arXiv:1907.11562}\ } (\bibinfo {year}
  {2019})}\BibitemShut {NoStop}%
\bibitem [{Dio()}]{Diode_footnote}%
  \BibitemOpen
  \href@noop {} {}\bibinfo {note} {{ \blue{ We note that employing
  diodes~\cite{Ezawa_elecirPRB19} is considered to be another approach to
  realize the skin effect } }}\BibitemShut {NoStop}%
\bibitem [{\citenamefont {Ezawa}(2019)}]{Ezawa_elecirPRB19}%
  \BibitemOpen
  \bibfield  {author} {\bibinfo {author} {\bibfnamefont {M.}~\bibnamefont
  {Ezawa}},\ }\href {\doibase 10.1103/PhysRevB.99.201411} {\bibfield  {journal}
  {\bibinfo  {journal} {Phys. Rev. B}\ }\textbf {\bibinfo {volume} {99}},\
  \bibinfo {pages} {201411} (\bibinfo {year} {2019})}\BibitemShut {NoStop}%
\bibitem [{\citenamefont {Hofmann}\ \emph
  {et~al.}(2019{\natexlab{b}})\citenamefont {Hofmann}, \citenamefont {Helbig},
  \citenamefont {Schindler}, \citenamefont {Salgo}, \citenamefont
  {Brzezi{\'n}ska}, \citenamefont {Greiter}, \citenamefont {Kiessling},
  \citenamefont {Wolf}, \citenamefont {Vollhardt}, \citenamefont {Kaba{\v{s}}i}
  \emph {et~al.}}]{Hofmann_ExpRecipSkin_19}%
  \BibitemOpen
  \bibfield  {author} {\bibinfo {author} {\bibfnamefont {T.}~\bibnamefont
  {Hofmann}}, \bibinfo {author} {\bibfnamefont {T.}~\bibnamefont {Helbig}},
  \bibinfo {author} {\bibfnamefont {F.}~\bibnamefont {Schindler}}, \bibinfo
  {author} {\bibfnamefont {N.}~\bibnamefont {Salgo}}, \bibinfo {author}
  {\bibfnamefont {M.}~\bibnamefont {Brzezi{\'n}ska}}, \bibinfo {author}
  {\bibfnamefont {M.}~\bibnamefont {Greiter}}, \bibinfo {author} {\bibfnamefont
  {T.}~\bibnamefont {Kiessling}}, \bibinfo {author} {\bibfnamefont
  {D.}~\bibnamefont {Wolf}}, \bibinfo {author} {\bibfnamefont {A.}~\bibnamefont
  {Vollhardt}}, \bibinfo {author} {\bibfnamefont {A.}~\bibnamefont
  {Kaba{\v{s}}i}},  \emph {et~al.},\ }\href@noop {} {\bibfield  {journal}
  {\bibinfo  {journal} {arXiv preprint arXiv:1908.02759}\ } (\bibinfo {year}
  {2019}{\natexlab{b}})}\BibitemShut {NoStop}%
\bibitem [{L0_()}]{L0_footnote}%
  \BibitemOpen
  \href@noop {} {}\bibinfo {note} {{ We note that inductors $L_0$ is essential
  for making one of the eigenvalues $j^{-1}_{n_0}$ becomes dominant
  }}\BibitemShut {NoStop}%
\end{thebibliography}
%


\clearpage

\renewcommand{\thesection}{S\arabic{section}}
\renewcommand{\theequation}{S\arabic{equation}}
\setcounter{equation}{0}
\renewcommand{\thefigure}{S\arabic{figure}}
\setcounter{figure}{0}
\renewcommand{\thetable}{S\arabic{table}}
\setcounter{table}{0}
\makeatletter
\c@secnumdepth = 2
\makeatother

\onecolumngrid
\begin{center}
 {\large \textmd{Supplemental Materials:} \\[0.3em]
 {\bfseries 
Mirror skin effect and its electric circuit simulation
 }
 }
\end{center}

\setcounter{page}{1}

\section{
Relation between a Hermitian system
}
\label{sec: rel yaoryu}

\blue{
The model describing topological properties of the point gap can be obtained from the corresponding Hermitian Hamiltonian with chiral symmetry. 
Namely, as the arbitrary Hermitian Hamiltonian $H_0(\bm{k})$ with chiral symmetry can be written as 
$H_0(\bm{k}):=
\left(
\begin{array}{cc}
0 & Q(\bm{k}) \\
Q^\dagger(\bm{k}) & 0
\end{array}
\right)_{\chi}
$, extracting matrix $Q(\bm{k})$ yields the non-Hermitian Hamiltonian showing nontrivial topological properties of the point gap.
}

\blue{
Specifically, \magenta{the} Hamiltonian~(\ref{eq: toy Hami}) \magenta{in the main text} is related with the Hermitian Hamiltonian in two dimensions discussed in Ref.~\onlinecite{Yao_ZtoZ8_PRB13} whose topology is protected by mirror symmetry.
The Hamiltonian for the Hermitian system reads
\begin{eqnarray}
H_{0}(\bm{k})&=& [2t(\cos k_x + \cos k_y) -\mu] \sigma_0\otimes \tau_3 +\Delta \sin k_x \sigma_0\otimes \tau_2  +\Delta \sin k_y \sigma_3\otimes \tau_1,
\end{eqnarray}
where $\sigma$'s and $\tau$'s are the Pauli matrices.
As well as the mirror symmetry defined as $M_x=i\sigma_1\otimes \tau_3 P_x$, this model also preserves the chiral symmetry with $\Gamma=\sigma_2\otimes\tau_1$; $\Gamma H_{0}(\bm{k}) \Gamma^{-1} =- H_{0}(\bm{k})$.
}

\blue{
Here, we choose the basis such that chiral symmetry is represented as $\Gamma=\rho_0\otimes\chi_3$ where $\rho$'s and $\chi$'s are the Pauli matrices.
In this basis, the Hamiltonian and the operator $M_x$ are written as
\begin{subequations}
\begin{eqnarray}
H_{0}(\bm{k})&=& 
\left(
\begin{array}{cc}
0 & Q(\bm{k}) \\
Q^\dagger(\bm{k}) & 0
\end{array}
\right)_{\chi}, \\
Q(\bm{k}) &=& 
[2t(\cos k_x + \cos k_y) -\mu] \rho_0 + i\Delta \sin k_x \rho_3  + i \Delta \sin k_y \rho_2,
\end{eqnarray}
and 
\begin{eqnarray}
M_x&=&i\rho_2\otimes \chi_0 P_x.
\end{eqnarray}
\end{subequations}
}

\blue{
The matrix $Q(\bm{k})$ appearing in the off-diagonal block corresponds to Hamiltonian~(\ref{eq: toy Hami}) \magenta{in the main text}.
}

\section{
Additional numerical data of Hamiltonian~(\ref{eq: toy Hami})
}
\label{sec: toy PBC app}

\subsection{
Amplitude of the right eigenstates
}
\label{sec: toy amp app}
In Fig.~\ref{fig: Amptoy_OBC}(a), we have seen that the eigenstates $|\Psi_{nR}\rangle$ $n=1,2,\cdots$ are extensively localized around the boundary.
Here, we show that the boundary condition is essential for the above extensive localization.

\begin{figure}[!h]
\begin{minipage}{0.44\hsize}
\begin{center}
\includegraphics[width=1\hsize,clip]{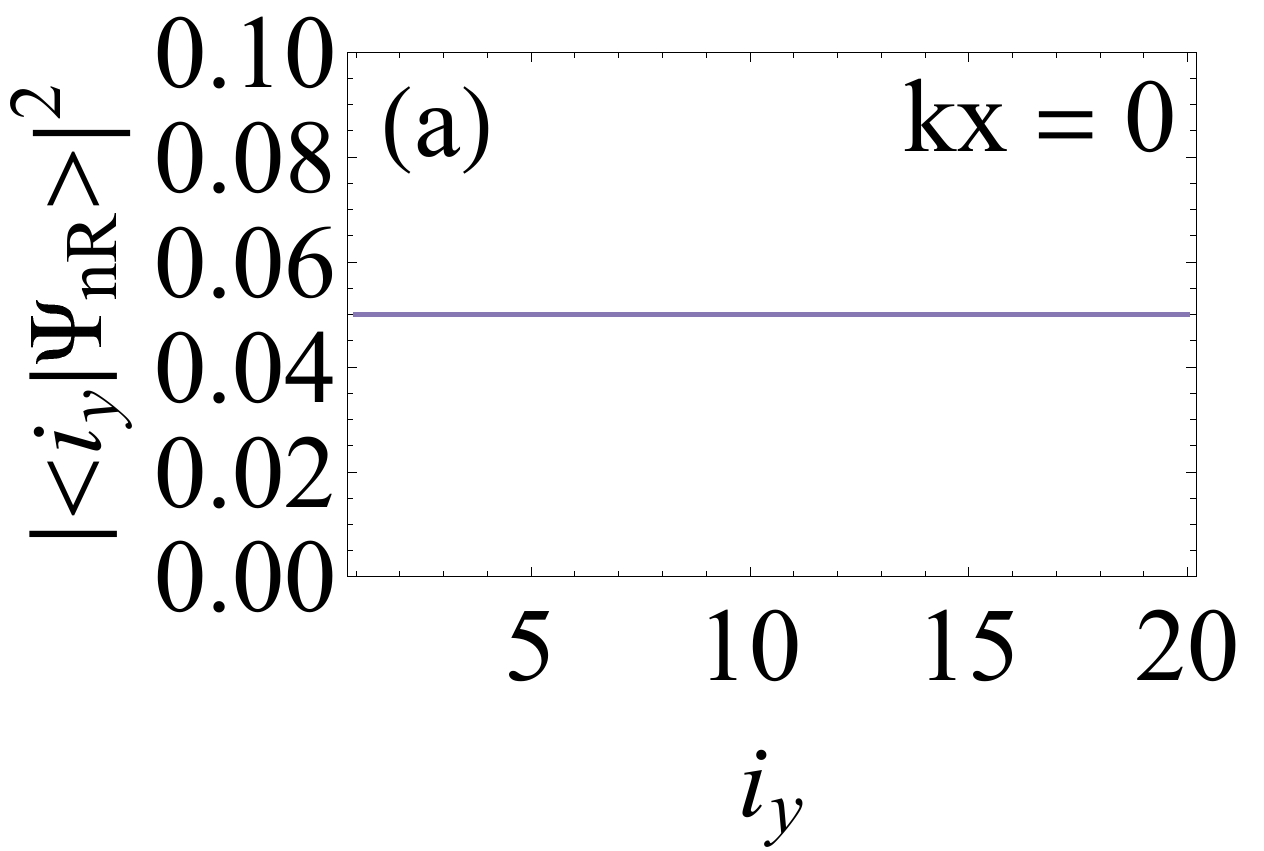}
\end{center}
\end{minipage}
\begin{minipage}{0.44\hsize}
\begin{center}
\includegraphics[width=1\hsize,clip]{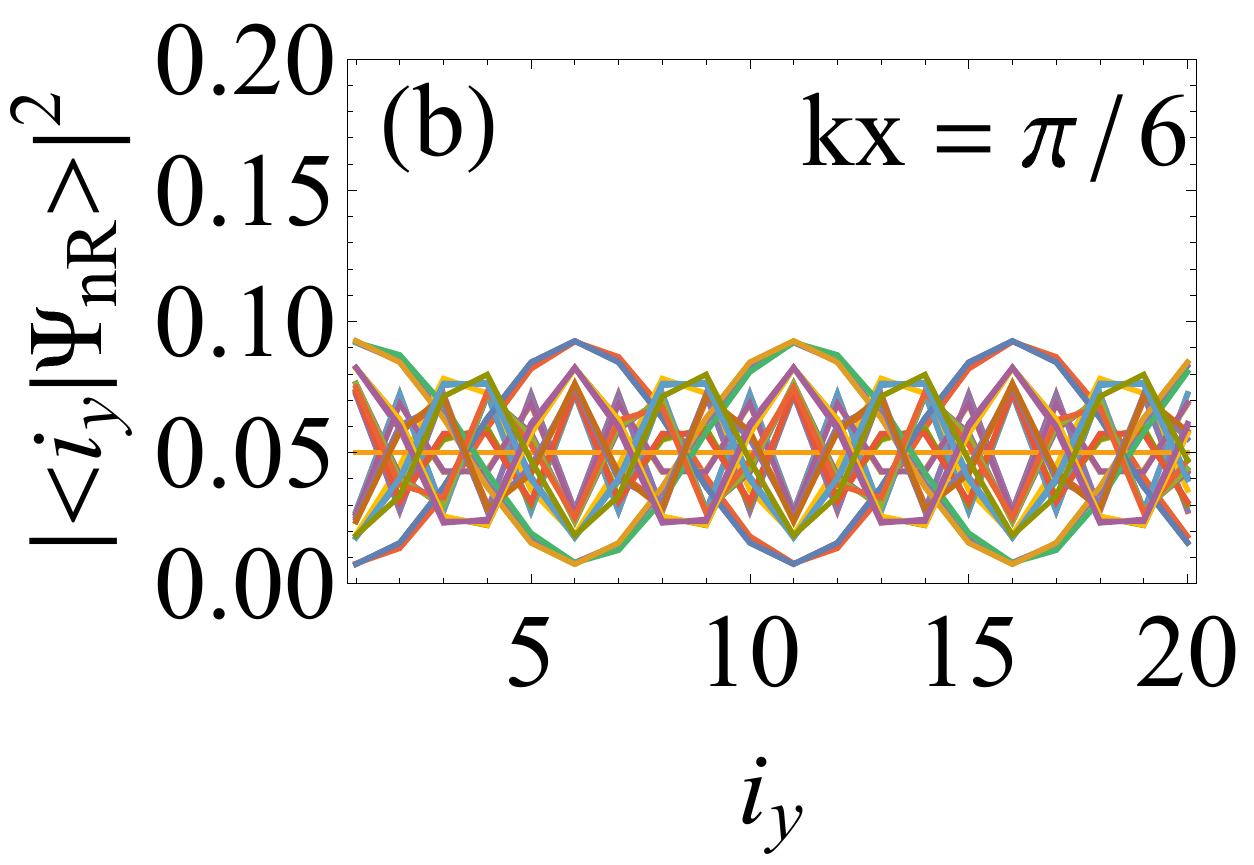}
\end{center}
\end{minipage}
\caption{(Color Online).
(a) [(b)]: Amplitude of the right eigenvectors of the Hamiltonian~(\ref{eq: toy Hami}) \magenta{in the main text} for $(t,\mu,\Delta)=(1,2,1.8)$ at $k_x=0$ $(\pi/6)$, respectively.
The data \blue{are} obtained under the PBC both for the $x$- and \blue{the} $y$-direction\magenta{s}. The number of the sites along the $y$-direction is set to $L_y=20$.
}
\label{fig: Amptoy PBC app}
\end{figure}

Figure~\ref{fig: Amptoy PBC app} plots the amplitude of the eigenstates under the PBC along the $y$-direction. 
Figure~\ref{fig: Amptoy PBC app}(a) [(b)] shows data for $k_x=0$ ($k_x=\pi/6$), respectively.
In these figures, we can see that the states are delocalized for the PBC.

\subsection{
\blue{
Energy spectrum for several cases of the boundary conditions
}
}
\label{sec: toy fullOBC app}
\blue{
Here, we show the energy spectrum obtained without applying the Fourier transformation. Directly diagonalizing the Hamiltonian [Eq.~(\ref{eq: toy Hami}) of the main text] in the real space allows us to systematically analyze the distinct cases of the boundary conditions.
}

\begin{figure}[!h]
\begin{minipage}{0.49\hsize}
\begin{center}
\includegraphics[width=1\hsize,clip]{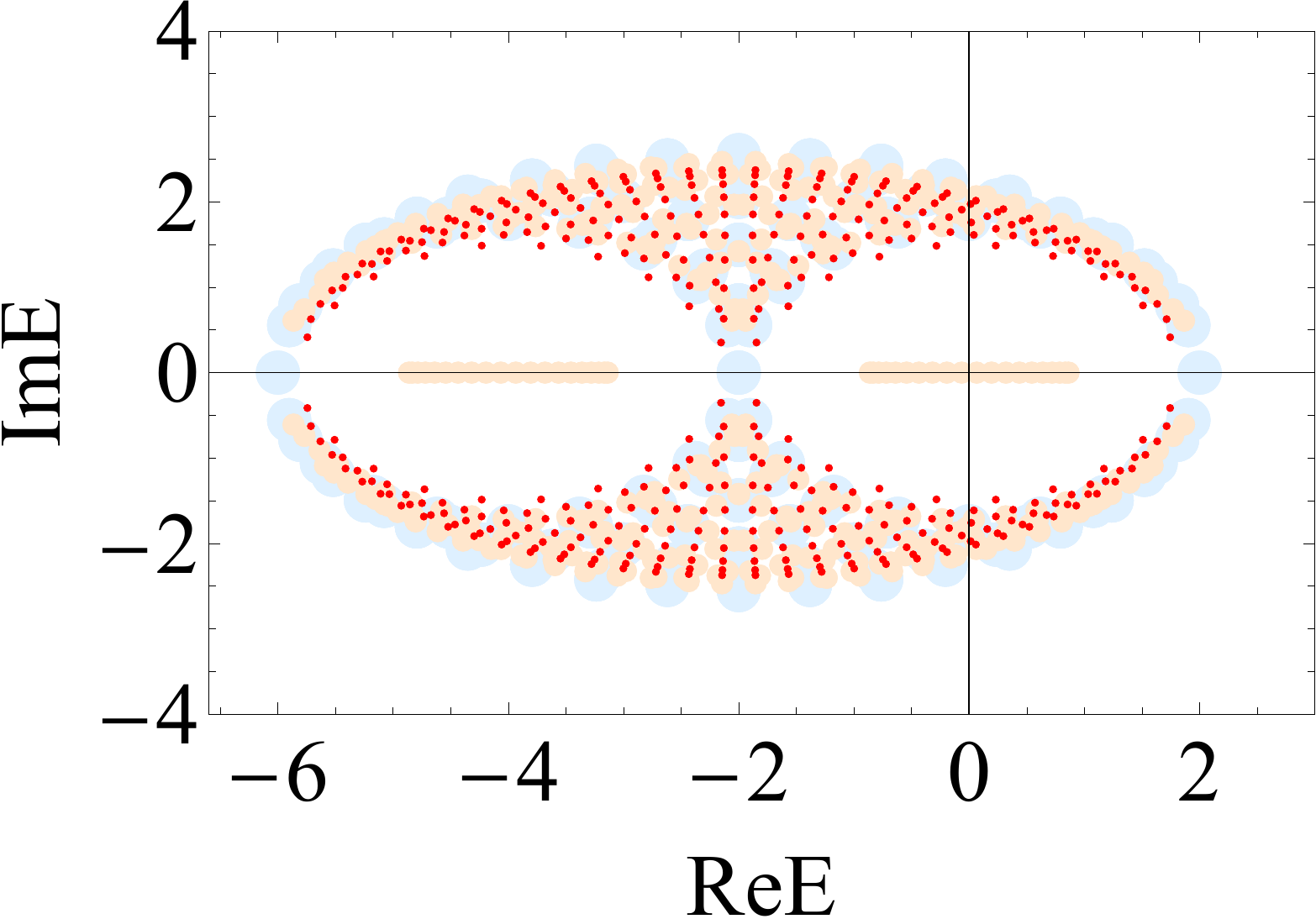}
\end{center}
\end{minipage}
\caption{(Color Online).
\blue{
Energy spectrum for several cases of boundary conditions. Data represented with large blue (small red) symbols are obtained by imposing the PBC (OBC) both for the $x$- and \blue{the} $y$-direction\magenta{s}, respectively. 
Medium orange symbols indicate the data obtained under the PBC (OBC) for \blue{the} $x$- ($y$-) direction, respectively.
}
\blue{
The data are obtained for $(t,\Delta,\mu)=(1,1.8,2)$ and $L_x=L_y=20$.
}
}
\label{fig: full_OBC}
\end{figure}

\blue{
Figure~\ref{fig: full_OBC} shows the spectrum for several cases of the boundary condition. 
The data for the full PBC (i.e., the PBC both for \blue{the} $x$- and \blue{the} $y$-direction\magenta{s}) are consistent with Fig.~\ref{fig: Engtoy} in the main text. 
Imposing the OBC for \blue{the} $y$-direction yields the significant change around the real axis, which is also consistent with the data shown in Fig.~\ref{fig: Engtoy} of the main text. 
As discussed in the main text, the eigenstates with these eigenvalues (i.e., $\mathrm{Im} E\sim0$) correspond to the anomalous localized states. 
Imposing the full OBC (i.e., the OBC both for \blue{the} $x$- and \blue{the} $y$-direction\magenta{s}), we can see that the eigenstates satisfying $\mathrm{Im} E\sim0$ vanish; the spectrum with the full OBC is similar to the one with the full PBC. 
A similar behavior is observed for the $\mathbb{Z}_2$ skin effect characterized by a two-dimensional strong topological invariant~\cite{Okuma_BECskin19}, which implies that the mirror skin effect is also characterized by the strong topological invariant, i.e., the mirror winding number.
}

\section{
Numerical results for other models
}
\label{sec: nu2 noM app}

\blue{
In this section, we show the following facts: 
(i) the mirror skin effect can also be observed for a model characterized with $\nu_{\mathrm{M}}=2$; 
(ii) breaking the mirror symmetry extinguishes the skin effect observed for Hamiltonian~(\ref{eq: toy Hami}) \magenta{in the main text}.
}

\subsection{
Results for a system characterized with $\nu_{\mathrm{M}}=2$
}

\blue{
Firstly, we show that the mirror skin effect can also be observed for a system whose topology is characterized by $\nu_{\mathrm{M}}=2$.
}

\blue{
Speficically, we diagonalize the following Hamiltonian: 
\begin{eqnarray}
\label{eq: toy Hami nu2}
H_{2}(\bm{k})&=& [2t(\cos k_x +\cos 2k_y )-\mu] \rho_0+i\Delta \sin k_x \rho_3 +i\Delta \sin 2k_y \rho_2.
\end{eqnarray}
Topological properties of this Hamiltonian are characterized by the mirror winding number $\nu_{\mathrm{M}}=2$, where the mirror operator is defined as $M_x=\rho_2P_x$.
}

\begin{figure}[!h]
\begin{minipage}{0.44\hsize}
\begin{center}
\includegraphics[width=1\hsize,clip]{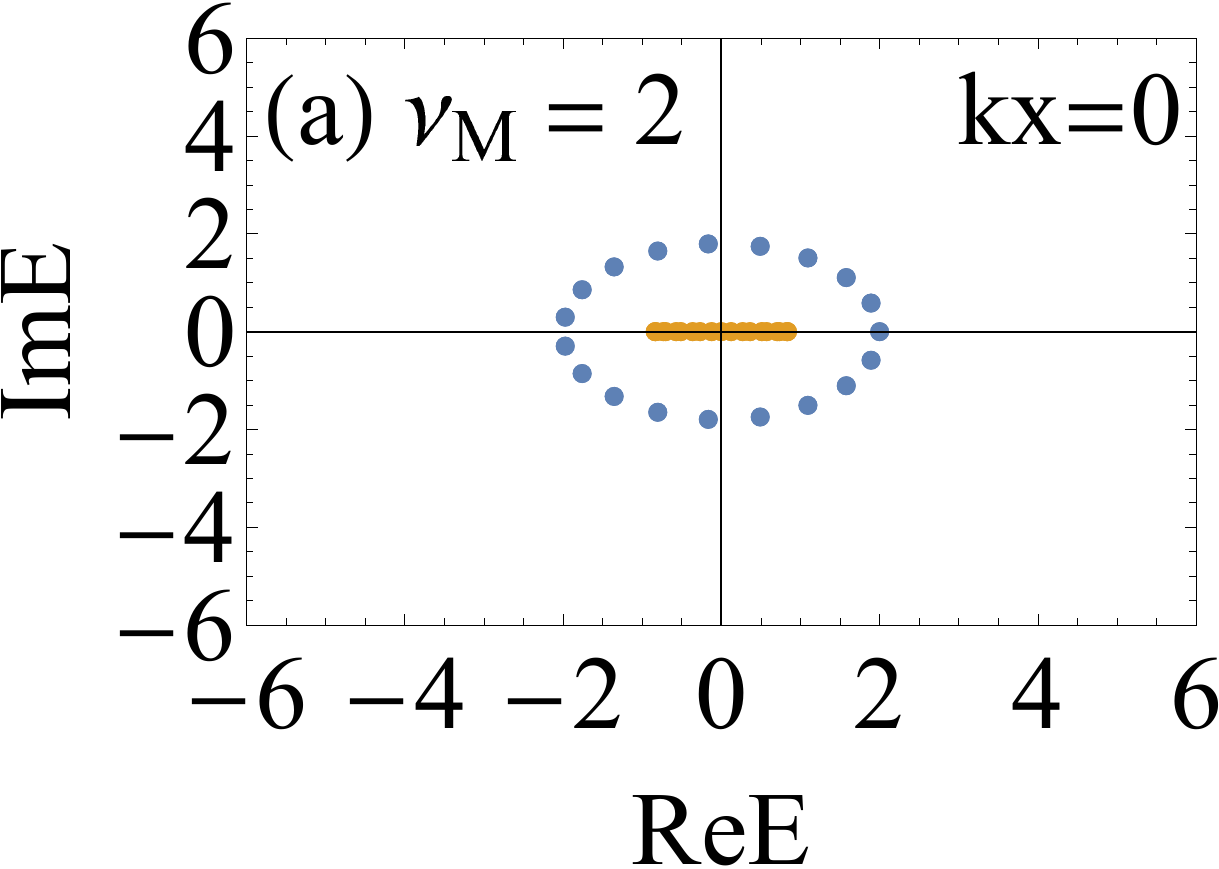}
\end{center}
\end{minipage}
\begin{minipage}{0.44\hsize}
\begin{center}
\includegraphics[width=1\hsize,clip]{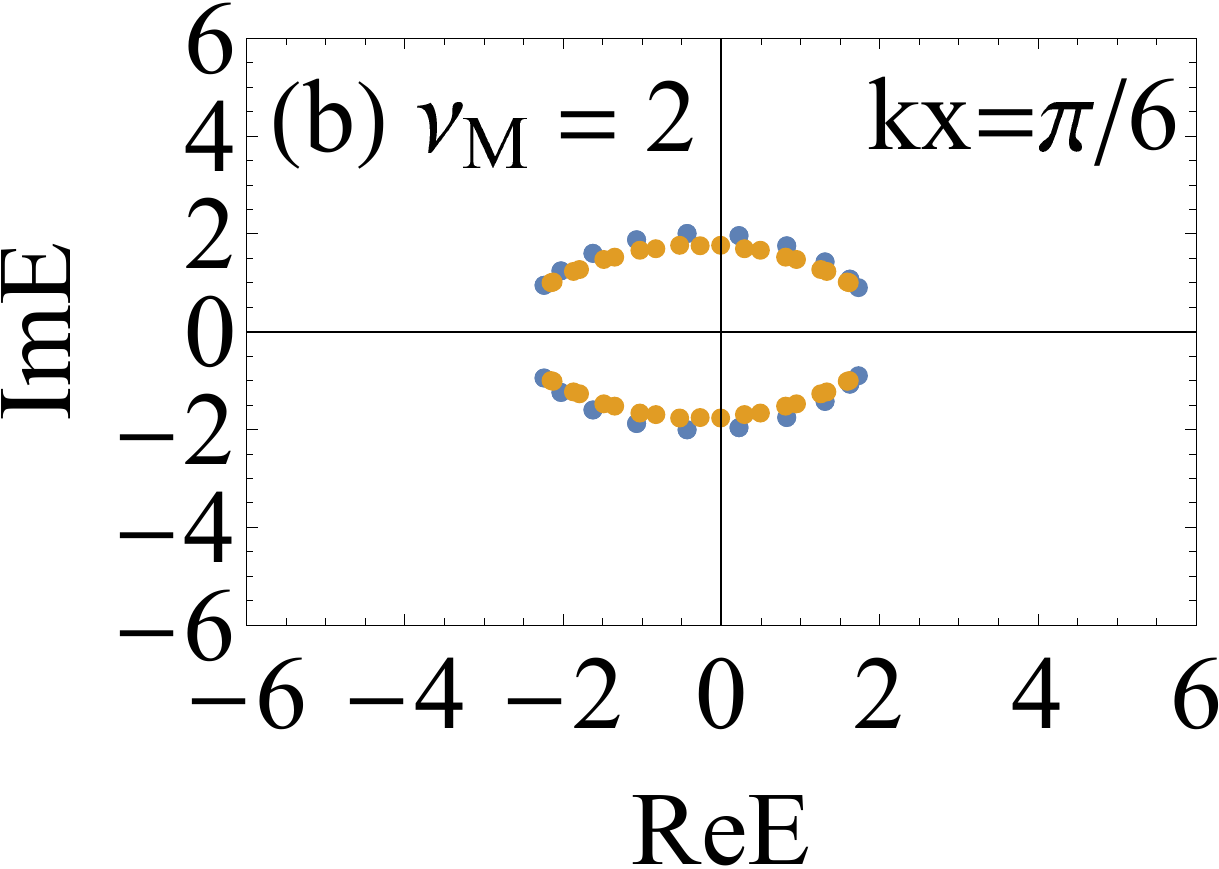}
\end{center}
\end{minipage}
\caption{(Color Online).
\blue{
Energy spectrum of the Hamiltonian~(\ref{eq: toy Hami nu2}) for $(t,\mu,\Delta)=(1,2,1.8)$. 
Data represented with blue (orange) symbols are obtained under the OBC (PBC) for the $y$-direction\blue{, respectively}. 
For \blue{the} $x$-direction, the PBC is imposed. The number of the sites for the $y$-direction is set to $L_y=20$.
}
}
\label{fig: nu2_En}
\end{figure}

\begin{figure}[!h]
\begin{minipage}{0.44\hsize}
\begin{center}
\includegraphics[width=1\hsize,clip]{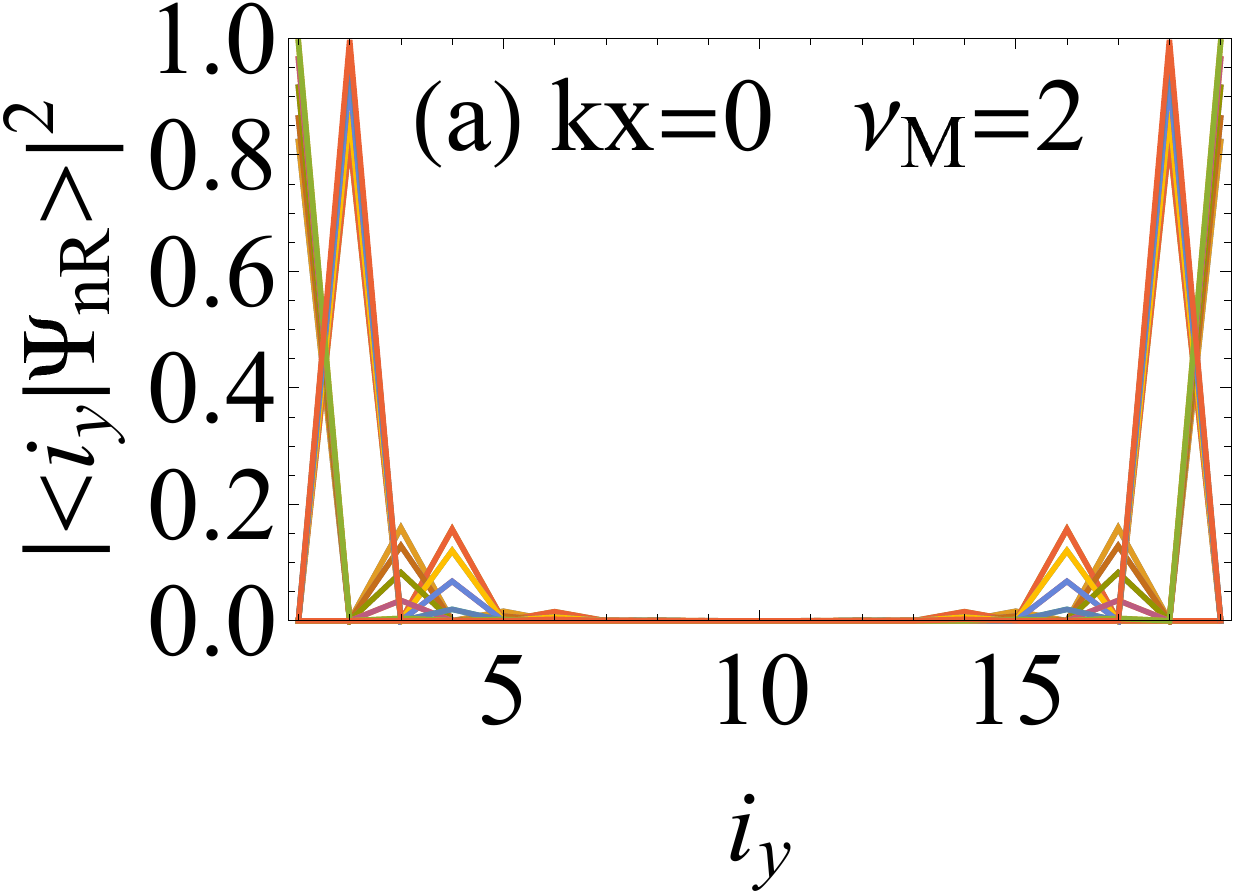}
\end{center}
\end{minipage}
\begin{minipage}{0.44\hsize}
\begin{center}
\includegraphics[width=1\hsize,clip]{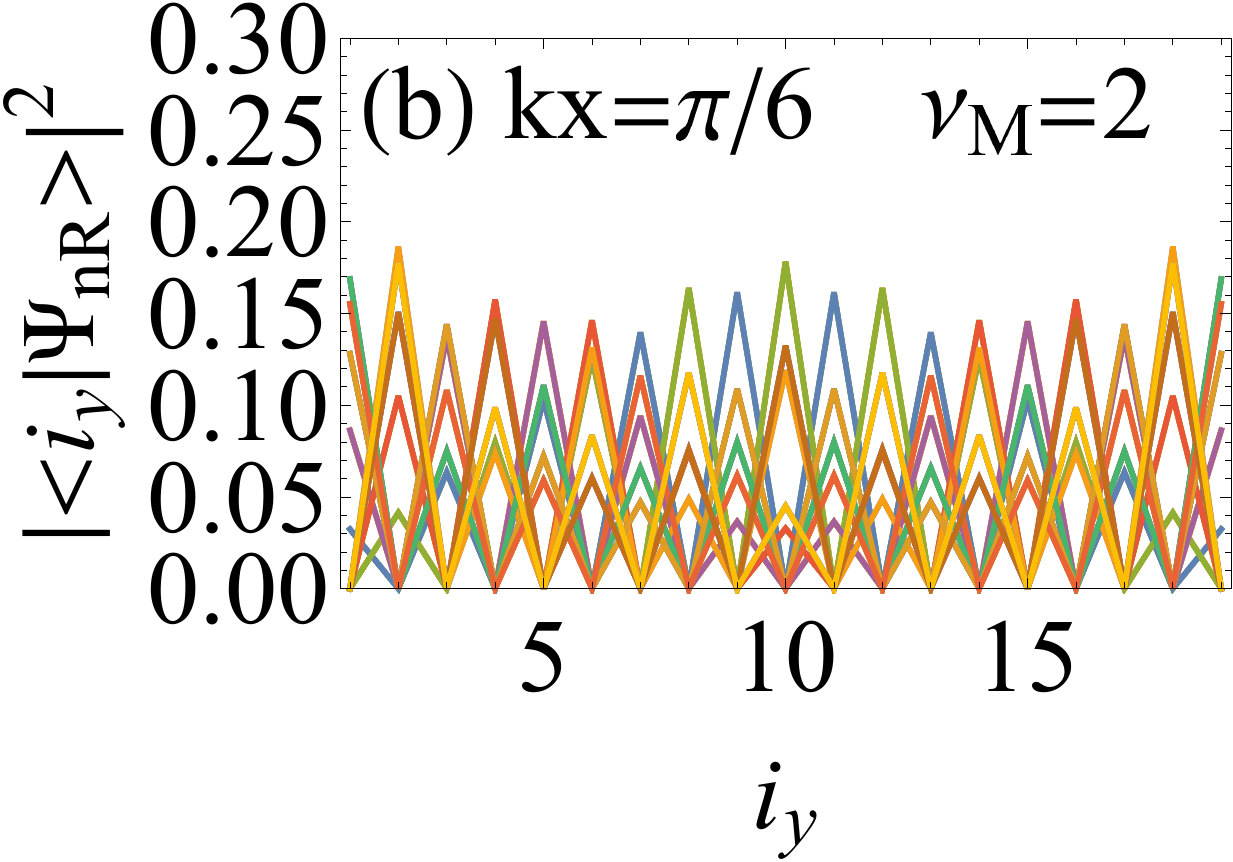}
\end{center}
\end{minipage}
\caption{(Color Online).
\blue{
Amplitude of the right eigenvectors of the Hamiltonian~(\ref{eq: toy Hami nu2}) for $(t,\mu,\Delta)=(1,2,1.8)$. 
The data are obtained under the PBC (OBC) for \blue{the} $x$- ($y$-) direction, respectively. 
The number of the sites along the $y$-direction is set to $L_y=20$.
}
}
\label{fig: nu2_Amp}
\end{figure}

\blue{
Energy spectra obtained under the PBC for the $x$-direction are shown in Fig.~\ref{fig: nu2_En}.
As is the case of the model~(\ref{eq: toy Hami}) \magenta{in the main text}, we can observe that switching the boundary condition for the $y$-direction results in the significant change of the energy spectrum at $k_x=0$. 
Such a significant difference cannot be observed away from $k_x=0$ or $\pi$.
Corresponding to the significant dependences of the spectrum on the boundary condition, we can also observe the anomalous localized states only the mirror invariant momenta (see Fig.~\ref{fig: nu2_Amp})
}

\blue{
The above results indicate that the mirror skin effect indeed occurs for the case of $\nu_{\mathrm{M}}=2$.
}

\subsection{
\blue{Results for a model without mirror symmetry}
}

\blue{
Here, we demonstrate that the mirror skin effect cannot be observed once the mirror symmetry is broken.
}

\blue{
Speficically, we diagonalize the following Hamiltonian: 
\begin{eqnarray}
\label{eq: toy Hami mb}
H_{mb}(\bm{k})&=& H(\bm{k}) + B \rho_1,
\end{eqnarray}
with $H(\bm{k})$ defined in Eq.~(\ref{eq: toy Hami}) \magenta{of the main text} and $B$ being a real number. 
The second term breaks the mirror symmetry defined with $M_x=\rho_2P_x$.
}

\begin{figure}[!h]
\begin{minipage}{0.44\hsize}
\begin{center}
\includegraphics[width=1\hsize,clip]{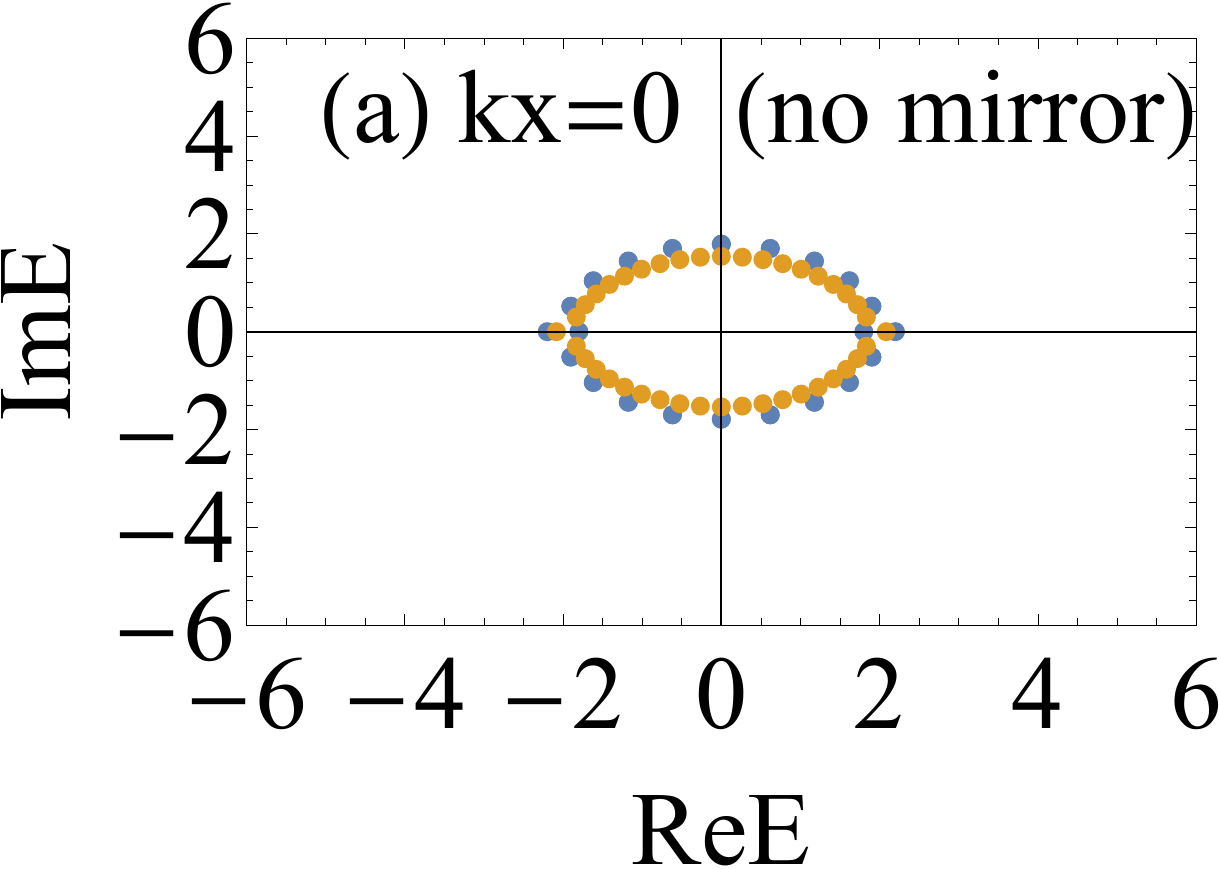}
\end{center}
\end{minipage}
\begin{minipage}{0.44\hsize}
\begin{center}
\includegraphics[width=1\hsize,clip]{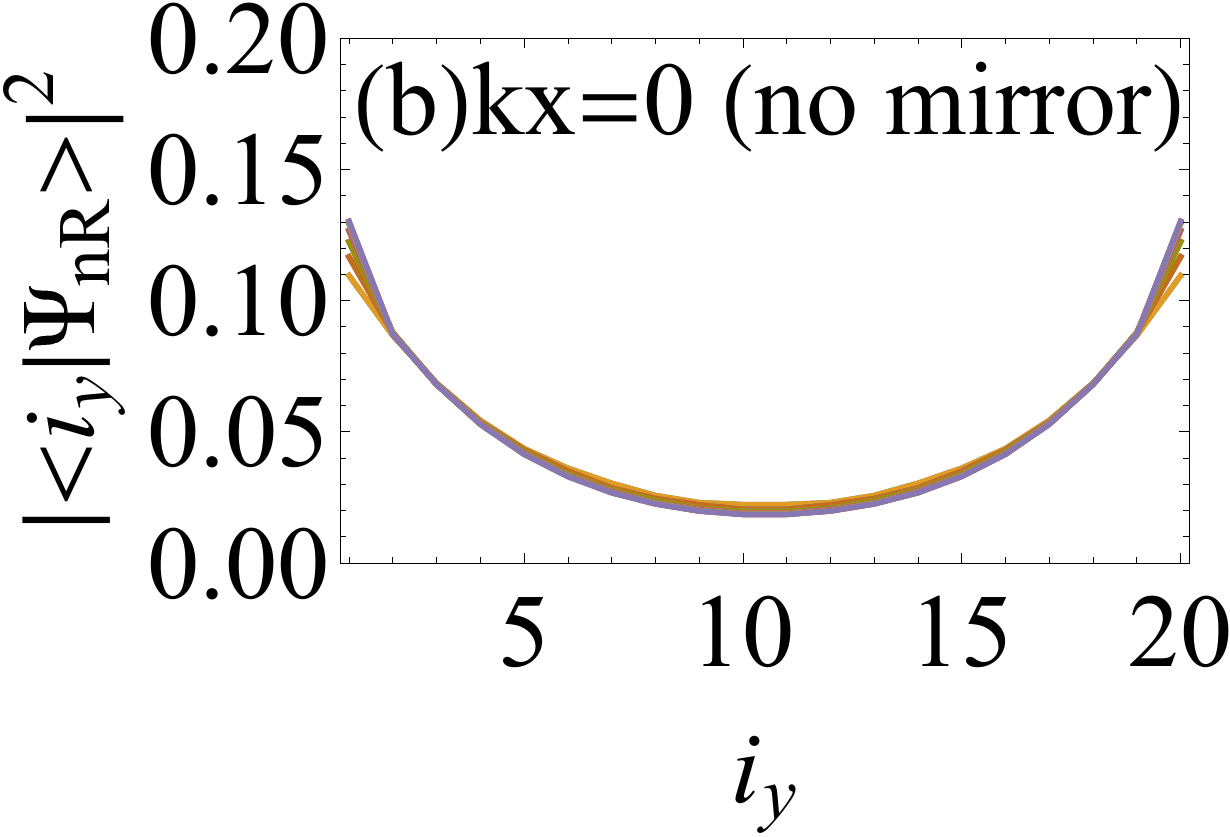}
\end{center}
\end{minipage}
\caption{(Color Online).
\blue{
Numerical data of the model [Eq.~(\ref{eq: toy Hami mb})] for $(t,\mu,\Delta,B)=(1,2,1.8,0.2)$ and $L_y=20$.
In panel (a), blue (orange) symbols represent energy spectrum obtained by imposing the PBC (OBC) for the $y$-direction, respectively. 
In panel (b), amplitude of the right eigenvectors of the Hamiltonian $H_{mb}$ is plotted.
These data are obtained by imposing the PBC for the $x$-direction.
}
}
\label{fig: EnAmp_nomirror}
\end{figure}

\blue{
The numerical data are shown in Fig.~\ref{fig: EnAmp_nomirror}. 
This figure shows that in the presence of the symmetry-breaking term, the significant dependence of the energy spectrum on the boundary condition cannot be observed even at $k_x=0$ [see Fig.~\ref{fig: EnAmp_nomirror}(a)] in contrast to the data for $B=0$. 
Correspondingly, anomalous localized edges cannot be observed [see Fig.~\ref{fig: EnAmp_nomirror}(b)]. 
}

\section{
Details of electric circuit
}
\label{sec: ele cir app}

\subsection{
Derivation of Eq.~(\ref{eq: Jskin})
}
\label{sec: derivation of J app}
\begin{figure}[!h]
\begin{minipage}{1\hsize}
\begin{center}
\includegraphics[width=0.7\hsize,clip]{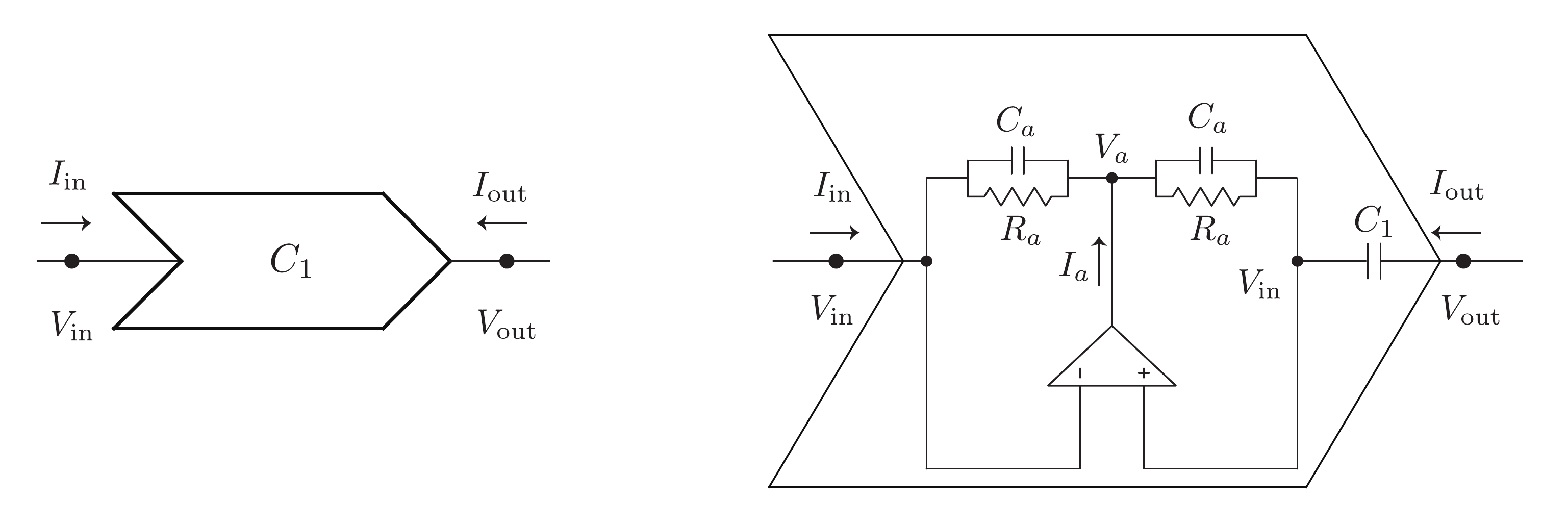}
\end{center}
\end{minipage}
\caption{(Color Online).
Sketch of the circuit elements.
We can tune the amplifier (denoted by triangle) so that the current does not flow into it. 
Here, the voltage at each nodes is also illustrated.
}
\label{fig: RampCamp app}
\end{figure}

We start with explaining how the negative impedance converters with current inversion induces the non-Hermitian terms.
The element shown in Fig.~\ref{fig: RampCamp app} responses as~\cite{Hofmann_EleCirChern_PRL19,Helbig_ExpSkin_19} 
\begin{eqnarray}
\label{eq: Camp_resp app}
\left(
\begin{array}{c}
I_{\mathrm{in}}  \\
I_{\mathrm{out}} 
\end{array}
\right)
&=&
i\omega C_1
\left(
\begin{array}{cc}
-1 & 1 \\
-1 & 1
\end{array}
\right)
\left(
\begin{array}{c}
V_{\mathrm{in}}  \\
V_{\mathrm{out}} 
\end{array}
\right),
\end{eqnarray}
where the vectors 
$
\left(
\begin{array}{cc}
I_{\mathrm{in}} & I_{\mathrm{out}}
\end{array}
\right)^T
$
and 
$
\left(
\begin{array}{cc}
V_{\mathrm{in}} & V_{\mathrm{out}}
\end{array}
\right)^T
$
represent the current and the voltage illustrated in Fig.~\ref{fig: RampCamp app}. $C_1$ denotes the capacitance.
This can be seen as follows. 
We can tune the current and voltage as shown in Fig.~\ref{fig: RampCamp app}.
In this case, based on the Kirchhoff's law, we obtain
\begin{subequations}
\begin{eqnarray}
I_{\mathrm{in}} &=&  (i\omega C_a + R^{-1}_a)(V_{\mathrm{in}}-V_a), \\ 
I_{\mathrm{out}} &=& (i\omega C_a + R^{-1}_a) (V_{\mathrm{in}}-V_a), \\ 
I_{\mathrm{out}} &=& i\omega C_1 (V_{\mathrm{out}}-V_{\mathrm{in}}),
\end{eqnarray}
\end{subequations}
where $R_a$ and $C_a$ represent the resistance and the capacitance, respectively. 
$V_a$ and $I_a$ denote the current and the voltage as illustrated in Fig.~\ref{fig: RampCamp app}, respectively. 
$\omega$ denotes the angular frequency.
Solving these equations yields Eq.~(\ref{eq: Camp_resp app}).
\blue{
We note that the definition of the current $I_\mathrm{out}$ differs from the one in Ref.~\onlinecite{Helbig_ExpSkin_19} by the minus sign [see sentences just below Eq.~(2) of Ref.~\onlinecite{Helbig_ExpSkin_19} and Eq.~(1) in the supplemental material of Ref.~\onlinecite{Hofmann_EleCirChern_PRL19}].
}
Here, we note that current $I_{\mathrm{in}}$ has the opposite sign; the ordinary capacitor responses as
\begin{eqnarray}
\left(
\begin{array}{c}
I_{\mathrm{in}}  \\
I_{\mathrm{out}} 
\end{array}
\right)
&=&
i\omega C_0
\left(
\begin{array}{cc}
 1 & -1 \\
-1 & 1
\end{array}
\right)
\left(
\begin{array}{c}
V_{\mathrm{in}}  \\
V_{\mathrm{out}} 
\end{array}
\right),
\end{eqnarray}
with capacitance $C_0$.
The above additional sign plays an essential role in the non-Hermitian hopping.

Now, let us consider the circuit illustrated in Fig.~\ref{fig: Emodel}.
Connecting intra-layer nodes by the negative impedance converters with current inversion [Eq.~(\ref{eq: Camp_resp app})] yields
\begin{subequations}
\begin{eqnarray}
I_{A \bm{R}_i} &=&  i\omega C_1 (-V_{ A \bm{R}_i-\bm{e}_x} +V_{ A \bm{R}_i+\bm{e}_x} ), \\
I_{B \bm{R}_i} &=& -i\omega C_1 (-V_{ B \bm{R}_i-\bm{e}_x} +V_{ B \bm{R}_i+\bm{e}_x} ),
\end{eqnarray}
\end{subequations}
where $I_{\alpha \bm{R}_i}$ ($V_{\alpha \bm{R}_i}$) denotes the current and the voltage of the node specified with $\alpha$ and $\bm{R}_i$, respectively.
$\bm{R}_i$ is the position vector at site $i$. $\alpha=A,B$ specifies the layer.
$\bm{e}_\mu$ ($\mu=x,y$) denotes the unit vector for each direction.
Applying the Fourier transformation, we obtain
\begin{eqnarray}
\label{eq: deri_J_ Camp intra app}
\left(
\begin{array}{c}
I_{A \bm{k}}  \\
I_{B \bm{k}}  \\
\end{array}
\right)
&=& -2 \omega C_1  \sin k_x \rho_3 
\left(
\begin{array}{c}
V_{A \bm{k}}  \\
V_{B \bm{k}}  \\
\end{array}
\right),
\end{eqnarray}
with
\begin{subequations}
\begin{eqnarray}
I_\alpha(\omega,\bm{k})&=&\frac{1}{\sqrt{L_xL_y}}\sum_a e^{i\bm{k}\cdot \bm{R}_a}I_{\alpha a} (\omega,\bm{k}),  \\
V_\alpha(\omega,\bm{k})&=& \frac{1}{\sqrt{L_xL_y}}\sum_a e^{i\bm{k}\cdot \bm{R}_a}V_{\alpha a}(\omega,\bm{k}).
\end{eqnarray}
\end{subequations}
Therefore, we can see that connecting intra-layer nodes by the elements defined in Eq.~(\ref{eq: Camp_resp app}) yields the third term of Eq.~(\ref{eq: Jskin}).
In a similar way, we can see that connecting inter-layer nodes with the elements~(\ref{eq: Camp_resp app}) yields the fourth term of Eq.~(\ref{eq: Jskin});
in the real space, we obtain
\begin{subequations}
\begin{eqnarray}
I_{A \bm{R}_i} &=&  i\omega C_1 (-V_{ B \bm{R}_i-\bm{e}_y} +V_{ B \bm{R}_i+\bm{e}_y} ), \\
I_{B \bm{R}_i} &=&  i\omega C_1 (-V_{ A \bm{R}_i-\bm{e}_y} +V_{ A \bm{R}_i+\bm{e}_y} ),
\end{eqnarray}
\end{subequations}
which yields
\begin{eqnarray}
\label{eq: deri_J_ Camp inter app }
\left(
\begin{array}{c}
I_{A \bm{k}}  \\
I_{B \bm{k}}  \\
\end{array}
\right)
&=& -2\omega C_1\sin k_y \rho_1 
\left(
\begin{array}{c}
V_{A \bm{k}}  \\
V_{B \bm{k}}  \\
\end{array}
\right).
\end{eqnarray}
%
Concerning the other circuit elements illustrated in Fig.~\ref{fig: Emodel}, we obtain
\begin{eqnarray}
I_{\alpha\bm{R}_i} &=& -i\omega C_0 \sum_{\mu=x,y} ( V_{\alpha\bm{R}_i-\bm{e}_\mu}- V_{\alpha\bm{R}_i} + V_{\alpha\bm{R}_i-\bm{e}_\mu} - V_{\alpha\bm{R}_i})+(i\omega L_0)^{-1}V_{\alpha \bm{R}_i},
\end{eqnarray}
which yields
\begin{eqnarray}
\label{eq: deri_J_hermi app}
I_{\alpha \bm{k}} &=& \left[ -2i\omega C_0 ( \cos k_x +\cos k_y -2  )   +(i\omega L_0)^{-1} \right] V_{\alpha \bm{k}}.
\end{eqnarray}

Summing up the contributions [Eqs.~(\ref{eq: deri_J_ Camp intra app}),~(\ref{eq: deri_J_ Camp inter app }),~and~(\ref{eq: deri_J_hermi app})], we end up with the admittance matrix~(\ref{eq: Jskin}).

\subsection{
Topological properties
}
\label{sec: vm of J app}

Here, we confirm that topological invariants take $(\nu_{\mathrm{tot}},\nu_{\mathrm{M}})=(0,-1)$ for $k_x=0,\pi$.
Firstly, we note that the winding number for each sector [Eq.~(\ref{eq: def vpm})] can be rewritten as
\begin{eqnarray}
\nu_{\pm} &=& \int \frac{dk_y}{2\pi} \sum_n \partial_{k_y} \mathrm{arg}(E_{\pm n(k_y)}-E_{\mathrm{pg}}),
\end{eqnarray}
where $E_{\pm n}$ ($n=1,2,\cdots, \mathrm{dim}H_{\pm}$) denotes the eigenvalue of the non-Hermitian Hamiltonian $H_{\pm}$.

Thus, by analyzing the momentum dependence of the argument of admittance eigenvalues, we obtain the topological invariants $\nu_{\mathrm{tot}}$ and $\nu_{M}$.

\begin{figure}[!h]
\begin{minipage}{0.44\hsize}
\begin{center}
\includegraphics[width=1\hsize,clip]{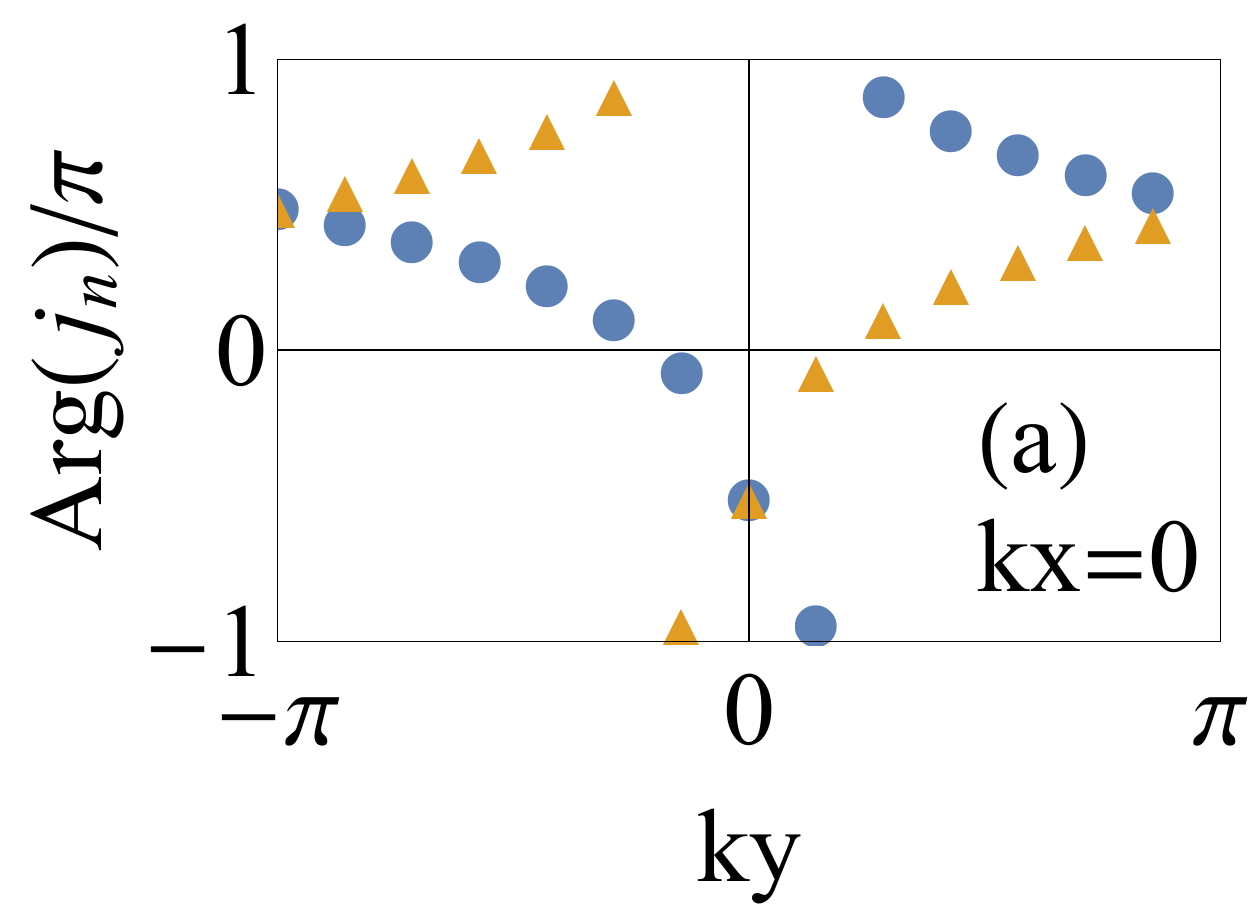}
\end{center}
\end{minipage}
\begin{minipage}{0.44\hsize}
\begin{center}
\includegraphics[width=1\hsize,clip]{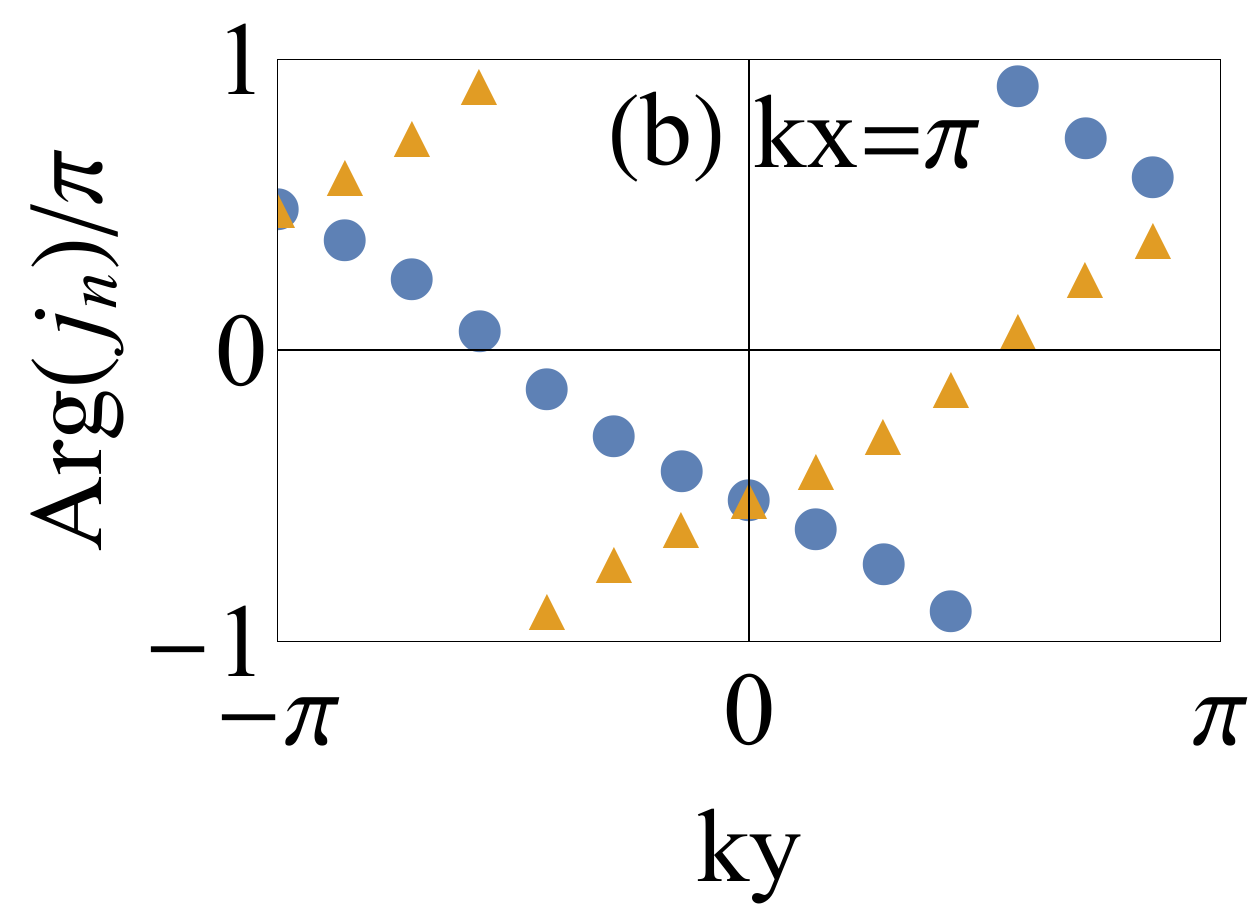}
\end{center}
\end{minipage}
\caption{(Color Online).
(a) [(b)]: Argument of the admittance eigenvalue $j(k_y)$ for $k_x=0$ ($k_x=\pi$).
Blue circles (orange triangles) represent the data for the plus and minus sector of the mirror operator $M_x$.
\green{
The data are obtained for $L_0=120\mathrm{\mu}\mathrm{H}$, $C_0=47\mathrm{n}\mathrm{F}$, $C_1=33\mathrm{n}\mathrm{F}$, and $f=117.4\mathrm{kHz}$, respectively.
}
}
\label{fig: ArgJ}
\end{figure}

In Fig.~\ref{fig: ArgJ}, the momentum dependence of the argument is plotted. Figure~\ref{fig: ArgJ}(a) indicates that the winding numbers take $(\nu_+,\nu_-)=(-1,1)$ for $k_x=0$ with $E_{\mathrm{pg}}=0$.
In a similar way, we can confirm that  the winding numbers take $(\nu_+,\nu_-)=(-1,1)$ for $k_x=0$ with $E_{\mathrm{pg}}=0.2i\mathrm{\Omega}^{-1}$.

\end{document}